%% file: NTT-survey-ACM.tex
\documentclass[]{acmart}

\setcopyright{none} 
\usepackage{amsmath,mathrsfs}
\usepackage{hyperref}
\hypersetup{hypertex=true,
	colorlinks=true,
	linkcolor=red,
	anchorcolor=blue,
	citecolor=blue}
\usepackage{bm}
\usepackage{enumitem,enumerate}
\usepackage{multirow,multicol,threeparttable,diagbox,makecell}
\usepackage{graphicx,float}
\usepackage{tikz,amsfonts,amstext}
\usepackage{subfigure}
\usepackage{fullpage}

\newcommand{\tabincell}[2]{\begin{tabular}{@{}#1@{}}#2\end{tabular}} 
\usepackage{threeparttable}

\begin{document}

\title{Number Theoretic Transform and Its Applications in Lattice-based Cryptosystems:\\A Survey}

\author{Zhichuang Liang}
\affiliation{
  \institution{Fudan University, Shanghai, China. } 
}

\author{Yunlei Zhao}
\affiliation{
	\institution{Fudan University, Shanghai, China; }  \institution{State Key Laboratory of Cryptology, Beijing, China.}
}

%
%
%
%

\begin{abstract}
	Number theoretic transform (NTT) is the most efficient method for multiplying two polynomials of high degree with integer coefficients, due to its series of advantages in terms of algorithm and implementation, and is consequently widely-used and particularly fundamental in the practical implementations of lattice-based cryptographic schemes. Especially, recent works have shown that NTT can be utilized in those schemes without NTT-friendly rings, and can outperform other multiplication algorithms. In this paper, we first review the basic concepts of polynomial multiplication, convolution and NTT. Subsequently, we systematically introduce basic radix-2 fast NTT algorithms in an algebraic way via Chinese Remainder Theorem. And then, we elaborate recent advances about the methods to weaken restrictions on parameter conditions of NTT. Furthermore, we systematically introduce how to choose appropriate strategy of NTT algorithms for the various given rings. Later, we introduce the applications of NTT in the lattice-based cryptographic schemes of NIST post-quantum cryptography standardization competition. Finally, we try to present some possible future research directions. 
\end{abstract}


\maketitle
\pagenumbering{Roman}
\textbf{Date}: \today

%
%
%
%

\newpage

\tableofcontents

%
%
%
%

\newpage
\pagenumbering{arabic}

\section{Introduction}

Most current public-key cryptographic schemes in use,  which are based on the hardness assumptions of factoring large integers and solving (elliptic curve) discrete logarithms, will suffer from quantum attack, if practical quantum computers are built. These cryptosystems play an important role in ensuring the confidentiality and authenticity of communications on the Internet. With the increasingly cryptographic security risks about quantum computing in recent years, post-quantum cryptography (PQC) has become a research focus for the crypto community. There are five main types of post-quantum cryptographic schemes: the hash-based, code-based, lattice-based, multivariable-based, and isogeny-based schemes, among which lattice-based cryptography is the most promising one due to its outstandingly balanced performance in security, communication bandwidth and computing efficiency. Most cryptographic primitives, such as public key encryption (PKE), key encapsulation mechanism (KEM), digital signature, key exchange, homomorphic encryption, etc, can be constructed based on lattices.

In the post-quantum cryptography standardization competition held by the US National Institute of Standards and Technology (NIST), lattice-based schemes account for 26 out of 64  schemes in the first round~\cite{nist-round-1-submissions}, 12 out of 26 in the second round~\cite{nist-round-2-submissions} and 7 out of 15 in the third round~\cite{nist-round-3-submissions}. NIST announced 4 candidates to be standardized~\cite{nist-to-be-standardized}, among which 3 schemes are based on lattices. Most of these lattice-based schemes are based on one of the following types: standard lattice, ideal lattice, NTRU lattice and module lattice. They are also instantiated based on the following hardness assumptions: Learning With Errors (LWE)~\cite{lwe-regev09} and its variants such as Ring-Learning With Errors (RLWE)~\cite{rlwe-LPR10} and Module-Learning With Errors (MLWE)~\cite{mlwe-LS15}, as well as the derandomized version of \{R,M\}LWE:  Learning With Rounding (LWR)~\cite{lwr-BPR12}, Ring-Learning With Rounding (RLWR)~\cite{lwr-BPR12} and Module-Learning With Rounding (MLWR)~\cite{mlwr-AA16,saber-DKR+18}. 

From a computational point of view, the fundamental and also time-consuming operation in many schemes based on standard lattices is matrix (vector) multiplication over a finite field, while that in schemes based on lattices with algebraic structures like the ideal, module, NTRU lattices is the multiplication of elements in the polynomial ring $\mathbb{Z}_q[x]/(\phi(x))$, where $\phi(x)$ is a polynomial with integer coefficients and $q$ is a positive integer. 
There are some schemes based on lattices with algebraic structures in NIST PQC competition: Kyber KEM~\cite{kyber-BDK+18,kyber-nist-round3}, Dilithium signature~\cite{dilithium-CHES18,dilithium-nist-round3} and Saber KEM~\cite{saber-DKR+18,saber-nist-round3} are based on module lattices, while Falcon signature~\cite{falcon-nist-round3}, NTRU KEM~\cite{ntru-nist-round3} and NTRU Prime KEM~\cite{ntru-prime-BCLV17,ntru-prime-nist-round3} are based on NTRU lattices. Among them, Kyber KEM, Dilithium signature and Falcon signature are standardized by NIST~\cite{nist-to-be-standardized}.

To compute polynomial multiplication, there are schoolbook algorithm~\cite{schoolbook}, Karatsuba~\cite{karatsuba62,karatsuba06}/ Toom-Cook~\cite{toom63,cook69} algorithm and DFT/FFT~\cite{fft-book-chu00,ct65,gs66}/NTT~\cite{ntt-Pol71,ntt-AB75} algorithm, among which the schoolbook algorithm is the most trivial and simplest, but with the quadratic complexity of $O(n^2)$, where $n$ is the length of polynomials. Karatsuba algorithm follows the ``divide and conquer'' technique, by dividing the original polynomial into two parts, resulting in the complexity of $O(n^{1.58})$. Toom-Cook algorithm is a generalization of the Karatsuba algorithm. Different from the Karatsuba algorithm, Toom-Cook algorithm divides the original polynomial into $k$ parts, giving the  complexity of $O(n^{r})$ where $r=\log_k(2k-1)$.

Discrete Fourier transform (DFT) can be utilized to multiply two polynomials, but the complexity of directly-computing DFT is $O(n^2)$, similar to that of schoolbook algorithm. In the past few decades, lots of fast algorithms to compute DFT have been proposed, which are called fast Fourier transform (FFT) at present. Actually, the idea of FFT was first developed by Carl Friedrich Gauss in his unpublished work in 1805~\cite{gauss1805}. It was not until 1965 that FFT attracted widespread attention, when Cooley and Tukey independently proposed Cooley-Tukey algorithm~\cite{ct65}, a fast algorithm of DFT. Number theoretic transform (NTT) is a special case of DFT over a finite field~\cite{ntt-Pol71}. Similarly, most FFT techniques can be applied to NTT, resulting with the corresponding fast algorithms for number theoretic transform. From an implementation point of view, FFT and NTT are the most efficient methods for computing polynomial multiplication of high degree, due to its quasilinear complexity $O(n \log n)$, which takes a significant advantage over all other multiplication algorithms. However, since FFT is performed over the complex field, the floating point operations during the computing process might cause errors about rounding precision when multiplying two polynomials with integer coefficients. In contrast, the operations during NTT computation are all performed with integers, which could avoid those precision errors. As most polynomials in lattice-based schemes are generated with integer coefficients, it is clear that NTT is especially suited for the implementation of those schemes. 


Most schemes based on lattices with algebraic structures prefer to NTT-based multiplication due to its many advantages (more details in section~\ref{sec-ntt-advantages}), apart from the quasilinear complexity and integer arithmetic. However, not all lattice-based schemes can use NTT directly, since NTT puts some restrictions on its parameter conditions. For example, NTT based on negative wrapped convolution (NWC) requires its length $n$ being a power of two and its modulus $q$ being a prime satisfying $q \equiv 1 \ (\bmod \ 2n)$. Among the standardized signature schemes in NIST PQC, both Dilithium and Falcon can meet the condition. The parameters of Kyber that is currently the only standardized KEM by NIST at most satisfy $q \equiv 1 \ (\bmod \ n)$, so it fails to utilize full NWC-based NTT. Neither does Saber, for the reason that Saber uses a power-of-two modulus $q$. The module reductions and rounding operations in Saber benefit from its power-of-two modulus, but its previous implementations also fail to use NTT, and Saber has to fall back to Karatsuba/Toom-Cook algorithm which is relatively less efficient in general. As for NTRU and NTRU Prime, they have NTT-unfriendly ring $\mathbb{Z}_q[x]/(\phi(x))$, where $\phi(x)$ has a prime degree, for which NWC-based NTT was not utilized for them at first. Therefore, it is very meaningful to generalize NTT methods to the case of NTT-unfriendly rings, so as to implement the underlying polynomial multiplication and obtain superior performance compared to the state-of-the-art implementations based on other multiplication algorithms. 

Moreover, as we observe, most NTT algorithms and their recent improvements contained in this paper have appeared in the literatures before, but until now they are not well summarized in other works and none has introduced their applications in lattice-based schemes. Besides, few literatures give systematic study of the mathematical principles, theoretic derivation, recent advances, and practical applications with respect to NTT. Those motivate our comprehensive work and systematization of NTT.  In this work, our contributions are listed as follows.

\begin{itemize}
	\item  We introduce the systematical knowledge of NTT, including the basic concepts, basic fast algorithms, computing advantages and computational complexity.
	
	\item  We review the recent advances about the methods to weaken restrictions on parameter conditions of NTT, present a taxonomy based on their strategies and make a comparison.
	
	\item  We propose high-level descriptions of some NTT techniques in a mathematically algebraic way for better understanding of the classification.
	
	\item  We classify the polynomial rings into three categories mainly, and systematically introduce how to choose appropriate strategy of NTT for the given ring.
	
	\item  As an important application, we introduce the recent advances of utilizing NTT in NIST PQC Round 3, especially the candidates with NTT-unfriendly rings. 
\end{itemize}

We hope that this paper will also provide researchers in relevant fields with basic but comprehensive mathematical knowledge of NTT. However, as for the limitations of this paper, we mainly focus on  those schemes based on lattices with algebraic structures from NIST PQC, because those schemes based on standard lattices don’t need NTT. In addition, recent optimized techniques for different platforms  (e.g., software/hardware) are not covered yet. But we leave it as a future work.


%
%
%
%

\section{Preliminaries}\label{sec-preliminaries}

In this section, we will define some notations, and give a brief introduction of polynomial multiplication and convolution.

\subsection{Notations and Definitions}

Let $ \mathbb{Z} $ be the ring of rational integers, $n$  and $q$ be some positive integers, and $ \mathbb{Z}_q \cong \{0,1,\ldots,q-1\}$. We write $ x' \equiv x$ $(\bmod$ $q)$ to mean that $x'-x$ is a multiple of $q$. We define $x'= x \bmod q$ to be the unique element $ x' $ in $\mathbb{Z}_q $ satisfying $ x' \equiv x$ $(\bmod$ $q)$. Polynomials are written as bold $\boldsymbol{a}$ or unbold ${a}(x)$. As for $\boldsymbol{a}$, $\boldsymbol{b}$ and $\boldsymbol{r}$, we also write $\boldsymbol{r} = \boldsymbol{a} \bmod \boldsymbol{b}$ to mean that $\boldsymbol{r}$ is the polynomial remainder of $\boldsymbol{a}$ divided by $\boldsymbol{b}$. For any $n_1,n_2 \in \mathbb{Z}$, $n_1>n_2$, and any $a_i$, define $\sum_{i=n_1}^{n_2}{a_i}=0$. The symbol ``$\circ$'' denotes the point-wise multiplication, i.e., multiplication of corresponding components.

\begin{definition}[Primitive and principal root of unity]
	Let $R$ be a commutative ring with multiplicative identity $1$, $k$ be a positive integer, and $\psi$ be an element in $R$. Define $\psi$ is the \textbf{primitive} $k$-th root of unity in $ R$, if and only if $\psi^k = 1$, and $\psi^i  \ne 1, i=1,2,\ldots,k-1$. Define $\psi$ is the \textbf{principal} $k$-th root of unity in $R$, if and only if $\psi^{k} = 1$, and $\sum_{j=0}^{k-1}{\psi^{jl}}=0, l=1,2,\ldots,k-1$. Notice that primitive and principle $k$-th root of unity coincide if $R=\mathbb{Z}_q$ where $q$ is a prime number. See~\cite{faster-integer-mul-Fur09,multi-moduli-NTT-ACC+21}  for more details. 		
\end{definition}

\begin{definition}[Bitreversal]
	Let $n$ be a power of two, and $b$ be a non-negative integer satisfying $ b<n $. The bitreversal of $b$ with respect to $n$ is defined as 
	$$\text{brv}_n ( b_{\log n -1} 2^{\log n -1} +  \ldots + b_{1} 2 + b_0 ) = b_0 2^{\log n -1} +  \ldots + b_{\log n -2} 2 + b_{\log n -1},$$
	where $b_i$ is the $i$-th bit of the binary expansion of $b$ . 	
\end{definition}

\subsection{Polynomial Rings}

Let $\mathbb{Z}[x]$ and $\mathbb{Z}_q[x]$ be the polynomial rings over $\mathbb{Z}$ and  $\mathbb{Z}_q$ respectively, with corresponding quotient rings $\mathbb{Z}[x]/(\phi(x))$ and $\mathbb{Z}_q[x]/(\phi(x))$, where $\phi(x)$ is a polynomial with integer coefficients. Especially, $\phi(x)$ is chosen to be a cyclotomic polynomial~\cite{cyclotomic-fields}. In this paper, we only take account of operations over $\mathbb{Z}_q$ unless otherwise noted, because the corresponding results over $\mathbb{Z}$ can be obtained easily. If $\deg \phi(x) =n$ holds, the element in $\mathbb{Z}_q[x]/(\phi(x))$, for example, $\boldsymbol{a}$, can be represented in the form of $\boldsymbol{a}=\sum_{i=0}^{n-1}{a_i x^i}$, or in the form of $\boldsymbol{a}=(a_0, a_1,\ldots,a_{n-1})$,  where $a_i \in \mathbb{Z}_q$. We will mainly focuses on $\mathbb{Z}_q[x]/(x^n-1)$ and $\mathbb{Z}_q[x]/(x^n+1)$, since they are widely used in lattice-based schemes. Let $n$ be a power of two, then $x^n+1$ is the $2n$-th cyclotomic polynomial.

\subsection{Polynomial Multiplication and Convolution}\label{sec-mul-over-ring}

Without loss of generality, we always considers $\boldsymbol{a}$ and $\boldsymbol{b}$ of degree $n-1$ in this paper. Pad them with zero if their lengths are less than $n$.

\begin{itemize}
	\item \textbf{Linear convolution.} Consider $\boldsymbol{c}= \boldsymbol{a} \cdot \boldsymbol{b} \in \mathbb{Z}_q[x]$, then $\boldsymbol{c} = \sum_{k=0}^{2n-2}{c_k x^k} \in \mathbb{Z}_q[x] $, where $c_k=\sum_{i+j=k}{a_i b_j} \bmod q, k=0,1,\ldots, 2n-2$. Here, $\boldsymbol{c}$ is referred to as the linear convolution of $\boldsymbol{a}$ and $\boldsymbol{b}$. Consider $\boldsymbol{c}= \boldsymbol{a} \cdot \boldsymbol{b} \in \mathbb{Z}_q[x]/(\phi(x))$. One can first compute $\boldsymbol{c}'= \boldsymbol{a} \cdot \boldsymbol{b}\in \mathbb{Z}_q[x] $, then $\boldsymbol{c}=\boldsymbol{c}' \bmod \phi(x)$. 

	\item \textbf{Cyclic convolution.} Consider $\boldsymbol{c}= \boldsymbol{a} \cdot \boldsymbol{b} \in \mathbb{Z}_q[x]/(x^n-1)$, then $\boldsymbol{c} = \sum_{k=0}^{n-1}{c_k x^k}$, where $c_k=\sum_{i=0}^{k}{a_i b_{k-i}} + \sum_{i=k+1}^{n-1}{a_i b_{k+n-i}}\bmod q, k=0,1,\ldots,n-1$. And $\boldsymbol{c}$ is referred to as the cyclic convolution (CC for short)\footnote{It is sometimes referred to as positive wrapped convolution.} of  $\boldsymbol{a}$ and $\boldsymbol{b}$. 

	\item \textbf{Negative wrapped convolution.} Consider $\boldsymbol{c}= \boldsymbol{a} \cdot \boldsymbol{b} \in \mathbb{Z}_q[x]/(x^n+1)$, then $\boldsymbol{c} = \sum_{k=0}^{n-1}{c_k x^k}$, where $c_k=\sum_{i=0}^{k}{a_i b_{k-i}} - \sum_{i=k+1}^{n-1}{a_i b_{k+n-i}}\bmod q, k=0,1,\ldots,n-1$. Here, $\boldsymbol{c}$ is referred to as their negative wrapped convolution (NWC for short)\footnote{It is sometimes referred to as negacyclic convolution, but here we refer to it as negative wrapped convolution to clearly distinguish it from cyclic convolution.}.
\end{itemize}

\section{Number Theoretic Transform (NTT)} \label{sec-ntt-definition}

In this section, we will introduce some basic concepts of number theoretic transform (NTT).
NTT is the special case of discrete Fourier transform (DFT) over a finite field~\cite{ntt-Pol71,ntt-AB75}.

\subsection{Cyclic Convolution-based NTT}

Here we introduce cyclic convolution-based NTT (CC-based NTT for short). $n$-point CC-based NTT has two parameters: the length or the point $n$, and the modulus $q$, where $n$ is a power of two and $q$ is a prime number satisfying $q \equiv 1 \ (\bmod \ n)$. It implies that the primitive $n$-th root of unity $\omega_n$ in $\mathbb{Z}_q$ exists. The forward transform, denoted by $\mathsf{NTT}$, is defined as: $\hat{\boldsymbol{a}} = \mathsf{NTT} (\boldsymbol{a} )$, where 
\begin{equation}\label{equ-classic-ntt}
	\hat{{a}}_{j} = \sum_{i=0}^{n-1}{{a}_{i} \omega_n^{ij}} \bmod q , j=0,1,\ldots,n-1.
\end{equation}

The inverse transform, denoted by $\mathsf{INTT}$, is defined as $\boldsymbol{a} = \mathsf{INTT} (\hat{\boldsymbol{a}})$, where 
\begin{equation}\label{equ-classic-intt}
	{{a}}_{i} = n^{-1} \sum_{j=0}^{n-1}{\hat{a}_{j} \omega_n^{-ij}} \bmod q , i=0,1,\ldots,n-1.
\end{equation}

Note that the inverse transform can be implemented by replacing the $\omega_n$ in $\mathsf{NTT}$ procedure with $\omega_n^{-1}$, followed by multiplying by a scale factor $n^{-1}$.  NTT and DFT share the same formula and similar properties, except that DFT has complex twiddle factors $\exp(-2 \pi i /n )$, while NTT uses integer primitive root of unity $\omega_n$. Some properties of $\mathsf{NTT}$ and $\mathsf{INTT}$ are listed as follows.

\begin{proposition}
	It always holds that $\boldsymbol{a} = \mathsf{INTT}(\mathsf{NTT} (\boldsymbol{a} ))$. 
\end{proposition}

\begin{proposition}[Cyclic convolution property~\cite{ntt-book-sun80}]\label{pro-cyclic-convolution}
	Let $\boldsymbol{c}$ be the cyclic convolution of $\boldsymbol{a}$ and $\boldsymbol{b}$, then it holds that
	$$\mathsf{NTT} (\boldsymbol{c})= \mathsf{NTT} \left( {\boldsymbol{a}} \right) \circ \mathsf{NTT} ( {\boldsymbol{b}} ).$$	
\end{proposition}

\subsection{Negative Wrapped Convolution-based NTT}

Here we introduce negative wrapped convolution-based NTT (NWC-based NTT for short). Moreover, the modulus $q$ is set to be a prime number satisfying $q \equiv 1 \ (\bmod \ 2n)$ such that the primitive $2n$-th root of unity $\psi_{2n}$ in $\mathbb{Z}_q$ exits. Take $\omega_n = \psi_{2n}^2 \bmod q$, and write $\boldsymbol{\psi}=(1,\psi_{2n},\psi_{2n}^{2},\ldots,\psi_{2n}^{n-1})$, $\boldsymbol{\psi}^{-1}=(1,\psi_{2n}^{-1},\psi_{2n}^{-2},\ldots,\psi_{2n}^{-(n-1)})$. Define $\bar{\boldsymbol{a}}=\boldsymbol{\psi} \circ \boldsymbol{a}$, where $\bar{a}_i = \psi_{2n}^{i} a_i$ in detail, which implies $\boldsymbol{a} = \boldsymbol{\psi}^{-1} \circ \bar{\boldsymbol{a}}$, where $a_i = \psi_{2n}^{-i} \bar{a}_i$. $n$-point NWC-based NTT is to integrate $\boldsymbol{\psi}$  (resp., $\boldsymbol{\psi}^{-1}$) into $\mathsf{NTT}$ (resp., $\mathsf{INTT}$), and denote them by $\mathsf{NTT}^{\psi}$  (resp., $\mathsf{INTT}^{\psi^{-1}}$), that is
\begin{align}
	\hat{\boldsymbol{a}}= & \mathsf{NTT}^{\psi}(\boldsymbol{a})= \mathsf{NTT}\left( \boldsymbol{\psi} \circ \boldsymbol{a}\right),\label{equ-classic-ntt-psi}                   \\
	\boldsymbol{a}=       & \mathsf{INTT}^{\psi^{-1}}(\hat{\boldsymbol{a}})= \boldsymbol{\psi}^{-1} \circ  \mathsf{INTT}(\hat{\boldsymbol{a}})\label{equ-classic-intt-psi}  .
\end{align}

More specifically, the forward transform $\hat{\boldsymbol{a}}=  \mathsf{NTT}^{\psi}(\boldsymbol{a})$ can be written as:
\begin{equation}
	\hat{{a}}_{j} = \sum_{i=0}^{n-1}{{a}_{i} \psi_{2n}^{i} \omega_n^{ij}} \bmod q , j=0,1,\ldots,n-1.
\end{equation}

The inverse transform $\boldsymbol{a}=\mathsf{INTT}^{\psi^{-1}}(\hat{\boldsymbol{a}})$ can be written as:
\begin{equation}
	{{a}}_{i} = n^{-1} \psi_{2n}^{-i} \sum_{j=0}^{n-1}{\hat{a}_{j} \omega_n^{-ij}} \bmod q , i=0,1,\ldots,n-1.
\end{equation}

Some properties of $\mathsf{NTT}^{\psi}$ and $\mathsf{INTT}^{\psi^{-1}}$ are listed as follows.
 
\begin{proposition}
	It always holds that $\boldsymbol{a} = \mathsf{INTT}^{\psi^{-1}}(\mathsf{NTT}^{\psi} (\boldsymbol{a} ))$. 
\end{proposition}

\begin{proposition}[Negative wrapped convolution property~\cite{fft-book-chu00}]\label{pro-negative-wrapped-convolution}	
	Let $\boldsymbol{c}$ be the negative wrapped convolution of  $\boldsymbol{a}$ and $\boldsymbol{b}$, then it holds that 
	$$\mathsf{NTT}^{\psi} ({\boldsymbol{c}})= \mathsf{NTT}^{\psi} \left( {\boldsymbol{a}} \right) \circ \mathsf{NTT}^{\psi} ( {\boldsymbol{b}} ).$$ 
\end{proposition}

\subsection{NTT-based Polynomial Multiplication}

NTT can be used to compute \{linear, cyclic, negative wrapped\} convolutions~\cite{convolution96}, which are equivalent to corresponding polynomial multiplications.

\begin{itemize}
	\item \textbf{Linear convolution-based polynomial multiplication.} To compute the linear convolution $\boldsymbol{c}= \boldsymbol{a} \cdot \boldsymbol{b} \in \mathbb{Z}_q[x] $, first, pad them to the length of $2n$ with zeros, resulting with ${\boldsymbol{a}'}=({a}_0, \ldots,{a}_{n-1},0,\ldots,0)$ and ${\boldsymbol{b}'}=({b}_0, \ldots,{b}_{n-1},0,\ldots,0)$. Second, use $2n$-point $\mathsf{NTT}/ \mathsf{INTT}$ for $\boldsymbol{c}= \mathsf{INTT} ( \mathsf{NTT} ( {\boldsymbol{a}'} ) \circ \mathsf{NTT} ( {\boldsymbol{b}'} ) )$. Moreover, to compute  $\boldsymbol{c}= \boldsymbol{a} \cdot \boldsymbol{b} \in \mathbb{Z}_q[x]/(\phi(x))$, one can compute $\boldsymbol{c}'= \boldsymbol{a} \cdot \boldsymbol{b}\in \mathbb{Z}_q[x] $ with  $2n$-point  $\mathsf{NTT}$/$\mathsf{INTT}$, followed by computing $\boldsymbol{c}=\boldsymbol{c}' \bmod \phi(x)$.

	
	\item \textbf{Cyclic convolution-based polynomial multiplication.} To compute the cyclic convolution  $\boldsymbol{c}= \boldsymbol{a} \cdot \boldsymbol{b} \in \mathbb{Z}_q[x]/(x^n-1) $, one can straightly use $n$-point $\mathsf{NTT}$/$\mathsf{INTT}$, according to cyclic convolution property:
	\begin{equation}\label{equ-cyclic-convolution-based-polymul}
		\boldsymbol{c}= \mathsf{INTT} \left( \mathsf{NTT} \left( {\boldsymbol{a}} \right) \circ \mathsf{NTT} ( {\boldsymbol{b}} ) \right).
	\end{equation}	
		
	\item \textbf{Negative wrapped convolution-based polynomial multiplication.} To compute the negative wrapped convolution $\boldsymbol{c}= \boldsymbol{a} \cdot \boldsymbol{b} \in \mathbb{Z}_q[x]/(x^n+1) $,  one can use the negative wrapped convolution property:
	\begin{equation} \label{equ-negative-wrapped-convolution-based-polymul}
		\boldsymbol{c}= \mathsf{INTT}^{\psi^{-1}} \left( \mathsf{NTT}^{\psi} \left( {\boldsymbol{a}} \right) \circ \mathsf{NTT}^{\psi} ( {\boldsymbol{b}} ) \right).
	\end{equation}
\end{itemize}

\subsection{Complexity}

The complexity of directly-computing NTT/INTT is $O(n^2)$. There are two $\mathsf{NTT}$s, one point-wise multiplication and one $\mathsf{INTT}$ for NTT-based multiplication. Therefore, the complexity of NTT-based multiplication without fast algorithms is $O(n^2)$.


\subsection{Advantages of NTT}\label{sec-ntt-advantages}

Here, let $\text{NTT}$/$\text{INTT}$ be any kind of forward/inverse transforms. 

\begin{itemize}
	\item Firstly, both $\text{NTT}$ and $\text{INTT}$ are linear transformations, based on which it can save $\text{INTT}$s in lattice-based schemes (e.g., Kyber~\cite{kyber-BDK+18,kyber-nist-round3} and Dilithium~\cite{dilithium-CHES18,dilithium-nist-round3}), i.e., 
	$$\sum_{i=0}^{l}{\boldsymbol{a}_i \boldsymbol{b}_i} = \sum_{i=0}^{l}{
		\text{INTT} \left( \text{NTT} \left( {\boldsymbol{a}_i} \right) \circ \text{NTT} ( {\boldsymbol{b}_i} ) \right) }=\text{INTT} (  \sum_{i=0}^{l}{
		\text{NTT} ( {\boldsymbol{a}_i} ) \circ \text{NTT} ( {\boldsymbol{b}_i} )  } ).$$	
	
	\item Additionally, consider $\boldsymbol{c}= \text{INTT} ( \text{NTT} ( {\boldsymbol{a}} ) \circ \text{NTT} ( {\boldsymbol{b}} ) )$, where $\boldsymbol{a}$ is random. Since the NTT transforms keep the randomness of a random polynomial, i.e., $\hat{\boldsymbol{a}}=\text{NTT} ({\boldsymbol{a}})$ is also random, one can directly generate a random polynomial, and view it as random $\hat{\boldsymbol{a}}$ already in the NTT domain, and compute  $\boldsymbol{c}= \text{INTT} ( \hat{\boldsymbol{a}} \circ \text{NTT} ( {\boldsymbol{b}} ) )$, thus eliminating the need for the forward transform.
	
	\item Besides, $\text{NTT}$ and $\text{INTT}$ preserve the dimension and bit length of all individual coefficients of a polynomial, i.e., $\hat{\boldsymbol{a}}$ and $\boldsymbol{a}$ share the same dimension and  bit length of any coefficient. Thus, $\hat{\boldsymbol{a}}$ can be stored where $\boldsymbol{a}$ is originally placed. 
	
	\item Finally, in some case where $\boldsymbol{a}$ involves in multiple multiplications, $\hat{\boldsymbol{a}}$ is computed once and stored for its use in subsequent multiplications, which can save forward transforms without any extra requirement of storage. 
\end{itemize}

\section{Some Tricks for Polynomial Multiplications}

In this subsection, we will show some useful tricks about polynomial multiplications.

\begin{definition}[One-iteration Karatsuba algorithm~\cite{karatsuba06}]\label{def-karatsuba-algorithm}
	Let $a,b,c,d$ be any numbers or polynomials. Briefly speaking, one-iteration Karatsuba algorithm implies that, to compute $t_1=a \cdot c$, $t_2=a \cdot d+b \cdot c$ and $ t_3=b \cdot d$, first compute $t_1$ and $ t_3$, and then compute $t_2$ by $t_2=(a+b) \cdot (c+d)-t_1-t_3$. One-iteration Karatsuba algorithm saves one multiplication at the cost of three extra additions (subtractions).
\end{definition}

\begin{definition}[Good's trick~\cite{ber01,good-trick51}]\label{def-good-trick}
	As for the ring $\mathbb{Z}_q[x]/(x^{h \cdot 2^k}-1)$ where $h$ is an odd number, Good's trick maps $\mathbb{Z}_q[x]/(x^{h \cdot 2^k}-1)$ to $(\mathbb{Z}_q[z]/(z^{2^k}-1))[y]/(y^{h}-1)$, where $\boldsymbol{a} = \sum_{l=0}^{h \cdot 2^k-1}a_l x^l$ is mapped to  $\sum_{l=0}^{{h \cdot 2^k}-1}a_l y^{(l \bmod h)} z^{(l \bmod 2^k)}=\sum_{i=0}^{h-1}\sum_{j=0}^{2^k-1}\tilde{a}_{i ,j} y^{i}z^{j}. $
	Write the coefficients $\tilde{a}_{i j}$ into a matrix $\tilde{A}=(\tilde{a}_{i ,j})_{h \times 2^k}$, and do $h$ parallel $2^k$-point NTT over $\mathbb{Z}_q[z]/(z^{2^k}-1)$ with each row. The corresponding point-wise multiplications are  $2^k$ parallel degree-($h-1$) polynomial multiplications in the ring $\mathbb{Z}_q[y]/(y^{h}-1)$ with each column from  $\tilde{A}$. Then we do $h$ parallel $2^k$-point INTT with each row. Denote the resulting matrix by $\tilde{C}=(\tilde{c}_{i ,j})_{h \times 2^k}$. Map $\sum_{i=0}^{h-1}\sum_{j=0}^{2^k-1}\tilde{c}_{i ,j} y^{i}z^{j}$ back to  $ \boldsymbol{c} = \sum_{l=0}^{h \cdot 2^k-1}c_l x^l \in \mathbb{Z}_q[x]/(x^{h \cdot 2^k}-1)$ according to the CRT formula $l=((2^k)^{-1}\bmod h)\cdot 2^k \cdot i+(h^{-1}\bmod 2^k)\cdot h\cdot j \bmod h \cdot 2^k$ to obtain $c_l x^l$ from $\tilde{c}_{i ,j} y^i z^j$.
\end{definition}

\begin{definition}[Sch$\ddot{\text{o}}$nhage’s trick~\cite{ber01,schonhage-trick77}]\label{def-schonhage-trick}
	Map the multiplicand $ \boldsymbol{a}=\sum_{i=0}^{2mn-1}a_i x^i   \in \mathbb{Z}_q[x]/(x^{2mn}-1)$  to  $\sum_{j=0}^{ 2n -1}{ (  \sum_{i=0}^{m-1}{a_{m \cdot j + i }  x^i } ) y^j}$ $\in (\mathbb{Z}_q[x][y]/(y^{2n}-1) )/(x^m-y)$ with $y=x^m$. To compute multiplication in $(\mathbb{Z}_q[x][y]/(y^{2n}-1) )/(x^m-y)$, one can first compute that in $\mathbb{Z}_q[x][y]/(y^{2n}-1)$, and then obtain the result modulo $(x^m-y)$. And to compute multiplication in $\mathbb{Z}_q[x]$ with multiplicands of degree less than $m$, we can do it in $\mathbb{Z}_q[x]/(x^{2m}+1)$ without modulo $(x^{2m}+1)$. Therefore, multiplication in $\mathbb{Z}_q[x][y]/(y^{2n}-1)$ can be computed in $(\mathbb{Z}_q[x]/(x^{2m}+1))[y]/(y^{2n}-1)$. Note that it is an NTT-friendly ring and $x$ is the primitive $4m$-th root of unity in $\mathbb{Z}_q[x]/(x^{2m}+1)$. 	
\end{definition}

\begin{definition}[Nussbaumer’s trick~\cite{ber01,nus80}]\label{def-nussbaumer-trick}
	Nussbaumer’s trick is similar to Sch$\ddot{\text{o}}$nhage’s trick. It maps $\boldsymbol{a}=\sum_{i=0}^{2mn-1}a_i x^i \in \mathbb{Z}_q[x]/(x^{2mn}+1)$ to $ \sum_{i=0}^{m-1}{ (  \sum_{j=0}^{ 2n -1}{a_{m\cdot j + i } y^j} )  x^i} \in (\mathbb{Z}_q[y]/(y^{2n}+1))[x]/(x^m-y)$ with $y=x^m$. To compute multiplication in $(\mathbb{Z}_q[y]/(y^{2n}+1))[x]/(x^m-y)$, first we compute multiplication in $(\mathbb{Z}_q[y]/(y^{2n}+1))[x]$, and then obtain the result modulo $(x^m-y)$. And to compute multiplication in $(\mathbb{Z}_q[y]/(y^{2n}+1))[x]$ with multiplicands of degree less than $n$, one can do it in $(\mathbb{Z}_q[y]/(y^{2n}+1))[x]/(x^{2n}-1)$ for $n\ge m$ without modulo $(x^{2n}-1)$. Note that it is an NTT-friendly ring  and  $y$ is the primitive $4n$-th  root of unity in $\mathbb{Z}_q[y]/(y^{2n}+1)$. 
\end{definition}

%
%
%
%

\section{Basic Radix-2 Fast Number Theoretic Transform} \label{sec-radix-2-NTT}

In this section, we will introduce the basic radix-2 fast number theoretic transform algorithms (radix-2 NTT for short), which are generalization of fast Fourier transform algorithms over a finite field. Here, ``radix-2'' means the length $n$ of NTT has a factor as a power of two, resulting that original algorithm can be divided into two parts of less length. During the practical computing process, there exists multiple different radix-2 NTT algorithms, among which we mainly focus on those widely used in lattice-based cryptographic schemes. All the radix-2 NTT algorithms in this section are summarized in Table~\ref{tab-raidx-2-fast-ntt} and their signal flows ($n=8$) are shown in Appendix~\ref{sec-signal-flow}. To describe NTT algorithms, there are two optional ways. One is from FFT perspectives, such as~\cite{ct65,gs66,ntt-merge-psi-RVM14,newhope-zhangneng20}. The other is from algebraical perspectives by Chinese Remainder Theorem, such as~\cite{ber01,ntt-in-saber-CHK+21,multi-moduli-NTT-ACC+21}. In fact, they are equivalent and the latter is the algebraic view of the former. We mainly follow algebraical perspectives in this section, since based on it we can describe recent improvements of NTT more succinctly. Interested readers can learn more details about FFT perspectives in Appendix~\ref{sec-radix-2-ntt-from-fft-perspectives}, or the works~\cite{fft-book-chu00,ber01,ct65,gs66,ntt-merge-psi-RVM14,newhope-zhangneng20}.

\input{table/tab-raidx-2-fast-ntt.tex}

\subsection{FFT Trick}\label{sec-fft-trick}

Bernstein~\cite{ber01} summarized and generalized FFT techniques from algebraical perspectives and used Chinese Remainder Theorem in ring form (or CRT for short, see Theorem~\ref{thm-crt-map-over-ring-init}) to describe FFT techniques that is referred to as FFT trick. 

\begin{theorem}[Chinese Remainder Theorem in ring form~\cite{ber01}]\label{thm-crt-map-over-ring-init}	
	Let $R$ be a commutative ring with multiplicative identity, $I_1,I_2,\ldots,I_k$ be ideals in $R$ that are pairwise co-prime, and $I$ be their intersection. Then there is a ring isomorphism:	
	\begin{equation}	
		\Phi:      R/I  \cong   R/I_1 \times R/I_2 \times \cdots \times R/I_k.
	\end{equation}	
\end{theorem}

In the work~\cite{ber01}, FFT trick means that according to Theorem~\ref{thm-crt-map-over-ring-init}, for polynomial rings $\mathbb{Z}_q[x]/(x^{2m} - \omega^2)$ where $m>0$ and invertible $\omega \in \mathbb{Z}_q$ , we have the following isomorphism: 
\begin{align*} 
		\begin{split}
			\Phi: \mathbb{Z}_q[x]/(x^{2m} - \omega^2  )  \  &\cong  \   \mathbb{Z}_q[x]/(x^{m} - \omega) \times \mathbb{Z}_q[x]/(x^{m} + \omega)\\
			\boldsymbol{a}  &\mapsto    \left(  \boldsymbol{a}'= \boldsymbol{a} \bmod x^{m} - \omega,   \boldsymbol{a}'' = \boldsymbol{a} \bmod x^{m} + \omega  \right) 
		\end{split}
\end{align*}
and the detailed mapping process:
\begin{align}
		&\Phi \left(  \sum\limits_{i=0}^{2m-1}{a_i x^i} \right) =\left( \sum\limits_{i=0}^{m-1}{ (a_i + \omega \cdot a_{i+m} ) x^i}, \sum\limits_{i=0}^{m-1}{ (a_i - \omega \cdot a_{i+m} ) x^i}    \right)  \label{equ-fft-trick-crt-map-level-0-forward}\\
		&\Phi^{-1} \left(  \sum\limits_{i=0}^{m-1}{ a_i' x^i},\sum\limits_{i=0}^{m-1}{ a_i'' x^i} \right) = \sum\limits_{i=0}^{m-1}{ \frac{1}{2}  (a_i'+a_i'') x^i} + \sum\limits_{i=0}^{m-1}{ \frac{ \omega^{-1} }{2}  (a_i' - a_i'')  x^{i+m}  }   \label{equ-fft-trick-crt-map-level-0-inverse}.
\end{align}

\begin{figure}[H]
	\centering
	\subfigure[Cooley-Tukey butterfly]{\label{fig-ct-gs-butterfly-a}
		\includegraphics[width=0.45\linewidth]{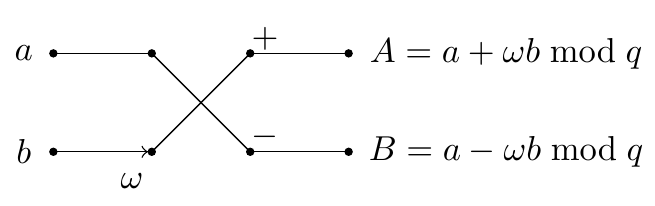}}
	\hspace{0.01\linewidth}
	\subfigure[Gentleman-Sande butterfly]{\label{fig-ct-gs-butterfly-b}
		\includegraphics[width=0.45\linewidth]{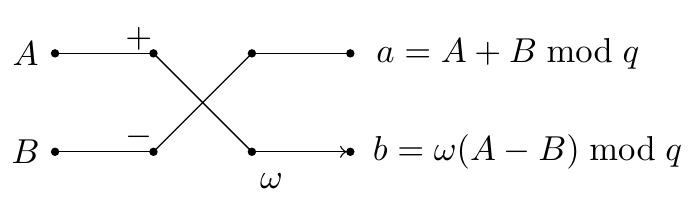}}
	\caption{Cooley-Tukey/Gentleman-Sande butterfly}
	\label{fig-ct-gs-butterfly}
\end{figure}

As for the forward FFT trick (see formula (\ref{equ-fft-trick-crt-map-level-0-forward})), it is very effective to compute $\boldsymbol{a}'$ and $\boldsymbol{a}'$. Their $i$-th coefficient can be computed via $a_i'=a_i + \omega \cdot a_{i+m},  a_i''=a_i - \omega \cdot a_{i+m} $, where $\omega \cdot a_{i+m} $ is computed once but used twice, $i=0,1,\ldots,m-1$. This type of operation is known as Cooley-Tukey butterfly or CT butterfly for short, which is illustrated in Figure~\ref{fig-ct-gs-butterfly-a}. It is the algorithm that first proposed by Cooley and Tukey~\cite{ct65}. The forward FFT trick totally takes $m$ multiplications, $m$ additions and $m$ subtractions. 

As for the inverse FFT trick  (see formula (\ref{equ-fft-trick-crt-map-level-0-inverse})), the $i$-th and ($i+\frac{n}{2}$)-th coefficient of $\boldsymbol{a}$ can be derived from the $i$-th coefficient of $\boldsymbol{a}'$ and  $\boldsymbol{a}''$ . The process is detailed below: $a_i= (a_i'+a_i'')/2 , a_{i+m} =   \omega^{-1} (a_i' - a_i'') /2$, $i=0,1,\ldots,m-1$. In the practical applications, the scale factor 2 can be omitted, with multiplying a total factor in the end (more details will be given below). This type of operation is known as Gentlemen-Sande butterfly or GS butterfly for short (see Figure~\ref{fig-ct-gs-butterfly-b}). It was first proposed by Gentlemen and Sande~\cite{gs66}. 

Based on FFT trick, we will introduce the fast algorithms to compute $\mathsf{NTT}$ and $\mathsf{INTT}$ over $\mathbb{Z}_q[x]/(x^n-1)$, as well as $\mathsf{NTT}^{\psi}$ and $\mathsf{INTT}^{\psi^{-1}}$ over $\mathbb{Z}_q[x]/(x^n+1)$ in the subsequent two subsections.

\subsection{FFT Trick for CC-based NTT over $\mathbb{Z}_q[x]/(x^n-1)$}

The work~\cite{ber01} introduced how to use FFT trick to compute $\mathsf{NTT}$ and $\mathsf{INTT}$ over $\mathbb{Z}_q[x]/(x^n-1)$, where $n$ is a power of two and $q$ is a prime number satisfying $q \equiv 1 \ (\bmod \ n)$. Denote by $\omega_{n}$ the primitive $n$-th root of unity in $\mathbb{Z}_q$. Here is the fast algorithm to compute $\mathsf{NTT}$ by using the forward FFT trick. It implies that for $\mathbb{Z}_q[x]/(x^n-1)$, we first have the following isomorphism: 
$$\mathbb{Z}_q[x]/(x^n-1)  \cong  \mathbb{Z}_q[x]/(x^{\frac{n}{2}} - 1) \times \mathbb{Z}_q[x]/(x^{\frac{n}{2}} + 1)$$

Notice that the forward FFT trick can be applied repeatedly to map $\mathbb{Z}_q[x]/(x^{\frac{n}{2}} \pm 1 ) $, according to the fact $x^{\frac{n}{2}} + 1=x^{\frac{n}{2}} - \omega_{n}^{\frac{n}{2}}$. In fact, $x^n-1$ has $n$ distinct roots in $\mathbb{Z}_q$, i.e.,  $\omega_{n}^{i},i=0,1,\ldots,n-1$. Therefore, forward FFT trick can be applied recursively from $\mathbb{Z}_q[x]/(x^n-1)$ all the way down to linear terms with the detailed mapping process of formula (\ref{equ-fft-trick-crt-map-level-0-forward}), which can be described via CRT mapping as:
\begin{equation}
	\begin{split}
		\mathbb{Z}_q[x]/(x^n -1) \cong \prod\limits_{i=0}^{n-1}{ \mathbb{Z}_q[x]/(x-\omega_{n}^{\text{brv}_n(i)} )} .
	\end{split}
\end{equation}

\begin{figure}[H]
	\centering
	\includegraphics[width=\linewidth]{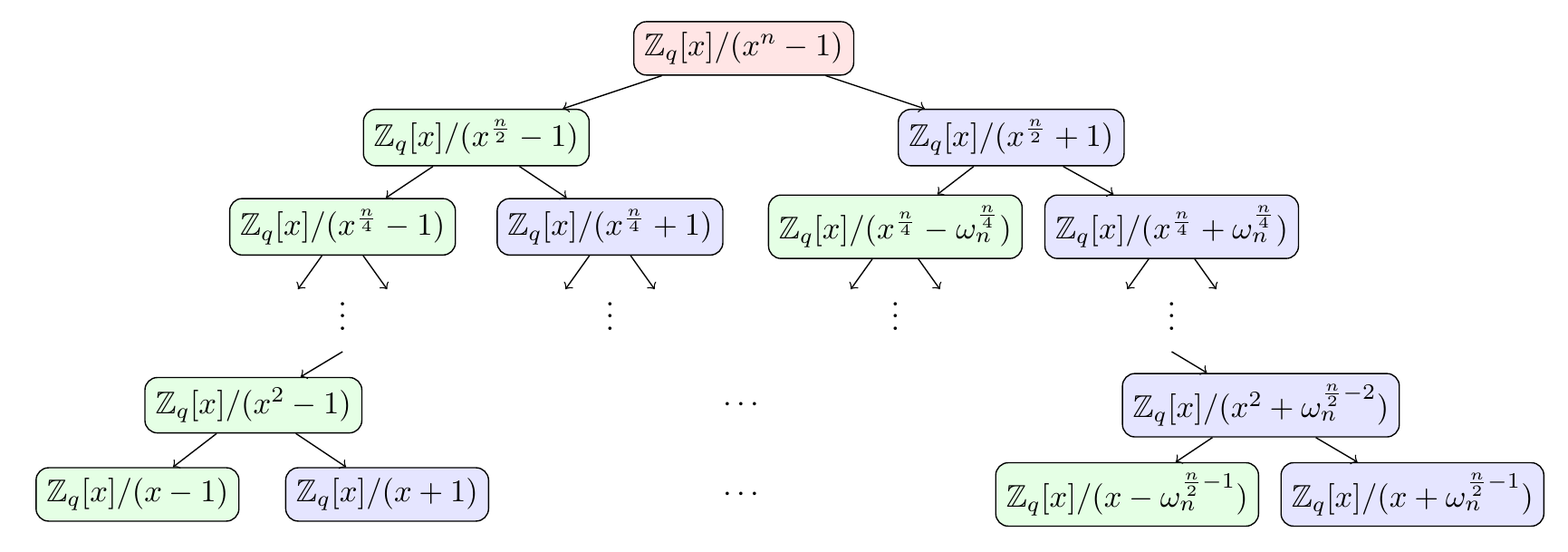}
	\caption{CRT map of FFT trick over $\mathbb{Z}_q[x]/(x^n-1)$}
	\label{fig-crt-map-complete-cyclic-ntt}
\end{figure}

Finally, $\boldsymbol{a}$ generates its images in $\mathbb{Z}_q[x]/(x - \omega_{n}^{\text{brv}_n(i)} )$, i.e.,  $\hat{{a}}_{\text{brv}_n(i)} $, which turns out to be the coefficient of $\hat{\boldsymbol{a}}$ indexed by $\text{brv}_n(i)$. Notice that using Cooley-Tukey algorithm, the coefficients of the input polynomials are indexed under natural order, while the coefficients of the output polynomials are indexed under bit-reversed order. In this paper, we follow the notations as used in~\cite{intt-merge-psi-POG15} which denotes this Cooley-Tukey $\mathsf{NTT}$ algorithm by $\mathsf{NTT}_{no \rightarrow bo}^{CT}$ where the subscripts $no \rightarrow bo$ indicates the input coefficients are under natural order and output coefficients are under bit-reversed  order. The signal flow of $\mathsf{NTT}_{no \rightarrow bo}^{CT}$ for $n=8$ can be seen in Figure~\ref{fig-classic-ntt-ct-gs-signal-flow-c} in Appendix~\ref{sec-signal-flow}. Adjust the input to bit-reversed order, then the Cooley-Tukey butterflies in the signal flow is changed elsewhere, as in Figure~\ref{fig-classic-ntt-ct-gs-signal-flow-a}. The output will be under natural order. This type of $\mathsf{NTT}$ is denoted by $\mathsf{NTT}_{bo \rightarrow no}^{CT}$. Figure~\ref{fig-crt-map-complete-cyclic-ntt} shows the detailed process of using FFT trick to map $\mathbb{Z}_q[x]/(x^n-1)$. The mapping process takes on the shape of a binary tree, with the root node being the $0$-th level and the leaf nodes being the ($\log n$)-th level. After the $k$-th level, $0 \le k < \log n $, there are $2^{k+1}$ nodes.

The fast algorithm to compute $\mathsf{INTT}$ can be obtained by iteratively inverting the CRT mappings mentioned above with formula (\ref{equ-fft-trick-crt-map-level-0-inverse}). In this case, the coefficients of the input polynomials are indexed under bit-reversed  order, i.e., $\hat{{a}}_{\text{brv}_n(i)}, i=0,1,\ldots,n-1$. Apply the inverse FFT trick (see formula (\ref{equ-fft-trick-crt-map-level-0-inverse})) to the computation from the ($k+1$)-th level to the $k$-th level, where $1 \le k < \log n $. Note that the scale factor 2 in each level of Gentleman-Sande butterfly can be omitted, with multiplying the final result by $n^{-1}$ in the end. The coefficients of the output polynomials are indexed under natural order. This type of Gentlemen-Sande $\mathsf{INTT}$ algorithm is denoted by $\mathsf{INTT}_{bo \rightarrow no}^{GS}$ (see  Figure~\ref{fig-classic-intt-ct-gs-signal-flow-d} for $n=8$). Adjust its input/output order and we can obtain $\mathsf{INTT}_{no \rightarrow bo}^{GS}$ (see Figure~\ref{fig-classic-intt-ct-gs-signal-flow-b}).

\subsection{FFT Trick for NWC-based NTT over $\mathbb{Z}_q[x]/(x^n+1)$}

The work~\cite{seiler18} proposed the way to use FFT trick to compute $\mathsf{NTT}^{\psi}$ and $\mathsf{INTT}^{\psi^{-1}}$ over $\mathbb{Z}_q[x]/(x^n+1)$, where $n$ is a power of two and $q$ is a prime number satisfying $q \equiv 1 \ (\bmod \ 2n)$. Since $\psi_{2n}^{n}=-1 \bmod q$, it holds that $x^n+1=x^n - \psi_{2n}^{n} =(x^{\frac{n}{2}} - \psi_{2n}^{\frac{n}{2}})(x^{\frac{n}{2}} + \psi_{2n}^{\frac{n}{2}}) \bmod q$. As for $\mathsf{NTT}^{\psi}$, the forward FFT trick implies that we have the following isomorphism: 
$$ \mathbb{Z}_q[x]/(x^n+1) \cong \mathbb{Z}_q[x]/(x^{\frac{n}{2}} - \psi_{2n}^{\frac{n}{2}}) \times \mathbb{Z}_q[x]/(x^{\frac{n}{2}} + \psi_{2n}^{\frac{n}{2}})$$

FFT trick can be applied repeatedly. Notice that $x^n+1$ has $n$ distinct roots in $\mathbb{Z}_q$, i.e.,  $\psi_{2n}^{2i+1},i=0,1,\ldots,n-1$. Therefore, there is a CRT isomorphism similarly.
\begin{equation}\label{equ-crt-map-level-all}
	\begin{split}
		\mathbb{Z}_q[x]/(x^n+1) \cong \prod_{i=0}^{n-1}{ \mathbb{Z}_q[x]/(x - \psi_{2n}^{2 \text{brv}_n(i)+1 } )  }   .
	\end{split}
\end{equation}

\begin{figure}[H]
	\centering
	\includegraphics[width=\linewidth]{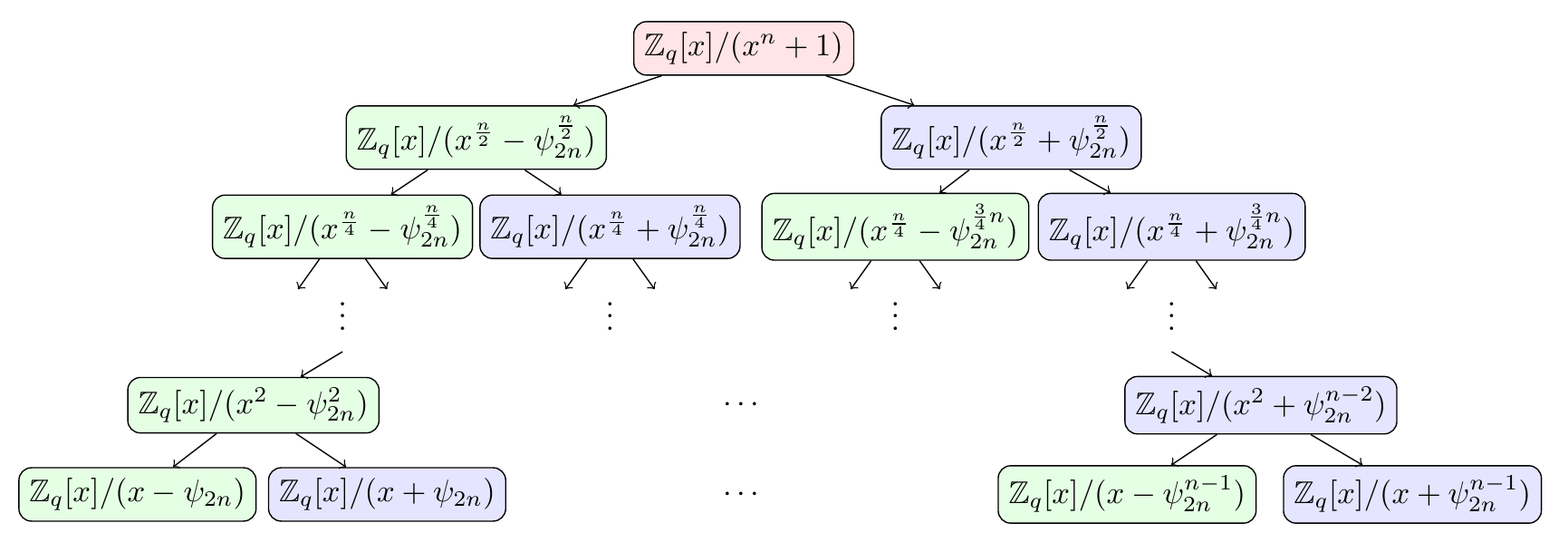}
	\caption{CRT map of FFT trick over $\mathbb{Z}_q[x]/(x^n+1)$}
	\label{fig-crt-map-complete-ntt}
\end{figure}

Figure~\ref{fig-crt-map-complete-ntt} shows the detailed process of using FFT trick to map $\mathbb{Z}_q[x]/(x^n+1)$. After the $k$-th level, where $0 \le k <\log n$, it produces $\mathbb{Z}_q[x]/ (x^{n/2^{k+1}}  \pm \psi_{2n}^{\text{brv}_n(2^k+i)} ) ,i =0,1,\ldots,2^k-1$ with pairs of rings. Such fast algorithm of $\mathsf{NTT}^{\psi}$ is denoted by $\mathsf{NTT}_{no \rightarrow bo}^{CT,\psi}$ (see Figure~\ref{fig-classic-ntt-intt-psi-signal-flow-c}). Similarly, by adjusting its input/output order, we can get $\mathsf{NTT}_{bo \rightarrow no}^{CT,\psi}$ (see Figure~\ref{fig-classic-ntt-intt-psi-signal-flow-a}).

Similarly, the fast algorithm to compute $\mathsf{INTT}^{\psi^{-1}}$ can be obtained by iteratively inverting the CRT mapping process with formula (\ref{equ-fft-trick-crt-map-level-0-inverse}). This type of fast algorithm for $\mathsf{INTT}^{\psi^{-1}}$ is denoted by $\mathsf{INTT}_{bo \rightarrow no}^{GS,\psi^{-1}}$. Adjust the input/output order and get $\mathsf{INTT}_{no \rightarrow bo}^{GS,\psi^{-1}}$. See Figure~\ref{fig-classic-ntt-intt-psi-signal-flow-b} and Figure~\ref{fig-classic-ntt-intt-psi-signal-flow-d}. Omitting the scale factor 2 in each level, we can multiply the final result by $n^{-1}$ in the end.

There is an alternative way to deal with the total factor $n^{-1}$. Zhang et al.~\cite{newhope-zhangneng20} noticed that $n^{-1}$ can both be integrated into  the computing process of each level, based on the fact that the scale factor 2 will be dealt with directly, by using addition and displacement (i.e., ``$>>$'') to compute $x/2 \bmod q$. When $x$ is even, $x/2 \equiv (x>>1) \bmod q$. When $x$ is odd, $x/2 \equiv (x>>1) + (q+1)/2 \bmod q$.

\subsection{Twisted FFT Trick}

The work~\cite{ber01} also summarized a variant of FFT trick, referred to as twisted FFT trick. It shows that Gentleman-Sande algorithm can be applied to compute $\mathsf{NTT}$, and Cooley-Tukey algorithm can be applied to compute $\mathsf{INTT}$. It means, mapping $\mathbb{Z}_q[x]/(x^n-1)  \cong  \mathbb{Z}_q[x]/(x^{\frac{n}{2}} - 1) \times \mathbb{Z}_q[x]/(x^{\frac{n}{2}} + 1)$ by CRT, followed by mapping $\mathbb{Z}_q[x]/(x^{\frac{n}{2}} + 1)$ via following isomorphism:
\begin{align}
	\begin{split}
		\Psi: \mathbb{Z}_q[x]/(x^{\frac{n}{2}} + 1) \  &\cong  \  \mathbb{Z}_q[x]/(x^{\frac{n}{2}} - 1)  \\
		x \  &\mapsto   \  \omega_{n} x
	\end{split}
\end{align}

Thus, as for $\mathbb{Z}_q[x]/(x^n-1)$, there is 
\begin{align}\label{equ-crt-map-twisted-level-0}
		\begin{split}
			\Psi\Phi: \mathbb{Z}_q[x]/(x^n-1)  \  &\cong  \  \mathbb{Z}_q[x]/(x^{\frac{n}{2}} - 1) \times \mathbb{Z}_q[x]/(x^{\frac{n}{2}} - 1 )\\
			\boldsymbol{a} \  &\mapsto   \  \left(  \boldsymbol{a}',\   \boldsymbol{a}''  \right) 
		\end{split}
\end{align}
and the detailed functioning process:
\begin{align}
		&(\Psi\Phi) \left(  \sum\limits_{i=0}^{n-1}{a_i x^i} \right) =\left( \sum\limits_{i=0}^{\frac{n}{2}-1}{ (a_i +  a_{i+\frac{n}{2}} ) x^i}, \sum\limits_{i=0}^{\frac{n}{2}-1}{ \omega_{n}^{i} \cdot (a_i - a_{i+\frac{n}{2}} ) x^i}    \right)  \label{equ-crt-map-twisted-level-0-forward}\\
		&(\Psi\Phi)^{-1} \left(  \sum\limits_{i=0}^{\frac{n}{2}-1}{ a_i' x^i},\sum\limits_{i=0}^{\frac{n}{2}-1}{ a_i'' x^i} \right) = \sum\limits_{i=0}^{\frac{n}{2}-1}{ \frac{a_i'+ \omega_{n}^{-i} \cdot a_i'' }{2} x^i} + \sum\limits_{i=0}^{\frac{n}{2}-1}{ \frac{ a_i' -  \omega_{n}^{-i} \cdot a_i'' }{2}  x^{i+\frac{n}{2}}  }   \label{equ-crt-map-twisted-level-0-inverse}.
\end{align}

\begin{figure}[H]
	\centering
	\includegraphics[width=\linewidth]{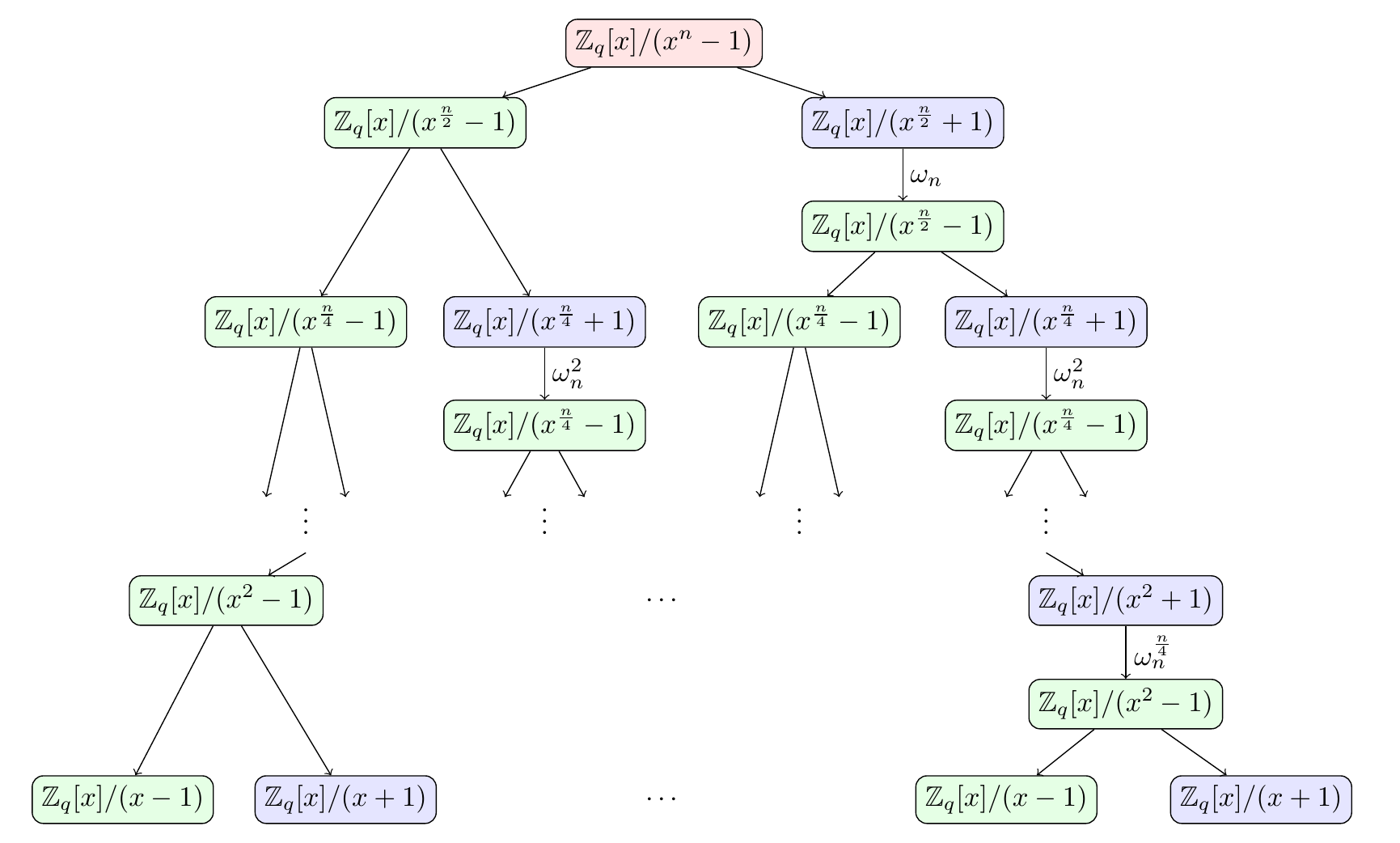}
	\caption{CRT map of twisted FFT trick over $\mathbb{Z}_q[x]/(x^n-1)$}
	\label{fig-crt-map-twisted-complete-ntt}
\end{figure}

As for the forward twisted FFT trick, for example, $\boldsymbol{a} \in \mathbb{Z}_q[x]/(x^n-1)$ in the $0$-th level generates $\boldsymbol{a}'$ and $\boldsymbol{a}''$, where $a_i'=a_i + a_{i+\frac{n}{2}},  a_i''=\omega_{n}^{i} \cdot( a_i - a_{i+\frac{n}{2}}) $, $i=0,1,\ldots,n/2-1$, where Gentleman-Sande algorithm are used. It can be applied repeatedly to map $ \mathbb{Z}_q[x]/(x^{\frac{n}{2}} - 1)$, and down to linear terms such as $\mathbb{Z}_q[x]/(x \mp 1)$. Specifically, in the $k$-th level, the similar isomorphism $\Psi: x \mapsto \omega_{n}^{2^{k-1}} x$ is applied from $ \mathbb{Z}_q[x]/(x^{n/2^{k}} + 1)$ to $ \mathbb{Z}_q[x]/(x^{n/2^{k}} - 1)$, $1 \le k <\log n$. The complete process of twisted FFT trick on mapping $\mathbb{Z}_q[x]/(x^n-1)$ is shown in Figure~\ref{fig-crt-map-twisted-complete-ntt}. Such GS $\mathsf{NTT}$ algorithm is denoted by $\mathsf{NTT}_{no \rightarrow bo}^{GS}$.  Adjust the input/output order, and we obtain  $\mathsf{NTT}_{bo \rightarrow no}^{GS}$. See Figure~\ref{fig-classic-ntt-ct-gs-signal-flow-d} and Figure~\ref{fig-classic-ntt-ct-gs-signal-flow-b}.

The inverse twisted FFT trick is computed in much the same way by iteratively inverting the above process, which is specified in  formula (\ref{equ-crt-map-twisted-level-0-inverse}). For example, the process of computing $\boldsymbol{a}$ in the $0$-th level from $\boldsymbol{a}'$ and $\boldsymbol{a}''$ in the first level with Cooley-Tukey butterflies is as follows: $a_i= (a_i'+ \omega_{n}^{-i} \cdot a_i'')/2 , a_{i+\frac{n}{2}} =  (a_i' - \omega_{n}^{-i} \cdot a_i'') /2$, $i=0,1,\ldots,n/2-1$. Such computing from the ($k+1$)-th level to the $k$-th level can be achieved in the same way, where $1 \le k < \log n $. The scale factor 2 in each level can be omitted, with multiplying the final result by $n^{-1}$ in the end. We denote this type of CT $\mathsf{INTT}$ by $\mathsf{INTT}_{bo \rightarrow no}^{CT}$ . Adjust its input/output order and the new transform is denoted by $\mathsf{INTT}_{no \rightarrow bo}^{CT}$. See Figure~\ref{fig-classic-intt-ct-gs-signal-flow-a} and Figure~\ref{fig-classic-intt-ct-gs-signal-flow-c}.

Notice that there only exists fast algorithm based on Cooley-Tukey butterfly for $\mathsf{NTT}^{\psi}$ , and that based on Gentleman-Sande butterfly for $\mathsf{INTT}^{\psi^{-1}}$. This is because, once we use Gentleman-Sande butterfly to compute $\mathsf{NTT}^{\psi}$ or use Cooley-Tukey butterfly to compute  $\mathsf{INTT}^{\psi^{-1}}$, the $\psi_{2n}$ term can not be further processed.

\subsection{In-place Operation, Reordering and Complexity} \label{sec-in-place-reorder-complexity}

\subsubsection{In-place operation} 
One can see from Figure~\ref{fig-ct-gs-butterfly-a} -~\ref{fig-ct-gs-butterfly-b}, Cooley-Tukey butterfly and Gentleman-Sande butterfly store the input and output data in the same address before and after the computing process, i.e., read data from some storage address for the computation, where the computing results are stored. This kind of operation is referred to as in-place operation. Obviously, there is no need of extra storage for in-place operation. In-place operation can be applied to all the radix-2 $\mathsf{NTT}$s/$\mathsf{INTT}$s as in Figure~\ref{fig-classic-ntt-ct-gs-signal-flow-a} -~\ref{fig-classic-ntt-intt-psi-signal-flow-d}. 

\subsubsection{Reordering} 
The input and output order of the polynomial coefficients have to be taken into consideration. Although the coefficients of the practical input polynomial are under natural order, the output after $\mathsf{NTT}_{no \rightarrow bo}^{CT}$/$\mathsf{NTT}_{no \rightarrow bo}^{GS}$ will end under bit-reversed order, which is the required order for input of $\mathsf{INTT}_{bo \rightarrow no}^{CT}$/$\mathsf{INTT}_{bo \rightarrow no}^{GS}$, but not the required one for input of $\mathsf{INTT}_{no \rightarrow bo}^{CT}$/$\mathsf{INTT}_{no \rightarrow bo}^{GS}$. In this case, extra reordering is needed from bit-reversed order to natural order. Besides, if $\mathsf{NTT}$ is conducted via $\mathsf{NTT}_{bo \rightarrow no}^{CT}$ and $\mathsf{NTT}_{bo \rightarrow no}^{GS}$, the input polynomial is supposed to be reordered from natural order to bit-reversed order. Similarly, there is a requirement on reordering the output polynomial of $\mathsf{INTT}_{no \rightarrow bo}^{CT}$ and $\mathsf{INTT}_{no \rightarrow bo}^{GS}$. In a word, there is a way for cyclic convolution-based polynomial multiplication without reordering, i.e., for  $\ddag,\natural \in \{ \text{CT,GS}\}$:
\begin{equation}
\boldsymbol{c}= \mathsf{INTT}_{bo \rightarrow no}^{\ddag} \left( \mathsf{NTT}_{no \rightarrow bo}^{\natural} \left( {\boldsymbol{a}} \right) \circ \mathsf{NTT}_{no \rightarrow bo}^{\natural} ( {\boldsymbol{b}} ) \right).
\end{equation}

Similarly, there is a way for NWC-based polynomial multiplication without extra reordering: 
\begin{equation}\label{equ-negative-wrapped-convolution-based-polymul-without-bitreversal}
	\boldsymbol{c}= \mathsf{INTT}_{bo \rightarrow no}^{GS,\psi^{-1}} \left( \mathsf{NTT}_{no \rightarrow bo}^{CT,\psi} \left( {\boldsymbol{a}} \right) \circ \mathsf{NTT}_{no \rightarrow bo}^{CT,\psi} ( {\boldsymbol{b}} ) \right).
\end{equation}

\subsubsection{Complexity} 
The complexities of NTT/INTT are given in Table~\ref{tab-ntt-complexity}. One can learn from Figure~\ref{fig-ct-gs-butterfly-a} that, each Cooley-Tukey butterfly consumes one multiplication and two additions (subtractions), where $\omega b$ is computed once and can be used twice. Similar analysis can also be applied to Gentleman-Sande butterfly. All the fast algorithms consist of $\log n$ levels, where there are $\frac{n}{2}$  butterfly operations on each level. As for the inverse transforms, their complexities require extra $n$ multiplications because of dealing with the scale factor $n^{-1}$. All the complexity of the polynomial multiplication based on these NTT fast algorithms is $O(n \log n)$, which has a significant advantage over that of polynomial multiplication based on directly-computing NTT/INTT, or any other polynomial multiplication algorithms such as the schoolbook algorithm and Karatsuba/Toom-Cook algorithm.

\input{table/tab-ntt-complexity.tex}

%
%
%
%

\section{Methods to Weaken Restrictions on Parameter Conditions of NTT}\label{sec-method-weaken-restrictions}

The full CC-based NTT requires that the parameter $n$ is a power of two and $q$ is prime satisfying $q \equiv 1 \ (\bmod \ n)$, while the full NWC-based NTT requires that the parameter $n$ is a power of two and $q$ is prime satisfying $q \equiv 1 \ (\bmod \ 2n)$. Traditionally, NTT puts some restrictions on its parameters. In recent years, many research efforts are made for NTT's restrictions on parameters and a series of methods have been proposed to weaken them. 

In this section, we mainly introduce the recent advances of weakening parameter restrictions with respect to $\mathbb{Z}_q[x]/(x^n \pm 1)$ where $n$ is a power of two. Those methods can be mainly classified into the following three categories: 

\begin{itemize}
	\item Method based on incomplete FFT trick;
	\item Method based on splitting polynomial ring;
	\item Method based on large modulus.
\end{itemize}

The first two methods are applied for the case that the modulus $q$ is an NTT-friendly prime of the form $q= q' \cdot 2^e+1$ but $q$ can not lead to a full NTT. The last method is applied for the case that the modulus $q$ is an NTT-unfriendly prime. The further classification  can be found in Figure~\ref{fig-weaken-restriction-methods}.

\subsection{Method Based on Incomplete FFT trick}\label{sec-incomplete-fft-trick}

We mainly introduce the method based on incomplete FFT trick over $\mathbb{Z}_q[x]/(x^n+1)$ where $n$ is a power of two and $q$ is NTT-friendly prime but does not satisfy  $q \equiv 1 \ (\bmod \ 2n)$. Then we extend the ring to $\mathbb{Z}_q[x]/(x^n-1)$ where $n$ is a power of two and $q$ is NTT-friendly prime but does not satisfy  $q \equiv 1 \ (\bmod \ n)$.

\input{figure/weaken-restriction-methods.tex}

\subsubsection{Method based on incomplete FFT trick over $\mathbb{Z}_q[x]/(x^n+1)$}\label{sec-incomplete-fft-trick-nwc}

Fully-mapping FFT trick means to map $\mathbb{Z}_q[x]/(x^n+1)$ down to linear terms, e.g., $\mathbb{Z}_q[x]/(x-\psi_{2n}^{2i+1} )$. See Figure~\ref{fig-crt-map-complete-ntt}. The condition $q \equiv 1 \ (\bmod \ 2n)$ is required such that the primitive $2n$-th root of unity $\psi_{2n}$ exits. Moenck~\cite{incomplete-ntt-moenck76} noticed that FFT trick does not have to map down to linear terms, and one can stop its mapping before the last $\beta$ level, $\beta=0,1,\ldots,\log n -1 $. Applying his method to FFT, Moenck named it mixed-basis FFT multiplications algorithm. Some recent researches introduced it to NTT~\cite{kyber-nist-round3,ntt-inscrypt20,newhope-compact,nttru,ntt-in-saber-CHK+21}. In detail, its CRT map of $\mathbb{Z}_q[x]/(x^n+1)$ is as follows:
$$   \mathbb{Z}_q[x]/(x^n+1)  \cong  \prod\limits_{i=0}^{n/2^{\beta}-1}{ \mathbb{Z}_q[x]/(x^{2^{\beta}} - \psi_{2n/{2^{\beta}}}^{2 \text{brv}_{n/{2^{\beta}}}(i)+1 } ) } .$$ 

It is referred to as ``Incomplete NTT'' in~\cite{ntt-in-saber-CHK+21} or ``Truncated-NTT'' in~\cite{ntt-inscrypt20}. The reason is that its CRT tree map is obtained by cropping the last  $\beta$ levels from fully-mapping ($\log n$)-level FFT trick tree map in Figure~\ref{fig-crt-map-complete-ntt}. Note that after forward transforms, ${\boldsymbol{a}}$'s images in $\mathbb{Z}_q[x]/(x^{2^{\beta}} - \psi_{2n/{2^{\beta}}}^{2 \text{brv}_{n/{2^{\beta}}}(i)+1 } )$ are degree-($2^{\beta}-1$) polynomials. The point-wise multiplication is performed about the corresponding degree-($2^{\beta}-1$) polynomials in $\mathbb{Z}_q[x]/(x^{2^{\beta}} - \psi_{2n/{2^{\beta}}}^{2 \text{brv}_{n/{2^{\beta}}}(i)+1 } )$. As for the inverse transforms, the scale factor 2 is omitted in every level, followed by multiplying by a total scalar ${(n/2^{\beta})}^{-1}$ in the end.

The forward/inverse transforms with $\beta$ levels cropped are denoted by $\mathsf{NTT}_{no \rightarrow bo,\beta}^{CT,\psi}$/$\mathsf{INTT}_{bo \rightarrow no,\beta}^{GS,\psi^{-1}}$ respectively, where $\beta=0,1,\ldots$, $\log n -1 $. Obviously, they are exactly $\mathsf{NTT}_{no \rightarrow bo}^{CT,\psi}$ and $\mathsf{INTT}_{bo \rightarrow no}^{GS,\psi^{-1}}$ if $\beta = 0$. The restriction on $n$ and $q$ can be weakened to $q \equiv 1 \ (\bmod \ \frac{2n}{2^{\beta}})$. The way to compute polynomial multiplication is the general form of formula  (\ref{equ-negative-wrapped-convolution-based-polymul-without-bitreversal}), that is 
\begin{equation}\label{equ-negative-wrapped-convolution-based-polymul-without-bitreversal-beta}
\boldsymbol{c}= \mathsf{INTT}_{bo \rightarrow no,\beta}^{GS,\psi^{-1}} \left( \mathsf{NTT}_{no \rightarrow bo,\beta}^{CT,\psi} \left( {\boldsymbol{a}} \right) \circ \mathsf{NTT}_{no \rightarrow bo,\beta}^{CT,\psi} ( {\boldsymbol{b}} ) \right).
\end{equation}

\textbf{A high-level description of incomplete FFT trick over $\mathbb{Z}_q[x]/(x^n+1)$.} Here, we propose a high-level description of incomplete FFT trick. We map $\mathbb{Z}_q[x]/(x^n+1)$ to  $\left( \mathbb{Z}_q [x] / (x^{2^\beta} - y ) \right)  [y]/(y^{\frac{n}{2^\beta}} + 1) $, along with rewriting $\boldsymbol{a} \in \mathbb{Z}_q[x]/(x^n+1)$ as $\boldsymbol{a} = \sum_{i=0}^{\frac{n}{2^{\beta}}-1} \tilde{\boldsymbol{a}}_i y^i$, where $y=x^{2^\beta}$ and $\tilde{\boldsymbol{a}}_i = \sum_{j=0}^{2^\beta-1} a_{2^\beta \cdot i + j}x^j \in \mathbb{Z}_q [x] / (x^{2^\beta} - y ) $. Thus, $\boldsymbol{a}$ can be seen as a polynomial of degree $(\frac{n}{2^{\beta}}-1)$ with respect to $y$. FFT trick will map
$$\left( \mathbb{Z}_q [x] / (x^{2^\beta} - y ) \right)  [y]/(y^{\frac{n}{2^\beta}} + 1)  \cong \prod\limits_{i=0}^{n/2^{\beta}-1}{ \left( \mathbb{Z}_q [x] / (x^{2^\beta} - y ) \right)  [y]/(y - \psi_{2n/{2^{\beta}}}^{2 \text{brv}_{n/{2^{\beta}}}(i)+1 } ) }.$$

Its forward transform (resp., inverse transform) is treated as radix-2 $\frac{n}{2^{\beta}}$-point full NWC-based $\mathsf{NTT}_{no \rightarrow bo}^{CT,\psi}$  (resp., $\mathsf{INTT}_{bo \rightarrow no}^{GS,\psi^{-1}}$) with respect to $y$. And the point-wise multiplication is performed in $\left( \mathbb{Z}_q [x] / (x^{2^\beta} - y ) \right)  [y]/(y - \psi_{2n/{2^{\beta}}}^{2 \text{brv}_{n/{2^{\beta}}}(i)+1 } )$.

\subsubsection{Method based on incomplete FFT trick over $\mathbb{Z}_q[x]/(x^n-1)$}
We follow the concepts from section~\ref{sec-incomplete-fft-trick-nwc}. The method based on incomplete FFT trick over $\mathbb{Z}_q[x]/(x^n-1)$ achieves its CRT map as follows:
$$   \mathbb{Z}_q[x]/(x^n-1)  \cong  \prod\limits_{i=0}^{n/2^{\beta}-1}{ \mathbb{Z}_q[x]/(x^{2^{\beta}} - \omega_{n/{2^{\beta}}}^{ \text{brv}_{n/{2^{\beta}}}(i) } ) } .$$ 
where $\beta=0,1,\ldots,\log n -1 $. Similarly, denote by $\mathsf{NTT}_{no \rightarrow bo,\beta}^{CT}$ and $\mathsf{INTT}_{bo \rightarrow no,\beta}^{GS}$ the forward and the inverse transform. The restriction on $q$ can be weakened to $q \equiv 1 \ (\bmod \ \frac{n}{2^{\beta}})$. 
Its way to compute $\boldsymbol{c}=\boldsymbol{a} \cdot \boldsymbol{b} \in \mathbb{Z}_q[x]/(x^n - 1)$  is 
$$\boldsymbol{c}= \mathsf{INTT}_{bo \rightarrow no,\beta}^{GS} \left( \mathsf{NTT}_{no \rightarrow bo,\beta}^{CT} \left( {\boldsymbol{a}} \right) \circ \mathsf{NTT}_{no \rightarrow bo,\beta}^{CT} ( {\boldsymbol{b}} ) \right).$$

Its high-level description can be written as: mapping $\mathbb{Z}_q[x]/(x^n-1)$ to  $\left( \mathbb{Z}_q [x] / (x^{2^\beta} - y ) \right)  [y]/(y^{\frac{n}{2^\beta}} - 1) $, followed by CRT isomorphism:
$$\left( \mathbb{Z}_q [x] / (x^{2^\beta} - y ) \right)  [y]/(y^{\frac{n}{2^\beta}} - 1)  \cong \prod\limits_{i=0}^{n/2^{\beta}-1}{ \left( \mathbb{Z}_q [x] / (x^{2^\beta} - y ) \right)  [y]/(y- \omega_{n/{2^{\beta}}}^{ \text{brv}_{n/{2^{\beta}}}(i) }) }.$$

\subsection{Method Based on Splitting Polynomial Ring}


We found that the computing strategies of Pt-NTT~\cite{ptntt-inscrypt18}, K-NTT~\cite{ptntt-eprint19,kntt-icics21} and H-NTT~\cite{ntt-inscrypt20}  are  similarly dependent on splitting initial polynomial ring, based on which we classify them into the category named the \textbf{method based on splitting polynomial ring}. And their basic idea can be traced back to Nussbaumer's trick~\cite{ber01,nus80} (see Definition~\ref{def-nussbaumer-trick}). Following the description of Nussbaumer's trick, they can essentially be described by the following isomorphism. Let $\alpha$ be a non-negative integer.
\begin{align}
\begin{split}
\Psi_\alpha: \mathbb{Z}_q[x]/(x^n \pm 1) \  &\cong  \  \left( \mathbb{Z}_q[y]/(y^{\frac{n}{2^\alpha}} \pm 1) \right)  [x] / (x^{2^\alpha} - y ) \\
\boldsymbol{a}=\sum\limits_{i=0}^{n-1}{a_i x^i} \  &\mapsto   \  \Psi_\alpha(\boldsymbol{a})= \sum\limits_{i=0}^{2^\alpha-1}{ \left(  \sum\limits_{j=0}^{\frac{n}{2^\alpha}-1}{a_{2^\alpha \cdot j + i } y^j} \right)  x^i}
\end{split}
\end{align}

Obviously, the isomorphism $\Psi_\alpha$ and its inverse $\Psi_\alpha^{-1}$ only perform simple reordering of the polynomial coefficients. $\Psi_\alpha$ is the identity mapping if $\alpha=0$. Briefly speaking, the general form of $\alpha$-round method based on splitting polynomial ring to compute $\boldsymbol{c}=\boldsymbol{a} \cdot \boldsymbol{b} \in \mathbb{Z}_q[x]/(x^n \pm 1)$ mainly contains the following three steps, where $n$ is a power of two and $q$ is a prime number (more details about $q$ can be seen below).

\begin{itemize}
	\item \textbf{Step 1, Splitting.} The polynomials $\boldsymbol{a}$ and $\boldsymbol{b}$ are split by $\Psi_\alpha$ into:
	$$\Psi_\alpha(\boldsymbol{a})=\sum\limits_{i=0}^{2^{\alpha}-1} \tilde{\boldsymbol{a}}_{i} \cdot x^i  ,\Psi_\alpha(\boldsymbol{b})=\sum\limits_{i=0}^{2^{\alpha}-1} \tilde{\boldsymbol{b}}_{i} \cdot x^i \in  \left( \mathbb{Z}_q[y]/(y^{\frac{n}{2^\alpha}} \pm 1) \right)  [x] / (x^{2^\alpha} - y ) ,$$ 
	where $y=x^{2^\alpha}$, and 
	$$\tilde{\boldsymbol{a}}_{i} =  \sum\limits_{j=0}^{\frac{n}{2^\alpha}-1}{a_{2^\alpha \cdot j + i } y^j}, \tilde{\boldsymbol{b}}_{i} =  \sum\limits_{j=0}^{\frac{n}{2^\alpha}-1}{b_{2^\alpha \cdot j + i } y^j} \in \mathbb{Z}_q[y]/(y^{\frac{n}{2^\alpha}} \pm 1).$$

	\item  \textbf{Step 2, Multiplication.} The product of $\Psi_\alpha(\boldsymbol{a})$ and $\Psi_\alpha(\boldsymbol{b})$ is obtained by 
	$ ( \sum\limits_{i=0}^{2^{\alpha}-1} \tilde{\boldsymbol{a}}_{i} \cdot x^i  ) ( \sum\limits_{i=0}^{2^{\alpha}-1} \tilde{\boldsymbol{b}}_{i} \cdot x^i  ) \bmod  x^{2^\alpha} - y ,$ 
	which means that one need to compute $\tilde{\boldsymbol{c}}_{i} \in \mathbb{Z}_q[y]/(y^{\frac{n}{2^\alpha}} \pm 1)$  for $ i= 0,1, \ldots , 2^{\alpha}-1$ as follows:
	\begin{align}\label{equ-nussbaumer-step-2-ci}
	\begin{split}
	\tilde{\boldsymbol{c}}_{i} &=  \sum_{l=0}^{i}\tilde{\boldsymbol{a}}_{l} \cdot \tilde{\boldsymbol{b}}_{i-l}+\sum_{l=i+1}^{2^{\alpha}-1} y \cdot \tilde{\boldsymbol{a}}_{l} \cdot \tilde{\boldsymbol{b}}_{2^{\alpha}+i-l} \in \mathbb{Z}_q[y]/(y^{\frac{n}{2^\alpha}} \pm 1).
	\end{split}
	\end{align}
	
	\item \textbf{Step 3, Gatheration.} Gather all the $\tilde{\boldsymbol{c}}_{i}$ by $\Psi_\alpha^{-1}$, and obtain $\boldsymbol{c}= \Psi_\alpha^{-1} \left(  \sum\limits_{i=0}^{2^{\alpha}-1}  \tilde{\boldsymbol{c}}_{i} \cdot x^i  \right) $.
\end{itemize}

Step 1 and Step 3 are simple and easy. Essentially, $\Psi_\alpha$ transforms NTT/INTT over $\mathbb{Z}_q[x]/(x^n \pm 1)$ into those over $\mathbb{Z}_q[y]/(y^{\frac{n}{2^\alpha}} \pm 1)$ which requires only $\frac{n}{2^\alpha}$-point NTT/INTT with arbitrary appropriate modulus $q$, but the point-wise multiplication needs to be adapted to NTT/INTT. There are three variants based on the method based on splitting polynomial ring, including Pt-NTT, K-NTT, H-NTT. The main difference between them is that they use different skills and NTTs to compute $\tilde{\boldsymbol{c}}_{i}$ in formula (\ref{equ-nussbaumer-step-2-ci}) of Step 2.

\subsubsection{Pt-NTT}

Preprocess-then-NTT (Pt-NTT) proposed by Zhou et al.~\cite{ptntt-inscrypt18} improves formula (\ref{equ-nussbaumer-step-2-ci}) as follows: 
\begin{align*}
	\begin{split}
	\tilde{\boldsymbol{c}}_{i} &=  \sum_{l=0}^{i}\tilde{\boldsymbol{a}}_{l} \cdot \tilde{\boldsymbol{b}}_{i-l}+\sum_{l=i+1}^{2^{\alpha}-1} \vec{\boldsymbol{a}}_{l} \cdot \tilde{\boldsymbol{b}}_{2^{\alpha}+i-l} \\
	&=  \text{INTT}\left(\sum_{l=0}^{i}\text{NTT}(\tilde{\boldsymbol{a}}_{l})\circ \text{NTT}(\tilde{\boldsymbol{b}}_{i-l})
	+\sum_{l=i+1}^{2^{\alpha}-1} \text{NTT}(\vec{\boldsymbol{a}}_{l})\circ \text{NTT}(\tilde{\boldsymbol{b}}_{2^{\alpha}+i-l}) \right),
	\end{split}
\end{align*}
where ``$\circ$'' is the corresponding point-wise multiplication, and for $\mathbb{Z}_q[y]/(y^{\frac{n}{2^\alpha}} + 1)$, there is
$$\vec{\boldsymbol{a}}_{l} = y \cdot \tilde{\boldsymbol{a}}_{l} = -a_{n - 2^\alpha + l } +  \sum_{j=0}^{\frac{n}{2^\alpha}-2}{a_{2^\alpha \cdot j + l } y^{j+1} } \in \mathbb{Z}_q[y]/(y^{\frac{n}{2^\alpha}} + 1),$$ 
or, for $\mathbb{Z}_q[y]/(y^{\frac{n}{2^\alpha}} - 1)$, there is
$$\vec{\boldsymbol{a}}_{l} = y \cdot \tilde{\boldsymbol{a}}_{l} = a_{n - 2^\alpha + l } +  \sum_{j=0}^{\frac{n}{2^\alpha}-2}{a_{2^\alpha \cdot j + l } y^{j+1} } \in \mathbb{Z}_q[y]/(y^{\frac{n}{2^\alpha}} - 1).$$ 

Here, Pt-NTT uses $\frac{n}{2^\alpha}$-point full NWC-based NTT/INTT over $\mathbb{Z}_q[y]/(y^{\frac{n}{2^\alpha}} + 1) $, or $\frac{n}{2^\alpha}$-point full CC-based NTT/INTT over $\mathbb{Z}_q[y]/(y^{\frac{n}{2^\alpha}} - 1) $.

\subsubsection{K-NTT}

Later, Zhu et al.~\cite{ptntt-eprint19,kntt-icics21} proposed Karatsuba-NTT (K-NTT) based on Pt-NTT, equipping with one-iteration Karatsuba algorithm (see Definition \ref{def-karatsuba-algorithm}). Its Step 2 is given as: 
\begin{align*}
	\begin{split}
	\tilde{\boldsymbol{c}}_{i} &=  \sum_{l=0}^{i}\tilde{\boldsymbol{a}}_{l} \cdot \tilde{\boldsymbol{b}}_{i-l}+\sum_{l=i+1}^{2^{\alpha}-1} y \cdot \tilde{\boldsymbol{a}}_{l} \cdot \tilde{\boldsymbol{b}}_{2^{\alpha}+i-l} \\
	&=  \text{INTT}\left(\sum_{l=0}^{i}\text{NTT}(\tilde{\boldsymbol{a}}_{l})\circ \text{NTT}(\tilde{\boldsymbol{b}}_{i-l})
	+\sum_{l=i+1}^{2^{\alpha}-1}\text{NTT}(y)\circ \text{NTT}(\tilde{\boldsymbol{a}}_{l})\circ \text{NTT}(\tilde{\boldsymbol{b}}_{2^{\alpha}+i-l}) \right).
	\end{split}
\end{align*}
	
Here, K-NTT uses $\frac{n}{2^\alpha}$-point full NWC-based NTT/INTT over $\mathbb{Z}_q[y]/(y^{\frac{n}{2^\alpha}} + 1) $, or $\frac{n}{2^\alpha}$-point full CC-based NTT/INTT over $\mathbb{Z}_q[y]/(y^{\frac{n}{2^\alpha}} - 1) $. 	
Since $y$ has been known, $\text{NTT}(y)$ can be computed and stored offline in advance. In Step 2, one-iteration Karatsuba algorithm is used in such a manner: First compute and store $\text{NTT}(\tilde{\boldsymbol{a}}_{i})\circ \text{NTT}(\tilde{\boldsymbol{b}}_{i})$ for any $i=j$, and then we have $ \text{NTT}(\tilde{\boldsymbol{a}}_{i})\circ \text{NTT}(\tilde{\boldsymbol{b}}_{j})+\text{NTT}(\tilde{\boldsymbol{a}}_{j})\circ \text{NTT}(\tilde{\boldsymbol{b}}_{i})
=(\text{NTT}(\tilde{\boldsymbol{a}}_{i})+ \text{NTT}(\tilde{\boldsymbol{a}}_{j}))\circ(\text{NTT}(\tilde{\boldsymbol{b}}_{i})+\text{NTT}(\tilde{\boldsymbol{b}}_{j})) -\text{NTT}(\tilde{\boldsymbol{a}}_{i})\circ \text{NTT}(\tilde{\boldsymbol{b}}_{i})- \text{NTT}(\tilde{\boldsymbol{a}}_{j})\circ \text{NTT}(\tilde{\boldsymbol{b}}_{j})$ for any $i \ne j$.

\subsubsection{H-NTT}

Furthermore, Liang et al.~\cite{ntt-inscrypt20} improved K-NTT and proposed a new variant of NTT referred to as Hybrid-NTT (H-NTT), by applying truncated-NTT in Step 2 and one-iteration Karatsuba algorithm in its point-wise multiplication. H-NTT uses truncated-NTT with $\beta$ levels cropped, instead of those full NTT in K-NTT. Its Step 2 is given as:
\begin{align*}
\begin{split}
	\tilde{\boldsymbol{c}}_{i} &=  \sum_{l=0}^{i}\tilde{\boldsymbol{a}}_{l} \cdot \tilde{\boldsymbol{b}}_{i-l}+\sum_{l=i+1}^{2^{\alpha}-1} y \cdot \tilde{\boldsymbol{a}}_{l} \cdot \tilde{\boldsymbol{b}}_{2^{\alpha}+i-l}  \\
	& = \mathsf{INTT}_{\beta} \bigg(\sum_{l=0}^{i}\mathsf{NTT}_{\beta}(\tilde{\boldsymbol{a}}_{l})\circ \mathsf{NTT}_{\beta}(\tilde{\boldsymbol{b}}_{i-l})+\sum_{l=i+1}^{2^{\alpha}-1} \mathsf{NTT}_{\beta}(y)\circ \mathsf{NTT}_{\beta}(\tilde{\boldsymbol{a}}_{l})\circ \mathsf{NTT}_{\beta}(\tilde{\boldsymbol{b}}_{2^{\alpha}+i-l}) \bigg),
\end{split}
\end{align*}	
where $\mathsf{NTT}_{\beta}$/$\mathsf{INTT}_{\beta}$ means NWC-based $\mathsf{NTT}_{no \rightarrow bo,\beta}^{CT,\psi}$/$\mathsf{INTT}_{bo \rightarrow no,\beta}^{GS,\psi^{-1}}$ over $\mathbb{Z}_q[y]/(y^{\frac{n}{2^\alpha}} + 1) $, or CC-based $\mathsf{NTT}_{no \rightarrow bo,\beta}^{CT}$/$\mathsf{INTT}_{bo \rightarrow no,\beta}^{GS}$ over $\mathbb{Z}_q[y]/(y^{\frac{n}{2^\alpha}} - 1) $. One-iteration Karatsuba algorithm is also used in its point-wise multiplication. For example, to compute $(\sum_{i=0}^{2^{\beta}-1}{ \hat{{a}}_{i} x^{i}  })(\sum_{i=0}^{2^{\beta}-1}{ \hat{{b}}_{i} x^{i}  })$ $\bmod \ x^{2^{\beta}} - \psi $, one can compute $\hat{{a}}_{i} \hat{{b}}_{j} $ for any $i=j$ first and then compute $\hat{{a}}_{i} \hat{{b}}_{j} + \hat{{a}}_{j} \hat{{b}}_{i} = (\hat{{a}}_{i}+\hat{{a}}_{j})(\hat{{b}}_{i}+\hat{{b}}_{j})- \hat{{a}}_{i} \hat{{b}}_{i} - \hat{{a}}_{j} \hat{{b}}_{j}$ for any $i \ne j$.

\subsubsection{Comparisons and discussions}

Based on the high-level description of incomplete FFT trick, we can see that it shares some similarities with the method based on splitting polynomial ring. In fact, there is an isomorphism between $( \mathbb{Z}_q [x] / (x^{2^\beta} - y ) )  [y]/(y^{\frac{n}{2^\beta}} \pm 1) $ and $( \mathbb{Z}_q[y]/(y^{\frac{n}{2^\alpha}} \pm 1) )  [x] / (x^{2^\alpha} - y)$ for any $\alpha=\beta$. And they have been proved computationally equivalent~\cite{ntt-inscrypt20}, which implies that their efficiencies are the same theoretically. 

These two methods can expand the value range of modulus $q $, because $q$ can only satisfy $q \equiv 1 \ (\bmod \frac{2n}{2^{\alpha+\beta}})$ for some $\alpha,\beta$, instead of $q \equiv 1 \ (\bmod \ 2n)$ when $n $ is fixed, for NWC-based NTT; besides, $q$ can only satisfy $q \equiv 1 \ (\bmod \frac{n}{2^{\alpha+\beta}})$ for some $\alpha,\beta$, instead of $q \equiv 1 \ (\bmod \ n)$ when $n $ is fixed, for CC-based NTT. 

However, the limitations on $q$ can not be ignored, since $q$ must be an NTT-friendly prime such that $\mathbb{Z}_q$ is a finite field and $x^n \pm 1$ can be split into polynomials of small degree over $\mathbb{Z}_q$. In addition, as for the method based on splitting polynomial ring, the polynomial operations over $\mathbb{Z}_q[x]/(x^n \pm 1)$ can be transformed into those over a smaller ring $\mathbb{Z}_q[y]/(y^{\frac{n}{2^\alpha}} \pm 1)$. It leads to more modular implementation, since NTT over $\mathbb{Z}_q[y]/(y^{\frac{n}{2^\alpha}} \pm 1)$ can be implemented as a ``black box'' (notice that we don't limit the types of NTT here), and be re-called when required in Step 2.

Both Pt-NTT and K-NTT use $\frac{n}{2^\alpha}$-point full NTT in their Step 2. It requires that the primitive $\frac{n}{2^{\alpha-1}}$-th root of unity exits for NWC-based NTT, or the primitive $\frac{n}{2^{\alpha}}$-th root of unity exits for CC-based NTT. Therefore, the parameter condition of Pt-NTT and K-NTT can be weakened to $q \equiv 1 \ (\bmod \frac{2n}{2^{\alpha}})$ for the initial ring $\mathbb{Z}_q[x]/(x^n + 1)$, or $q \equiv 1 \ (\bmod \frac{n}{2^{\alpha}})$ for the initial ring $\mathbb{Z}_q[x]/(x^n - 1)$. H-NTT can further weaken the parameter condition to  $q \equiv 1 \ (\bmod \frac{2n}{2^{\alpha+\beta}})$  for the ring $\mathbb{Z}_q[x]/(x^n + 1)$ by using truncated-NTT, or $q \equiv 1 \ (\bmod \frac{n}{2^{\alpha+\beta}})$  for the ring $\mathbb{Z}_q[x]/(x^n - 1)$ . By comparison, it can be seen that $\alpha$-round K-NTT and  truncated-NTT are the special cases of H-NTT when $\beta=0$ and $\alpha=0$ respectively. More specially, $\mathsf{NTT}_{no \rightarrow bo}^{CT,\psi}$ and $\mathsf{INTT}_{bo \rightarrow no}^{GS,\psi^{-1}}$ are the cases of H-NTT over $\mathbb{Z}_q[x]/(x^n + 1)$ when $\alpha=0,\beta=0$ , and $\mathsf{NTT}_{no \rightarrow bo}^{CT}$ and $\mathsf{INTT}_{bo \rightarrow no}^{GS}$ are the cases of H-NTT over $\mathbb{Z}_q[x]/(x^n - 1)$ when $\alpha=0,\beta=0$ .

\subsection{Method Based on Large Modulus}\label{sec-method-large-modulus}

Here we consider $\mathbb{Z}_q[x]/(x^n \pm 1)$, where $n$ is a power of two and  the modulus $q$ is an NTT-unfriendly prime, but $q$ can be actually any positive integer. Recent researches~\cite{ntt-in-saber-CHK+21,ntt-in-saber-FSS20,ntt-in-saber-FVR+21,multi-moduli-NTT-ACC+21} indicate that one can use NTTs in this case. In detail, the process of $\boldsymbol{c}=\boldsymbol{a} \cdot \boldsymbol{b} \in \mathbb{Z}_q[x]/(x^n \pm 1)$ is divided into two steps.

\begin{itemize}
	\item \textbf{Step 1.} Compute $\boldsymbol{c}'= \boldsymbol{a} \cdot \boldsymbol{b} \in \mathbb{Z}_N[x]/(x^n \pm 1)$, where $N$ is a positive integer and larger than the maximum absolute value of the coefficients during the computation over $\mathbb{Z}$. 
	
	\item \textbf{Step 2.} One can recover the result in $\mathbb{Z}_q[x]/(x^n \pm 1)$ through reduction module $q$, i.e., $\boldsymbol{c}=\boldsymbol{c}' \bmod q$.
\end{itemize}

It is called the \textbf{method based on large modulus} in this paper, because its key step is to select a large enough modulus $N$ in Step 1. Obviously, $N$ could be chosen such that $N>nq^2$. When $N$ is large enough, the product of $\boldsymbol{a} $ and $ \boldsymbol{b} $ in $\mathbb{Z}[x]/(x^n \pm 1)$ is identical to that in $\mathbb{Z}_N[x]/(x^n \pm 1)$. In order to apply NTTs, we consider two sub-cases of the values of $N$. One is that $N$ is an NTT-friendly prime. The other is that $N$ is the product of some NTT-friendly primes. When $N$ is the product of some NTT-friendly primes, i.e., $N = \prod_{i=1}^{l}{ q_i }$ where each $q_i$ is an NTT-friendly prime, in this case, it can further be classified into two sub-methods. One is the method based on residue number system (RNS). The other is the method based on composite-modulus ring.

\subsubsection{Method based on NTT-friendly large prime}
The works~\cite{ntt-in-saber-CHK+21,ntt-in-saber-FSS20,ntt-in-saber-FVR+21} show that, when $N$ is an NTT-friendly prime, NTT can be performed in $\mathbb{Z}_N[x]/(x^n \pm 1)$ directly in Step 1. Notice that if $N$ is set sufficiently large and NTT-friendly, this method always works regardless of the value of the original modulus $q$. 

For example, if $N$ is a prime satisfying $N>nq^2$ and $N \equiv 1 \ (\bmod \ 2n)$, $n$-point full NWC-based NTT of modulus $N$ can always be used over $\mathbb{Z}_N[x]/(x^n + 1)$. Besides, if $N$ is a prime satisfying $N>nq^2$ and $N \equiv 1 \ (\bmod \ n)$, $n$-point full CC-based NTT of modulus $N$ can always be used over $\mathbb{Z}_N[x]/(x^n - 1)$.

\subsubsection{Method based on residue number system} 
Residue number system (RNS) is widely used in the context of homomorphic encryption, e.g.,~\cite{fhe-bgv12,fhe-bra12,fhe-fv12}, for computing NTTs over many primes. The works~\cite{ntt-in-saber-CHK+21,ntt-in-saber-FSS20,multi-moduli-NTT-ACC+21} show that, based on RNS, negative wrapped convolution with a modulus that is the product of some primes, can be transformed into ones with smaller moduli by Chinese Remainder Theorem, that is
\begin{equation*}
\mathbb{Z}_N[x]/(x^n \pm 1)  \cong  \prod_{i=1}^{l}{ \mathbb{Z}_{q_i}[x]/(x^n \pm 1 ) } .
\end{equation*}

Denote by $\boldsymbol{c}_i$  the product of $\boldsymbol{a} $ and $ \boldsymbol{b}$ in $\mathbb{Z}_{q_i}[x]/(x^n \pm 1 ),i=1,\ldots,l$. After using NTTs to compute $\boldsymbol{c}_i,i=1,\ldots,l$, the original product $\boldsymbol{c}$ in $\mathbb{Z}_{N}[x]/(x^n \pm 1 )$ can be recovered from $\boldsymbol{c}_i,i=1,\ldots,l$, by Chinese Remainder Theorem in number theoretic form~\cite{ntt-in-saber-CHK+21}.

\subsubsection{Method based on composite-modulus ring}The work~\cite{multi-moduli-NTT-ACC+21} shows that NTT can be performed over a polynomial ring with a composite modulus directly. Specifically, the work~\cite{multi-moduli-NTT-ACC+21} generalizes the concepts of NTT from a finite field to an integer ring, the basic idea of which was first developed in terms of FFT by F$\ddot{\text{u}}$rer~\cite{faster-integer-mul-Fur09}. 

Consider $\mathbb{Z}_{N}[x]/(x^n + 1 )$, where $N = \prod_{i=1}^{l}{ q_i }$ and each $q_i$ is NTT-friendly prime. If $2n| \gcd(q_1-1,\ldots,q_l-1)$, there exits a principal $2n$-th root of unity $\psi_{2n}$ in $\mathbb{Z}_{N}$. $\psi_{2n}$ is a principle $2n$-th root of unity in $\mathbb{Z}_{N}$, iff $(\psi_{2n} \bmod q_i)$ is a principle $2n$-th root of unity in $\mathbb{Z}_{q_i}$ for any $i$. FFT trick over a composite-modulus polynomial ring is similar to that mentioned in section~\ref{sec-fft-trick}, i.e., $$\mathbb{Z}_N[x]/(x^n+1) \cong \prod\limits_{i=0}^{n-1}{ \mathbb{Z}_N[x]/(x - \psi_{2n}^{2 \text{brv}_n(i)+1 } )  }.$$
Its forward transform and inverse transform are illustrated as $\mathsf{NTT}_{no \rightarrow bo}^{CT,\psi}$ and $\mathsf{INTT}_{bo \rightarrow no}^{GS,\psi^{-1}}$  over $\mathbb{Z}_N[x]/(x^n+1)$. The CRT map of truncated-NTT with $\beta$ levels cropped is as follows: $$\mathbb{Z}_N[x]/(x^n+1)  \cong  \prod\limits_{i=0}^{n/2^{\beta}-1}{ \mathbb{Z}_N[x]/(x^{2^{\beta}} - \psi_{2n/{2^{\beta}}}^{2 \text{brv}_{n/{2^{\beta}}}(i)+1 } ) },$$
where $\psi_{2n/{2^{\beta}}}$ is a principal $2n/{2^{\beta}}$-th root of unity in $\mathbb{Z}_{N}$. Its forward transform and inverse transform are illustrated as $\mathsf{NTT}_{no \rightarrow bo,\beta}^{CT,\psi}$ and $\mathsf{INTT}_{bo \rightarrow no,\beta}^{GS,\psi^{-1}}$ over $\mathbb{Z}_N[x]/(x^n+1)$, respectively.

As for $\mathbb{Z}_{N}[x]/(x^n - 1 )$, where $N = \prod_{i=1}^{l}{ q_i }$ and each $q_i$ is NTT-friendly prime. If $n| \gcd(q_1-1,\ldots,q_l-1)$, there exits a principal $n$-th root of unity $\omega_{n}$ in $\mathbb{Z}_{N}$. $\omega_{n}$ is a principle $n$-th root of unity in $\mathbb{Z}_{N}$, iff $(\omega_{n} \bmod q_i)$ is a principle $n$-th root of unity in $\mathbb{Z}_{q_i}$ for any $i$. The CRT map of full-mapping FFT trick and truncated-NTT with $\beta$ levels cropped  over $\mathbb{Z}_{N}[x]/(x^n - 1 )$ can be derived similarly to those over $\mathbb{Z}_{N}[x]/(x^n + 1 )$.

\subsubsection{Comparisons and discussions}

We compare the method based on large modulus and those based on incomplete FFT trick/splitting polynomial ring. The three sub-methods of the method based on large modulus are valid for any original modulus $q$ (including NTT-friendly ones), and can completely remove the restriction on $q$. But, their shortcomings are also obvious. 

Specifically, the method based on incomplete FFT trick and based on splitting polynomial ring still choose original $q$ as the modulus. But, the modulus $N$ used in the method based on NTT-friendly large prime and method based on composite-modulus ring, is much larger than original modulus $q$. For examples, $N$ could be chosen as $nq^2$. Although $N$ be smaller if one of the multiplicands is small, it will still be several orders of magnitude larger than $q$. It causes that the storage of coefficients and the computing-resource consume will be more than the cases of $q$ during the computation. Besides, as for the method based on residue number system, there are more than one NTT computation needed to be computed. Traditionally, it is more time-consuming and resource-consuming, because full NWC-based NTT, the method based on incomplete FFT trick and the method based on splitting polynomial ring only need one NTT computation. Therefore, if $q$ is an NTT-friendly prime number, the methods based on incomplete FFT trick and based on splitting polynomial ring are recommended strongly for an efficient implementation, instead of the method based on large modulus. But, when $q$ is an NTT-unfriendly prime number, one can turn to the method based on large modulus.

Note that, there are some connections between the method based on residue number system and composite-modulus ring. Here we review $\mathbb{Z}_N[x]/(x^n \pm 1)$. Both of them could choose $N = \prod_{i=1}^{l}{ q_i }$ where each $q_i$ is NTT-friendly prime. If we split $N$ via CRT (Theorem~\ref{thm-crt-map-over-ring-init}) and keep $x^n\pm 1$ unchanged, it implies the method based on residue number system. If we split $x^n \pm 1$ via CRT and keep $N$ unchanged, it implies the method based on composite-modulus ring. Therefore, these two methods are derived from different splitting forms of moduli via CRT.

%
%
%
%

\section{Choosing Number Theoretic Transform for Given Rings}\label{sec-choose-ntt-for-rings}

In this section, we will introduce how to choose appropriate NTT for the given polynomial ring. We mainly classify the rings into three categories for the convenience of understanding:

\begin{itemize}
	\item $\mathbb{Z}_q[x]/(x^n \pm 1)$ with respect to power-of-two $n$;
	\item $\mathbb{Z}_q[x]/(x^n \pm 1)$ with respect to non-power-of-two $n$;
	\item $\mathbb{Z}_q[x]/( \phi(x))$ with respect to general $\phi(x)$ of degree $n$.
\end{itemize}

We mainly focus on the special rings of the form $\mathbb{Z}_q[x]/( (x^n \pm 1 )$ in the first two categories, and then we extend the results to the general rings of the form $\mathbb{Z}_q[x]/( \phi(x))$ with respect to general $\phi(x)$ of degree $n$.

\subsection{$\mathbb{Z}_q[x]/(x^n \pm 1)$ with Respect to Power-of-two $n$}\label{sec-choose-ntt-for-rings-power-of-two-n}

\subsubsection{NTT-friendly $q$}

In this case, $q$ is an NTT-friendly prime of the form  $q= q' \cdot 2^e+1$. Furthermore, we classify it into two sub-cases. The first sub-case is that $q$ can lead to a full NTT. The another sub-case is that $q$ can not lead to a full NTT.

We first consider the first sub-case of $n$ being a power of two and $q$ leading to a full NTT. As for $\mathbb{Z}_q[x]/(x^n+1)$, there exits $q \equiv 1 \ (\bmod \ 2n)$, then full NWC-based NTT (e.g., $\mathsf{NTT}_{no \rightarrow bo}^{CT,\psi}$ and $\mathsf{INTT}_{bo \rightarrow no}^{GS,\psi^{-1}}$) can be used to multiply two polynomials in $\mathbb{Z}_q[x]/(x^n+1)$.
Restate that one can compute $\boldsymbol{c}= \boldsymbol{a} \cdot \boldsymbol{b} \in \mathbb{Z}_q[x]/(x^n+1) $ by
\begin{equation*} 
\boldsymbol{c}= \mathsf{INTT}_{bo \rightarrow no}^{GS,\psi^{-1}} \left( \mathsf{NTT}_{no \rightarrow bo}^{CT,\psi} \left( {\boldsymbol{a}} \right) \circ \mathsf{NTT}_{no \rightarrow bo}^{CT,\psi} ( {\boldsymbol{b}} ) \right).
\end{equation*}

As for $\mathbb{Z}_q[x]/(x^n-1)$, there exits $q \equiv 1 \ (\bmod \ n)$, then full CC-based NTT (e.g., $\mathsf{NTT}_{no \rightarrow bo}^{CT}$ and $\mathsf{INTT}_{bo \rightarrow no}^{GS}$) can be used to multiply two polynomials in $\mathbb{Z}_q[x]/(x^n-1)$.
Restate that one can compute $\boldsymbol{c}= \boldsymbol{a} \cdot \boldsymbol{b} \in \mathbb{Z}_q[x]/(x^n-1) $ by
\begin{equation*} 
\boldsymbol{c}= \mathsf{INTT}_{bo \rightarrow no}^{GS} \left( \mathsf{NTT}_{no \rightarrow bo}^{CT} \left( {\boldsymbol{a}} \right) \circ \mathsf{NTT}_{no \rightarrow bo}^{CT} ( {\boldsymbol{b}} ) \right).
\end{equation*}

We then consider the another sub-case of $n$ being a power of two and $q$ not leading to a full NTT. It means that, for $\mathbb{Z}_q[x]/(x^n+1)$, $q$ does not satisfy $q \equiv 1 \ (\bmod \ 2n)$, or, for $\mathbb{Z}_q[x]/(x^n-1)$, $q$ does not satisfy $q \equiv 1 \ (\bmod \ n)$. In this sub-case, the method based on incomplete FFT trick and the method splitting polynomial ring are highly recommended.

For example, as for $\mathbb{Z}_q[x]/(x^n+1)$, if $q$ only satisfies $q \equiv 1 \ (\bmod \frac{2n}{2^{\alpha}})$ (resp., $q \equiv 1 \ (\bmod \frac{2n}{2^{\beta}})$), then one can use $\alpha$-round method based on splitting polynomial ring (resp., method based on incomplete FFT trick with $\beta$ levels cropped) can be used to  compute $\boldsymbol{c}= \boldsymbol{a} \cdot \boldsymbol{b} \in \mathbb{Z}_q[x]/(x^n+1) $. Besides, as for $\mathbb{Z}_q[x]/(x^n-1)$, if $q$ only satisfies $q \equiv 1 \ (\bmod \frac{n}{2^{\alpha}})$ (resp., $q \equiv 1 \ (\bmod \frac{n}{2^{\beta}})$), then one can use $\alpha$-round method based on splitting polynomial ring (resp., method based on incomplete FFT trick with $\beta$ levels cropped) can be used to  compute $\boldsymbol{c}= \boldsymbol{a} \cdot \boldsymbol{b} \in \mathbb{Z}_q[x]/(x^n-1) $.

Furthermore, if $q$ only satisfies $q \equiv 1 \ (\bmod \frac{2n}{2^{\alpha+\beta}})$ for $\mathbb{Z}_q[x]/(x^n+1)$, one can use H-NTT with $\alpha$-round splitting and $\beta$ levels cropped; similarly, if $q$ only satisfies $q \equiv 1 \ (\bmod \frac{n}{2^{\alpha+\beta}})$ for $\mathbb{Z}_q[x]/(x^n-1)$, one can use H-NTT with $\alpha$-round splitting and $\beta$ levels cropped.

\subsubsection{NTT-unfriendly $q$}

In this case, $q$ is an NTT-unfriendly prime. In order to compute $\boldsymbol{c}= \boldsymbol{a} \cdot \boldsymbol{b} \in \mathbb{Z}_q[x]/(x^n \pm 1) $, one can use the method based on large modulus. Firstly, compute $\boldsymbol{c}'= \boldsymbol{a} \cdot \boldsymbol{b} \in \mathbb{Z}_N[x]/(x^n \pm 1)$, where $N$ is a positive integer and larger than the maximum absolute value of the coefficients during the computation over $\mathbb{Z}$, and then compute $\boldsymbol{c}=\boldsymbol{c}' \bmod q$.

To compute $\boldsymbol{c}'= \boldsymbol{a} \cdot \boldsymbol{b} \in \mathbb{Z}_N[x]/(x^n \pm 1)$, one can use the method based on NTT-friendly large prime, the method based on residue number system or the method based on composite-modulus ring, as described in section~\ref{sec-method-large-modulus}.

\subsection{$\mathbb{Z}_q[x]/(x^n \pm 1)$ with Respect to Non-power-of-two $n$}\label{sec-choose-ntt-for-rings-non-power-of-two-n}

Here we turn to consider the case of non-power-of-two $n$, but actually $n$ can be a general integer. The works~\cite{ntt-in-saber-CHK+21,ntt-in-saber-FVR+21} show that as for radix-2 fast NTT algorithms, there are some useful technical methods. The essential idea about the computation of $\boldsymbol{c}=\boldsymbol{a} \cdot \boldsymbol{b} \in \mathbb{Z}_q[x]/(x^n \pm 1)$  is described through two steps as follows.

\begin{itemize}
	
	\item \textbf{Step 1.} Compute $\boldsymbol{c}'= \boldsymbol{a} \cdot \boldsymbol{b} \in \mathbb{Z}_q[x]/( x^{n'} \pm 1)$, where $n'$ is a larger integer and $ {n'} \ge 2n $.
	
	\item \textbf{Step 2.} One can recover the result in $\mathbb{Z}_q[x]/(x^n \pm 1)$ through reduction modulo $x^{n'} \pm 1$, i.e., $\boldsymbol{c}=\boldsymbol{c}' \bmod x^{n'} \pm 1$.
	
\end{itemize}

As for the ring $\mathbb{Z}_q[x]/( x^{n'} \pm 1)$, the works~\cite{ntt-in-saber-CHK+21,ntt-in-saber-FVR+21} furthermore classify it into two cases. The first case is that $n'$ is a power of two, which is denoted by $2^k$. The other case is that $n'$ is of the form $h \cdot 2^k$, where $h$ is an odd number.

\subsubsection{Power-of-two $n'$}

As for the first case, the computation is transformed into that with respect to $\boldsymbol{c}'= \boldsymbol{a} \cdot \boldsymbol{b} \in \mathbb{Z}_q[x]/( x^{n'} \pm 1)$, where $n'$ is a power of two. The methods to compute polynomial multiplication over $\mathbb{Z}_q[x]/( x^{n'} \pm 1)$ can be referred to section~\ref{sec-choose-ntt-for-rings-power-of-two-n}.

\subsubsection{Use Good's trick}

As for the other case, the computation is transformed into that with respect to $\boldsymbol{c}'= \boldsymbol{a} \cdot \boldsymbol{b} \in \mathbb{Z}_q[x]/( x^{n'} \pm 1)$, where $n'=h \cdot 2^k$ and $h$ is an odd number. One can use Good's trick (see Definition~\ref{def-good-trick}) to compute polynomial multiplication over $\mathbb{Z}_q[x]/( x^{h \cdot 2^k}  - 1 )$.

In this paper, Good's trick is recommended, since there are more freedom to select an odd number $h$ and power-of-two $2^k$. If we expand the length to power-of-two  $n'$, it is inconvenient to find a suitable polynomial of some particular degree up to the next power of two.

\subsection{$\mathbb{Z}_q[x]/( \phi(x))$ with Respect to General $\phi(x)$ of Degree $n$}

Here we consider the more general ring $\mathbb{Z}_q[x]/( \phi(x))$ where $\phi(x)$ is an arbitrary polynomial of degree $n$. In order to apply fast NTT algorithm, there are complex but efficient methods. The computation of $\boldsymbol{c}=\boldsymbol{a} \cdot \boldsymbol{b} \in \mathbb{Z}_q[x]/(\phi(x) )$ can be described through two steps as follows.

\begin{itemize}
	\item \textbf{Step 1.} Compute $\boldsymbol{c}'= \boldsymbol{a} \cdot \boldsymbol{b} \in \mathbb{Z}_N[x]/( x^{n'} \pm 1)$, where $N$ is a positive integer and larger than the maximum absolute value of the coefficients during the computation over $\mathbb{Z}$, $n'$ is a larger integer and $ {n'} \ge 2n $.
	
	\item \textbf{Step 2.} And then compute $\boldsymbol{c}= \left( \boldsymbol{c}' \bmod \phi(x) \right) \bmod q$.
\end{itemize}

Therefore, the computation of $\boldsymbol{c}=\boldsymbol{a} \cdot \boldsymbol{b} \in \mathbb{Z}_q[x]/(\phi(x) )$  is transformed into that of $\boldsymbol{c}'=\boldsymbol{a} \cdot \boldsymbol{b} \in \mathbb{Z}_N[x]/( x^{n'} \pm 1)$. In this case, one can use the methods in section~\ref{sec-choose-ntt-for-rings-non-power-of-two-n} to compute polynomial multiplications over $\mathbb{Z}_N[x]/( x^{n'} \pm 1)$.

Besides, the works~\cite{ntt-in-ntru-ACC+21,ntt-in-ntru-BBC21} also use Sch$\ddot{\text{o}}$nhage’s trick and Nussbaumer’s trick (see Definition~\ref{def-schonhage-trick} and~\ref{def-nussbaumer-trick}) to compute polynomial multiplications over $\mathbb{Z}_N[x]/( x^{n'} \pm 1)$.

%
%
%
%

\section{Number Theoretic Transform in NIST PQC}\label{sec-ntt-in-nist}

In this section, we will introduce NTT and its variants for NIST PQC Round 3 candidates. Their parameter sets can be seen in Table~\ref{tab-parameter-in-cryptosystem}. Among the schemes, key encapsulation mechanisms (KEM) include Kyber~\cite{kyber-BDK+18,kyber-nist-round3}, Saber~\cite{saber-DKR+18,saber-nist-round3}, NTRU~\cite{ntru-nist-round3}, NTRU Prime~\cite{ntru-prime-BCLV17,ntru-prime-nist-round3}. Digital signature schemes include Dilithium~\cite{dilithium-CHES18,dilithium-nist-round3} and Falcon~\cite{falcon-nist-round3}.  Among them, Kyber, Dilithium and Falcon are standardized by NIST~\cite{nist-to-be-standardized}. All these KEMs are constructed from passive secure (IND-CPA secure or OW-CPA secure) PKEs via tweak variants of Fujisaki-Okamoto transform~\cite{fo-transform-FO99,fo-transform-FO13,fo-transform-HHK17}. 

We mainly focus on their NTT-based implementations over their underlying rings $\mathbb{Z}_q[x]/(\phi(x))$, where $\mathbb{Z}_q[x]/(\phi(x))$ is instantiated as in Table~\ref{tab-parameter-in-cryptosystem}. Notice that not all the schemes can directly use full NTT to multiply polynomials. For example, full NWC-based NTT over $\mathbb{Z}_q[x]/(x^n+1)$ further requires that the prime $q$ satisfies $q \equiv 1 \ (\bmod \ 2n)$. Only Dilithium and Falcon can meet the situation. Besides, full NTT can not be directly used in those lattice-based schemes with power-of-two moduli, such as Saber and NTRU.

\input{table/tab-parameter-in-cryptosystem.tex}

\subsection{Kyber}

Kyber~\cite{kyber-nist-round1,kyber-nist-round2,kyber-nist-round3,kyber-BDK+18} is an IND-CCA secure MLWE-based KEM from the lattice-based cryptography suite called ``Cryptographic Suite for Algebraic Lattices'' (CRYSTALS for short). It is currently the only standardized KEM by NIST PQC~\cite{nist-to-be-standardized}. 

The Kyber submission in NIST PQC Round 1, named Kyber Round 1~\cite{kyber-nist-round1}, uses $n=256$ and $q=7681$ which satisfies $q \equiv 1 \ (\bmod \ 2n)$. Its way to compute $\boldsymbol{c}=\boldsymbol{a} \cdot \boldsymbol{b} \in \mathcal{R}_q$ in C reference implementation is elaborated as 
$$\boldsymbol{c}= \boldsymbol{\psi}^{-1} \circ  \mathsf{INTT}_{bo \rightarrow no}^{GS} \left( \mathsf{NTT}_{no \rightarrow bo}^{CT,\psi} \left( {\boldsymbol{a}} \right) \circ \mathsf{NTT}_{no \rightarrow bo}^{CT,\psi} ( {\boldsymbol{b}} ) \right),$$ 
where the inverse transform is $\boldsymbol{\psi}^{-1} \circ  \mathsf{INTT}_{bo \rightarrow no}^{GS}$, the output of which is the same as that of $\mathsf{INTT}_{bo \rightarrow no}^{GS,\psi^{-1}}$ according to formula (\ref{equ-classic-intt-psi}). As for its AVX2 optimized implementation, Kyber Round 1 uses $\mathsf{NTT}_{no \rightarrow bo}^{CT,\psi}$ and $\mathsf{INTT}_{bo \rightarrow no}^{GS,\psi^{-1}}$, and the computation is restated as
\begin{equation*}
\boldsymbol{c}= \mathsf{INTT}_{bo \rightarrow no}^{GS,\psi^{-1}} \left( \mathsf{NTT}_{no \rightarrow bo}^{CT,\psi} \left( {\boldsymbol{a}} \right) \circ \mathsf{NTT}_{no \rightarrow bo}^{CT,\psi} ( {\boldsymbol{b}} ) \right).
\end{equation*}

The Kyber submissions in the second and third round of NIST PQC competition use a smaller prime number $q=3329$ which no longer satisfies $q \equiv 1 \ (\bmod \ 2n)$, but $q \equiv 1 \ (\bmod \ n)$.  Kyber Round 2 and Round 3 use truncated-NTT with one level cropped: $\mathsf{NTT}_{no \rightarrow bo,\beta=1}^{CT,\psi}$ and $\mathsf{INTT}_{bo \rightarrow no,\beta=1}^{GS,\psi^{-1}}$. The point-wise multiplication is treated as the corresponding polynomial multiplications in $\mathbb{Z}_q[x]/(x^2 - \omega_{n}^{2 \text{brv}_{n/2}(i)+1 })$. Its way to compute $\boldsymbol{c}=\boldsymbol{a} \cdot \boldsymbol{b} \in \mathcal{R}_q$ is restated as
\begin{equation*}
	\boldsymbol{c}= \mathsf{INTT}_{bo \rightarrow no,\beta=1}^{GS,\psi^{-1}} \left( \mathsf{NTT}_{no \rightarrow bo,\beta=1}^{CT,\psi} \left( {\boldsymbol{a}} \right) \circ \mathsf{NTT}_{no \rightarrow bo,\beta=1}^{CT,\psi} ( {\boldsymbol{b}} ) \right).
\end{equation*}

Further, Liang et al.~\cite{ntt-inscrypt20} improve its truncated-NTT by using H-NTT with $\alpha=0$, which can be seen as $\mathsf{NTT}_{no \rightarrow bo,\beta=1}^{CT,\psi}$ and $\mathsf{INTT}_{bo \rightarrow no,\beta=1}^{GS,\psi^{-1}}$ with one-iteration Karatsuba algorithm in its point-wise multiplication. Actually, the idea of decreasing the modulus $q$ of Kyber Round 1 from $q = 7681$ to $q = 3329$  was first proposed by Zhou et al.~\cite{ptntt-inscrypt18}, and they denoted it by ``small-Kyber''. Different from truncated-NTT used in Kyber Round 2 and Round 3, they used 1-round Pt-NTT with $\frac{n}{2}$-point full NWC-based NTT for  ``small-Kyber''. But, their implementation had a slightly worser performance than the initial one.

\subsection{Dilithium}

Dilithium~\cite{dilithium-nist-round3} is a signature scheme based on module lattice and is one of the algorithms from CRYSTALS. It is one of the standardized signature scheme in NIST PQC~\cite{nist-to-be-standardized}. Its parameter sets satisfy the condition $q \equiv 1 \ (\bmod \  2n)$ such that  $n$-point full NWC-based $\mathsf{NTT}_{no \rightarrow bo}^{CT,\psi}$ and $\mathsf{INTT}_{bo \rightarrow no}^{GS,\psi^{-1}}$ can be utilized. Its way to compute $\boldsymbol{c}=\boldsymbol{a} \cdot \boldsymbol{b} \in \mathcal{R}_q$ is restated as
\begin{equation*}
	\boldsymbol{c}= \mathsf{INTT}_{bo \rightarrow no}^{GS,\psi^{-1}} \left( \mathsf{NTT}_{no \rightarrow bo}^{CT,\psi} \left( {\boldsymbol{a}} \right) \circ \mathsf{NTT}_{no \rightarrow bo}^{CT,\psi} ( {\boldsymbol{b}} ) \right).
\end{equation*}

\subsection{Falcon}

Falcon~\cite{falcon-nist-round3} is a signature scheme based on NTRU lattice. It is the another standardized signature scheme in NIST PQC~\cite{nist-to-be-standardized}. Falcon also uses $n$-point full NWC-based $\mathsf{NTT}_{no \rightarrow bo}^{CT,\psi}$ and $\mathsf{INTT}_{bo \rightarrow no}^{GS,\psi^{-1}}$ to compute polynomial multiplication in its verification algorithm. Its way to compute $\boldsymbol{c}=\boldsymbol{a} \cdot \boldsymbol{b} \in \mathcal{R}_q$ is also written as
\begin{equation*}
	\boldsymbol{c}= \mathsf{INTT}_{bo \rightarrow no}^{GS,\psi^{-1}} \left( \mathsf{NTT}_{no \rightarrow bo}^{CT,\psi} \left( {\boldsymbol{a}} \right) \circ \mathsf{NTT}_{no \rightarrow bo}^{CT,\psi} ( {\boldsymbol{b}} ) \right).
\end{equation*}

\subsection{Saber}

Saber~\cite{saber-DKR+18,saber-nist-round3} is an IND-CCA secure KEM based on MLWR, and was one of the Finalists in NIST PQC Round 3. Since the modulus $q=2^{13}$ chosen by Saber is not a prime number, but a power of two, it does not satisfy the parameter conditions of NTT. The previous works~\cite{saber-DKR+18} used Toom-Cook and Karatsuba algorithm, instead of NTT. Recent researches~\cite{ntt-in-saber-CHK+21,ntt-in-saber-FSS20,ntt-in-saber-FVR+21,multi-moduli-NTT-ACC+21} successfully applies NTT in Saber via the method based on large modulus.

The polynomial multiplications in Saber mainly contain matrix-vector polynomial multiplication $\mathbf{{A}} \mathbf{{s}} $ and vector-vector polynomial multiplication $\mathbf{{b}}^{T} \mathbf{{s}} $ where $\mathbf{{A}} \in \mathcal{R}_q^{k \times k}, \mathbf{{b}}, \mathbf{{s}}\in \mathcal{R}_q^{k \times 1}$, $|s_i| \le \mu $ and $\mu$ is the parameter of the central binomial distribution. In~\cite{ntt-in-saber-CHK+21,ntt-in-saber-FSS20,ntt-in-saber-FVR+21,multi-moduli-NTT-ACC+21}, the coefficients of $\mathbf{{A}}$ and $\mathbf{{b}}$ are represented in $[-q/2, q/2)$, and the coefficients of $ \mathbf{{s}} $ are represented in $[-\mu/2, \mu/2]$. The coefficients obtained by matrix-vector and vector-vector multiplication range in $ \left[ -{knq\mu/4},{knq\mu/4} \right]$. Therefore, the modulus $N$ used in the method based on large modulus, must be chosen carefully such that $N > knq\mu/2 $ holds. There are three parameter sets, referred to as LightSaber-KEM, Saber-KEM and FireSaber-KEM, corresponding to $k=2,3,4$ and $\mu=10,8,6$, respectively.

Chung et al.~\cite{ntt-in-saber-CHK+21} use different $N$ and NTT methods for their ARM Cortex-M4 and AVX2 implementations in Saber. Specifically, they use the method based on NTT-friendly large prime in ARM Cortex-M4 implementations, and choose $\mathsf{NTT}_{no \rightarrow bo,\beta=2}^{CT,\psi}$ and $\mathsf{INTT}_{bo \rightarrow no,\beta=2}^{GS,\psi^{-1}}$. LightSaber-KEM chooses $N=20972417$, while Saber-KEM and Firesaber-KEM choose $N=25166081$. All of them are NTT-friendly primes and satisfy the condition $N \equiv 1 \ (\bmod \frac{n}{2})$. As for their AVX2 implementations, they use the method based on residue number system. $\mathsf{NTT}_{no \rightarrow bo}^{CT,\psi}$ and $\mathsf{INTT}_{bo \rightarrow no}^{GS,\psi^{-1}}$ are used in all the three parameter sets. The modulus $N=q_1 q_2$ is chosen where $q_1=7681$ and $q_2=10753$ are NTT-friendly primes such that $q_i \equiv 1 \ (\bmod \ 2n),i=1,2$ holds. 

Abdulrahman et al.~\cite{multi-moduli-NTT-ACC+21} use the method based on composite-modulus ring in their ARM Cortex-M3 and Cortex-M4 implementations of Saber. They choose a large composite number $N=q_1 q_2$, where $q_1=7681$ and $q_2=3329$. In order to achieve speed and memory optimized implementation, the NTT method they use $\mathsf{NTT}_{no \rightarrow bo,\beta=2}^{CT,\psi}$ and $\mathsf{INTT}_{bo \rightarrow no,\beta=2}^{GS,\psi^{-1}}$ over $\mathbb{Z}_N[x]/(x^n+1)$. 

\subsection{NTRU}

NTRU~\cite{ntru-nist-round3} ia an IND-CCA secure KEMs based on NTRU cryptosystem~\cite{ntru-HPS98}, and was one of the Finalists in NIST PQC Round 3. NTRU actually includes two suit of schemes named NTRU-HRSS and NTRUEncrypt, which two teams independently submitted to the first round of NIST PQC competition. But they merged their schemes, and gave a new name ``NTRU'' after the first round. NTRU uses the polynomial ring $\mathbb{Z}_q[x]/( x^n  - 1 )$ where $n$ is a prime number and $q$ is a power of two. There are two ways to utilize NTT in NTRU. The first way was applied by Fritzmann et al.~\cite{ntt-in-saber-FVR+21}. To allow a modular NTT arithmetic architecture with RISC-V instruction set extensions, they first map $\mathbb{Z}_q[x]/( x^n  - 1 )$ to $\mathbb{Z}_q[x]/( x^{n'}  - 1 )$ where $n'$ is a power of two and $ {n'} \ge 2n $. $n'$ could be set to be $2048$, for $n \in \{ 509,677,821,701\}$. Then they utilize the method based on NTT-friendly large prime over $\mathbb{Z}_q[x]/( x^{n'}  - 1 )$. Specifically, they choose an NTT-friendly sufficiently large prime $N = 549755809793$, such that $N \equiv 1 \ (\bmod \ n')$  holds. Firstly, compute $\boldsymbol{c}'=\boldsymbol{a} \cdot \boldsymbol{b} \in \mathbb{Z}_N[x]/( x^{n'}  - 1 )$ by using $n'$-point full cyclic convolution-based NTT, and then compute  $\boldsymbol{c} =  (\boldsymbol{c}' \bmod x^n-1) \bmod q$. 

The other way is to use Good's trick in Chung et al.'s work~\cite{ntt-in-saber-CHK+21}.  As for $n\in \{677,701\}$, they choose $h=3$ and $k=9$ for Good's trick in their ARM Cortex-M4 implementation. For $h$ parallel $2^k$-point NTT over $\mathbb{Z}_q[z]/(z^{2^k}-1)$, they apply the method based NTT-friendly large prime, where it maps $\mathbb{Z}_q[z]/(z^{2^k}-1)$ to $\mathbb{Z}_N[z]/(z^{2^k}-1)$, and use $N=5747201$  and $N=1389569$ for the cases of $n=701$ and $n=677$ respectively.

\subsection{NTRU Prime}

NTRU Prime~\cite{ntru-prime-BCLV17,ntru-prime-nist-round3} is an IND-CCA secure KEMs based on NTRU crytosystem~\cite{ntru-HPS98}, and was one of the Alternates in NIST PQC Round 3. It was proposed for the aim ``an efficient implementation of high security prime-degree large-Galois-group inert-modulus ideal-lattice-based cryptography''~\cite{ntru-prime-BCLV17}. The NTRU Prime submission to the NIST PQC competition~\cite{ntru-prime-nist-round3} offers two KEMs: Streamlined NTRU Prime and NTRU LPRime. NTRU Prime tweaks the classic NTRU scheme to use rings with less special structures, i.e., $\mathbb{Z}_q[x]/(x^n - x -1)$, where both $n$ and $q$ are primes. 

For radix-2 NTT-based implementation of NTRU Prime, one can map $\mathbb{Z}_q[x]/(x^n - x -1)$ to $\mathbb{Z}_N[x]/( x^{n'}  - 1 )$ where $n' \ge 2n$, and choose an NTT-friendly sufficiently large prime $N$ such that $N> 2n q^2$. After computing NTT over $\mathbb{Z}_N[x]/( x^{n'}  - 1 )$, the final result can be obtained module $q$ and $x^n - x -1$. 

Besides, the works~\cite{ntt-in-ntru-ACC+21,ntt-in-ntru-prime-PMT21} show that Good's trick can also be utilized over $\mathbb{Z}_q[x]/(x^n - x -1)$ in NTRU Prime. It maps $\mathbb{Z}_q[x]/(x^n - x -1)$ to $\mathbb{Z}_N[x]/(x^{n'} -1)$ with $n'=1536$ for the case of $n=761$. The work~\cite{ntt-in-ntru-ACC+21} chooses $N=6984193$, while~\cite{ntt-in-ntru-prime-PMT21} chooses $N=q_1 q_2 q_3$ where $q_1=7681, q_2=12289, q_3 =15361$. As for the parameters in Good's trick, they use $h=3$ and $k=9$.

The work~\cite{ntt-in-ntru-ACC+21} also presented an optional idea that one can manufacture roots of unity for radix-2 NTT in $\mathbb{Z}_q[x]/(x^n - x -1)$ by using Sch$\ddot{\text{o}}$nhage’s trick and Nussbaumer’s trick, which was later implemented with an improvement by Bernstein et al.~\cite{ntt-in-ntru-BBC21}. 
More specifically, since the original polynomials of NTRU Prime have degree $n<1024$, Bernstein et al.~\cite{ntt-in-ntru-BBC21} map $\mathbb{Z}_q[x]/(x^n - x -1)$ to $\mathbb{Z}_q[x]/(x^{2048}-1)$. They use  Sch$\ddot{\text{o}}$nhage’s trick in order to transform the operations to multiplications in $\mathbb{Z}_q[x]/(x^{64}+1)$. Then, they use Nussbaumer’s trick to compute the multiplications in $\mathbb{Z}_q[x]/(x^{64}+1)$.

\section{Conclusion and Future Work}\label{sec-conclusion}

This paper makes a mathematically systematic study of NTT, including its history, basic concepts, basic radix-2 fast computing algorithms, methods to weaken restrictions on parameter conditions, selections for the given rings and applications in the lattice-based schemes of NIST PQC. 

Here are some future works listed as follows.

\begin{itemize}
	
	\item Notice that the concepts of fast NTT algorithms and incomplete FFT trick, are first invented in terms of FFT. In most cases, one can learn from skills and techniques of FFT, and apply them into NTT methods. There are still numerous FFT techniques, such as radix-$2^l$, mixed-radix, split-radix algorithms, Stockham butterfly FFT~\cite{sk-fft}, WFTA~\cite{wfta} etc. Therefore, one could choose a best one for the NTT method from the implementation point of view.
	
	\item For efficient and secure implementation of NTT, resistance against implementation
	attacks such as side-channel attacks have been increasingly considered as an
	important criteria for NIST PQC. Side-channel attack can make use of the information leaked from the target devices to recover some secrets of cryptographic schemes. For example, in order to avoiding timing attack~\cite{sca-timing-attack-Kocher96}, all the operations should be under the strategy of constant implementation. There are some types of attacks for the implementation of NTT, including single trace attack, simple power analysis, fault attack and so on~\cite{sca-ntt-RPBC20,sca-sok-ct-rsa-HPA21,pqc-implementations-a-survey-NDRRB19}. The correlative strategies against side-channel attack on NTT have been under development. We leave it as a future work of comprehensive survey on implementing NTT against side-channel attacks.
	
\end{itemize}

%
%
%
%
%

\section*{Acknowledgments}

We are grateful to all the anonymous referees of CT-RSA 2022 and AsiaCCS 2022 for their constructive and insightful review comments. We would like to thank Yu Liu and Jieyu Zheng for their help on the writing and useful discussions on this work.

%
%
%
%
%
%
%

\bibliographystyle{alpha} 
\bibliography{NTT-ref}

%
%
%
%

\newpage

\appendix

\section{Signal Flow of Radix-2 NTT}\label{sec-signal-flow}

\begin{figure}[H]
	\centering
	\subfigure[$\mathsf{NTT}_{bo \rightarrow no}^{CT}$]{\label{fig-classic-ntt-ct-gs-signal-flow-a}
		\includegraphics[width=0.43\linewidth]{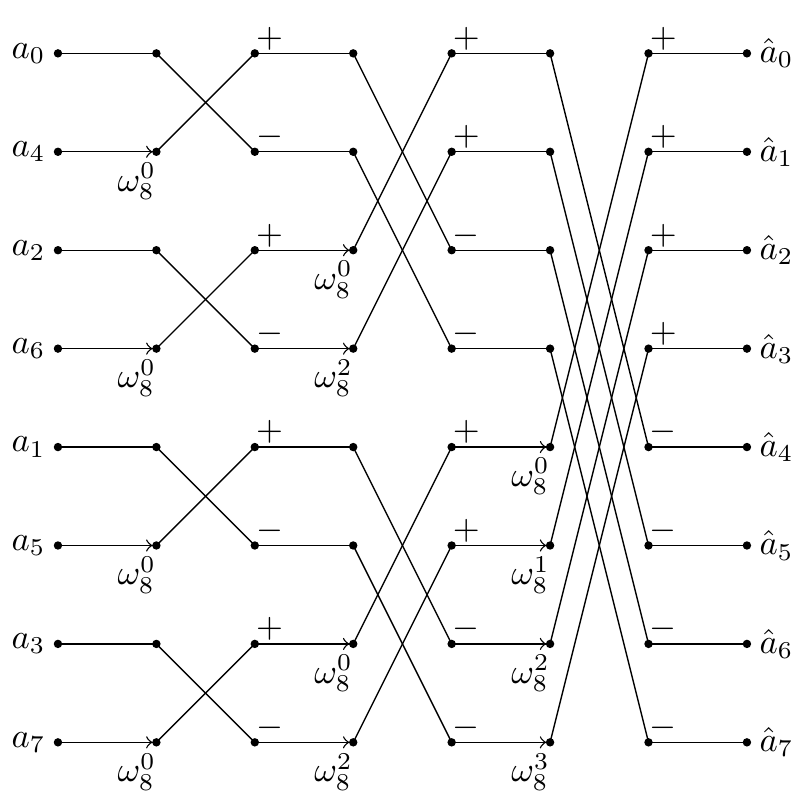}}
	\hspace{0.01\linewidth}
	\subfigure[$\mathsf{NTT}_{bo \rightarrow no}^{GS}$]{\label{fig-classic-ntt-ct-gs-signal-flow-b}
		\includegraphics[width=0.43\linewidth]{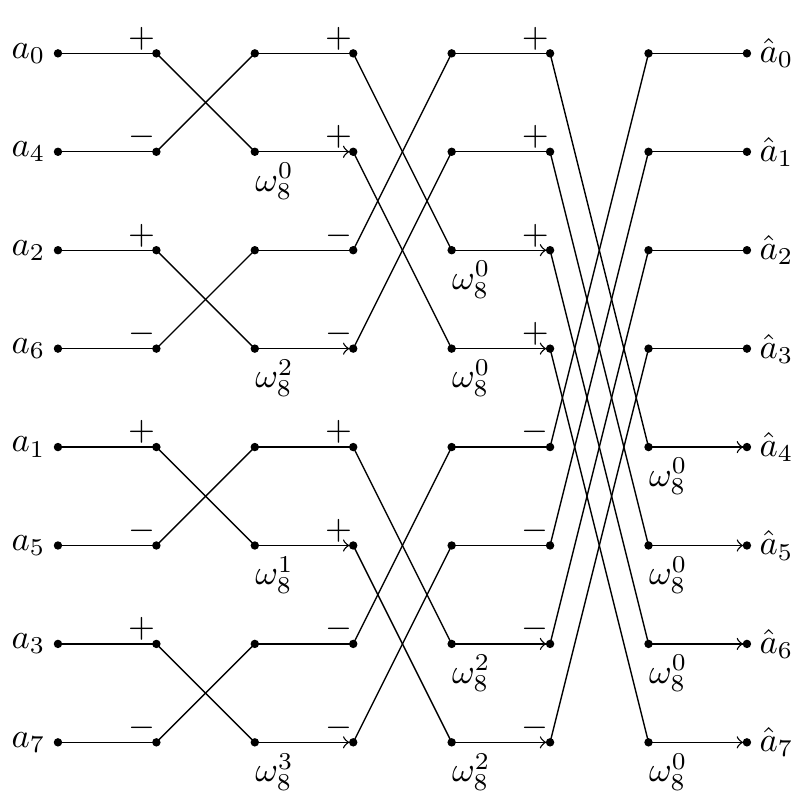}}
	\vfill
	\subfigure[$\mathsf{NTT}_{no \rightarrow bo}^{CT}$]{\label{fig-classic-ntt-ct-gs-signal-flow-c}
		\includegraphics[width=0.43\linewidth]{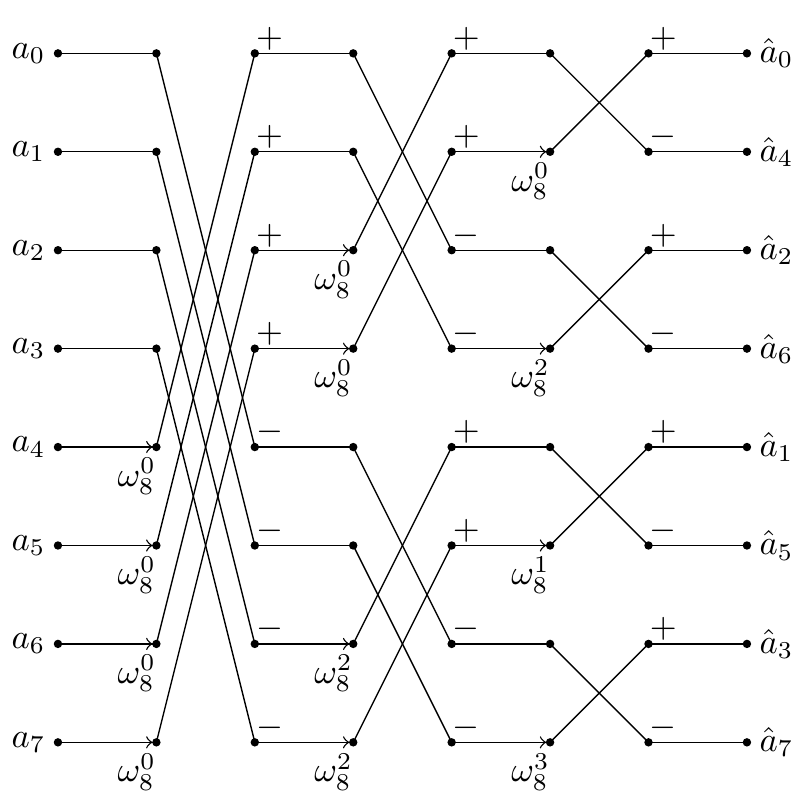}}
	\hspace{0.01\linewidth}
	\subfigure[$\mathsf{NTT}_{no \rightarrow bo}^{GS}$]{\label{fig-classic-ntt-ct-gs-signal-flow-d}
		\includegraphics[width=0.43\linewidth]{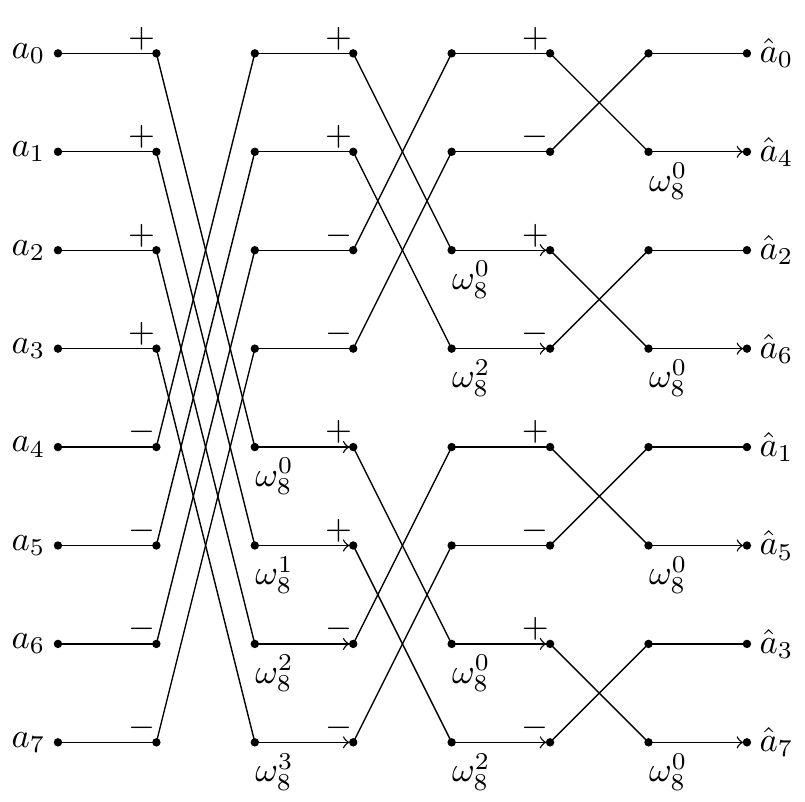}}
	\caption{Signal flow of radix-2 $\mathsf{NTT}$ for $n=8$}
	\label{fig-classic-ntt-ct-gs-signal-flow}
\end{figure}

\begin{figure}[H]
	\centering
	\subfigure[$\mathsf{INTT}_{bo \rightarrow no}^{CT}$]{\label{fig-classic-intt-ct-gs-signal-flow-a}
		\includegraphics[width=0.43\linewidth]{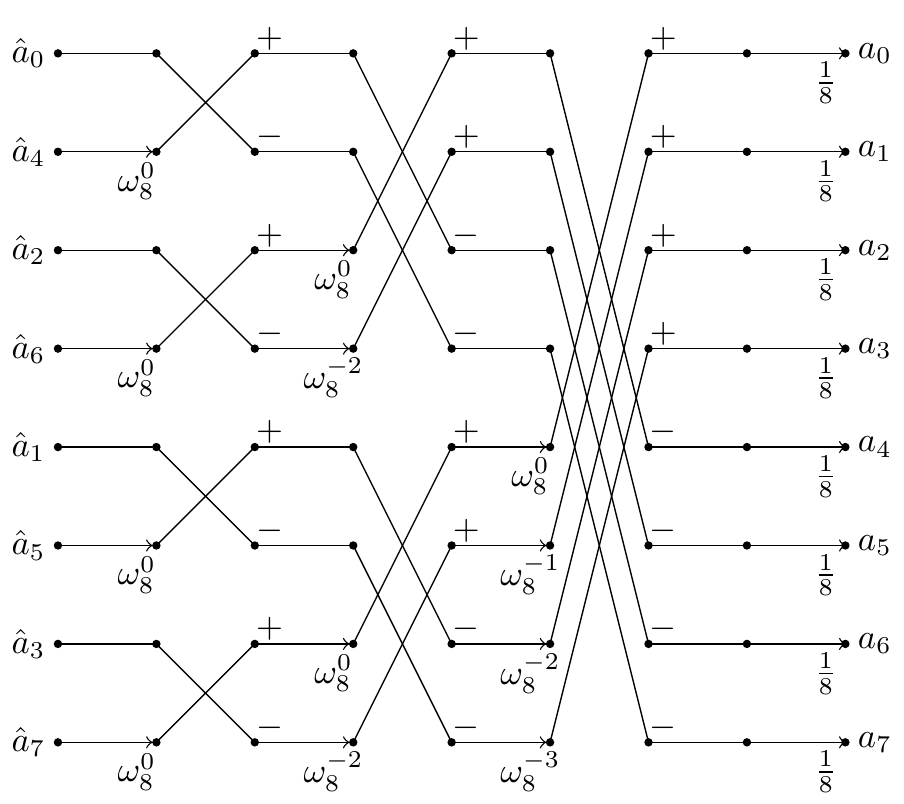}}
	\hspace{0.01\linewidth}
	\subfigure[$\mathsf{INTT}_{bo \rightarrow no}^{GS}$]{\label{fig-classic-intt-ct-gs-signal-flow-b}
		\includegraphics[width=0.43\linewidth]{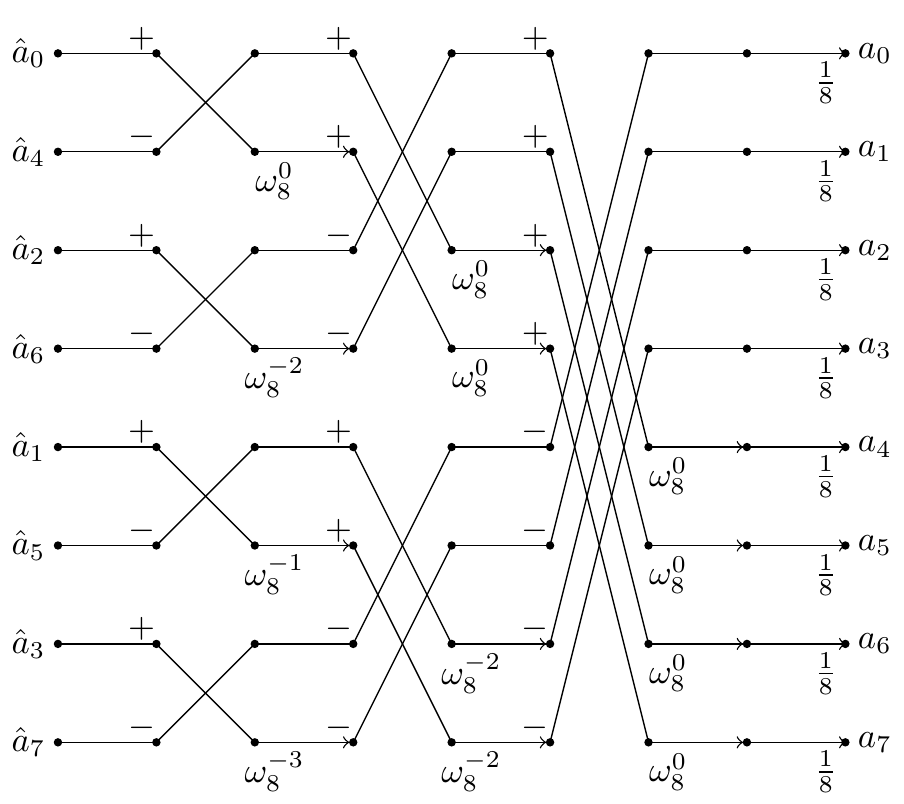}}
	\vfill
	\subfigure[$\mathsf{INTT}_{no \rightarrow bo}^{CT}$]{\label{fig-classic-intt-ct-gs-signal-flow-c}
		\includegraphics[width=0.43\linewidth]{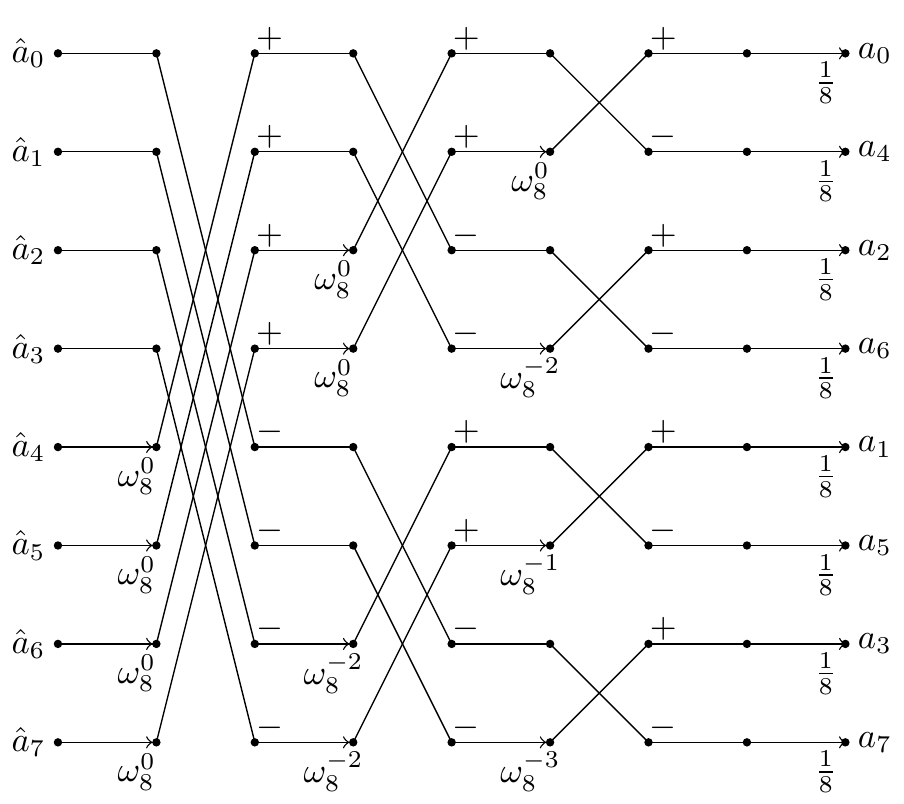}}
	\hspace{0.01\linewidth}
	\subfigure[$\mathsf{INTT}_{no \rightarrow bo}^{GS}$]{\label{fig-classic-intt-ct-gs-signal-flow-d}
		\includegraphics[width=0.43\linewidth]{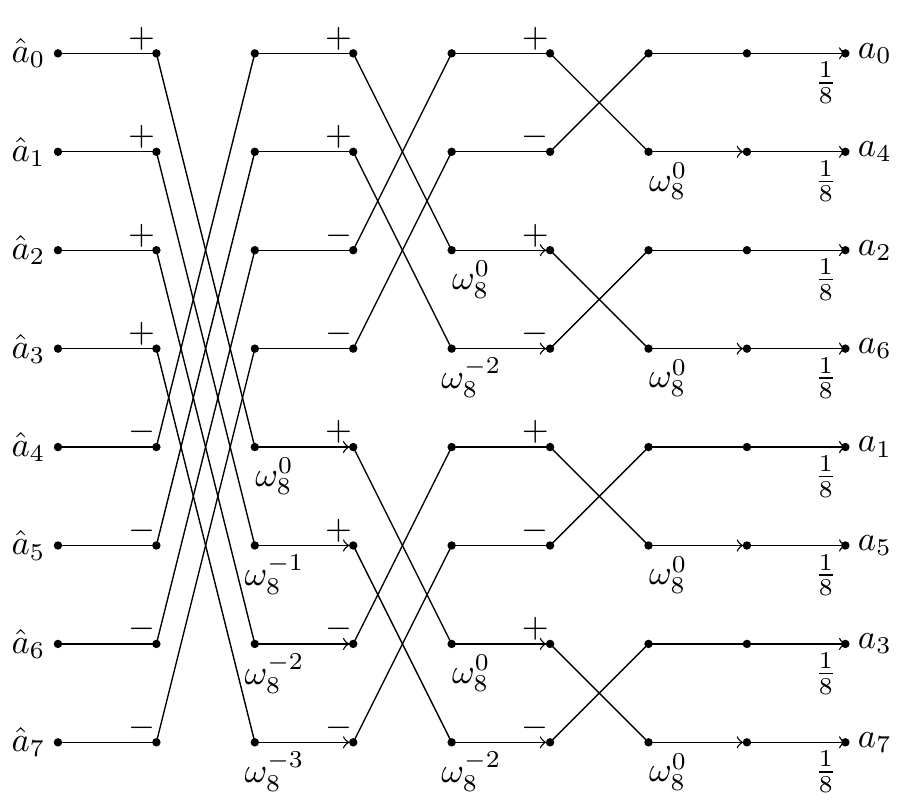}}
	\caption{Signal flow of radix-2 $\mathsf{INTT}$ for $n=8$}
	\label{fig-classic-intt-ct-gs-signal-flow}
\end{figure}

\begin{figure}[H]
	\centering
	\subfigure[$\mathsf{NTT}_{bo \rightarrow no}^{CT,\psi}$]{\label{fig-classic-ntt-intt-psi-signal-flow-a}
		\includegraphics[width=0.43\linewidth]{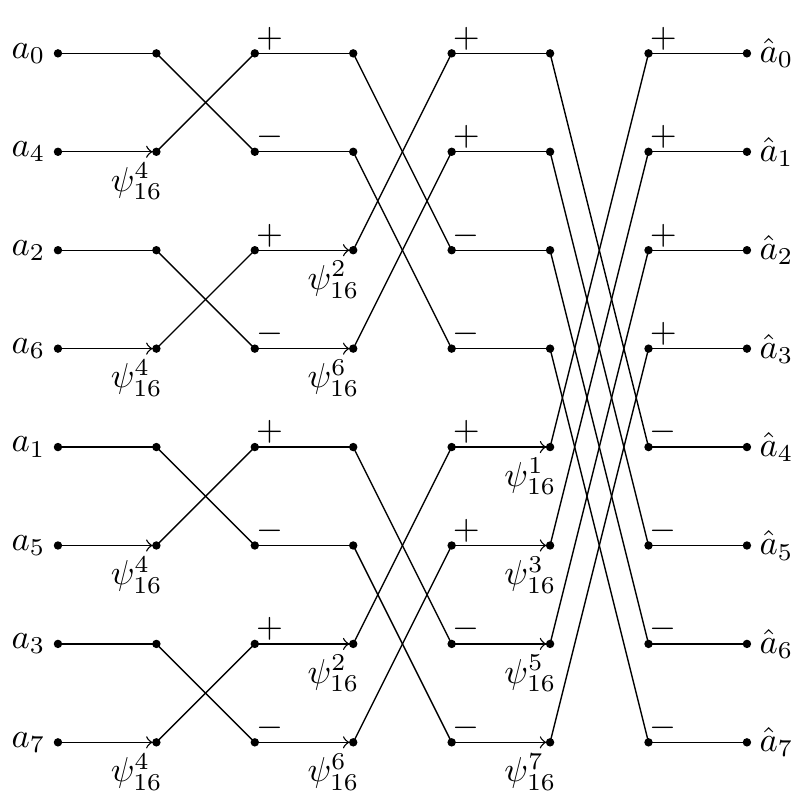}}
	\hspace{0.01\linewidth}
	\subfigure[$\mathsf{INTT}_{bo \rightarrow no}^{GS,\psi^{-1}}$]{\label{fig-classic-ntt-intt-psi-signal-flow-b}
		\includegraphics[width=0.48\linewidth]{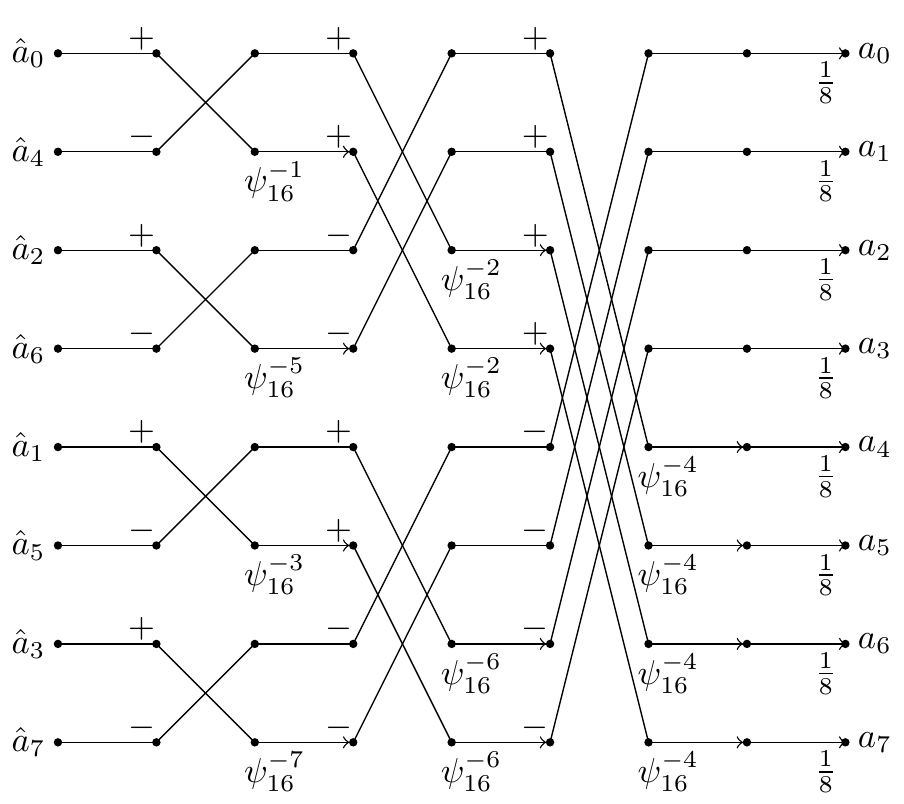}}
	\vfill
	\subfigure[$\mathsf{NTT}_{no \rightarrow bo}^{CT,\psi}$]{\label{fig-classic-ntt-intt-psi-signal-flow-c}
		\includegraphics[width=0.43\linewidth]{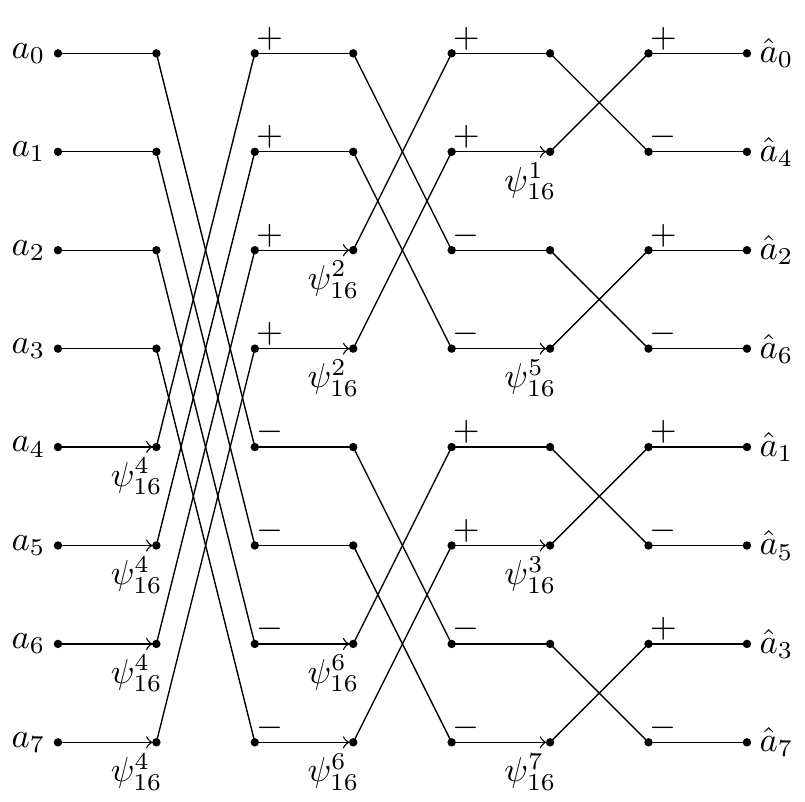}}
	\hspace{0.01\linewidth}
	\subfigure[$\mathsf{INTT}_{no \rightarrow bo}^{GS,\psi^{-1}}$]{\label{fig-classic-ntt-intt-psi-signal-flow-d}
		\includegraphics[width=0.48\linewidth]{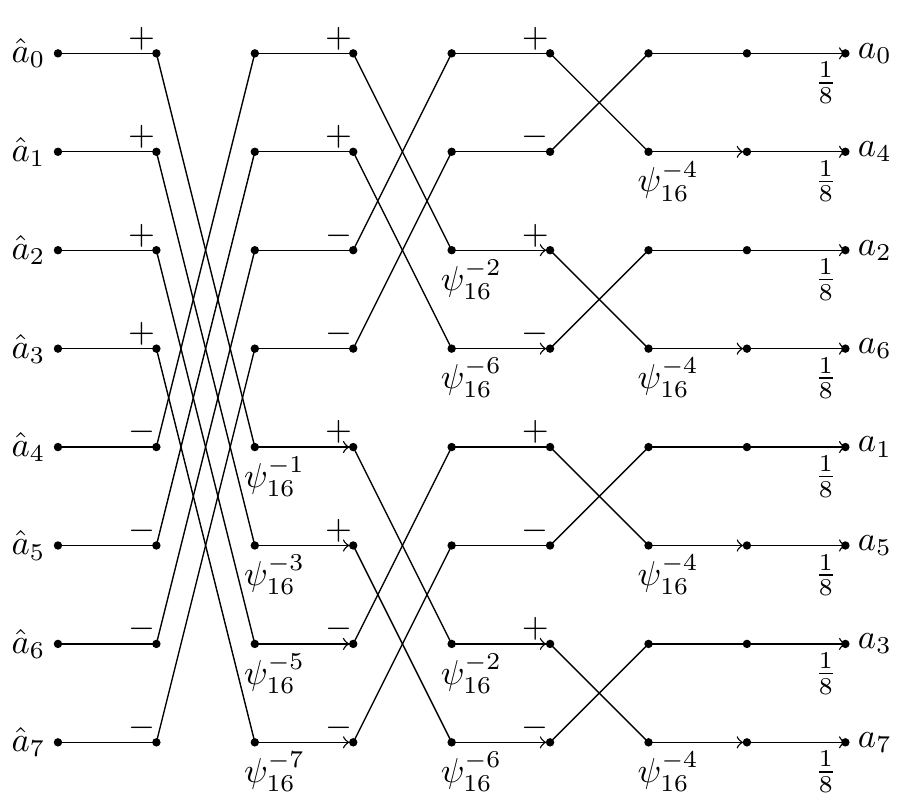}}
	\caption{Signal flow of radix-2 $\mathsf{NTT}^{\psi}$/$\mathsf{INTT}^{\psi^{-1}}$ for $n=8$}
	\label{fig-classic-ntt-intt-psi-signal-flow}
\end{figure}

\section{Number Theoretic Transform over $\mathbb{Z}_q[x]/(x^n - x^{n/2} + 1)$}\label{sec-ntt-over-non-power-of-two}

In this section, we introduce some progresses about relaxing the requirement of $n$ being power-of-two such that NTT can be utilized over non-power-of-two rings.

\subsection{Incomplete FFT trick over $\mathbb{Z}_q[x]/(x^n - x^{n/2} + 1)$}

Some progresses about relaxing the requirement of $n$ being power-of-two are made by Lyubashevsky and Seiler~\cite{nttru}. They first introduced a special incomplete NTT to lattice-based cryptographic schemes, by offering a new non-power-of-two ring structure $\mathbb{Z}_q[x]/(x^n - x^{n/2} + 1)$ where $n=3\cdot 2^e$ instead of a power of two and $q$ is a prime number satisfying $q \equiv 1 \ (\bmod \ n)$ such that $\psi_{n}$ exits. Actually, $x^n - x^{n/2} + 1$ is the $3n$-th cyclotomic polynomial of degree $n$ where $n=3^l\cdot 2^e$, $l \ge 0, e \ge 1$, not a power-of-two cyclotomic polynomial any more. The main observation they use is the CRT map as follows:
$$\mathbb{Z}_q[x]/(x^n - x^{n/2} + 1) \cong \mathbb{Z}_q[x]/( x^{n/2} - \zeta_{1}) \times \mathbb{Z}_q[x]/( x^{n/2} - \zeta_{2}),$$
where $\zeta_{1} + \zeta_{2}=1$ and $ \zeta_{1} \cdot \zeta_{2}=1$. In their instantiation, they choose $\zeta_{1} = \psi_{n}^{n/6}$ and $ \zeta_{2} = \zeta_{1}^5$.

As for its forward transform, ${\boldsymbol{a}} \in \mathbb{Z}_q[x]/(x^n - x^{n/2} + 1) $ from the $0$-th level generates its images $ {\boldsymbol{a}}_{l} = {\boldsymbol{a}} \bmod x^{n/2} - \zeta_{1} $  and $ {\boldsymbol{a}}_{r} = {\boldsymbol{a}} \bmod x^{n/2} -( 1 - \zeta_{1}) $ in $\mathbb{Z}_q[x]/( x^{n/2} - \zeta_{1}) $ and $ \mathbb{Z}_q[x]/( x^{n/2} - \zeta_{2}) $ respectively in the first level, by using the fact that $\zeta_{2}=1-\zeta_{1}$. In order to get the coefficients, one can compute ${{a}}_{l,i}={{a}}_{i} + \zeta_{1}{{a}}_{i+n/2},{{a}}_{r,i}={{a}}_{i} + {{a}}_{i+n/2} - \zeta_{1}{{a}}_{i+n/2},i=0,1,\ldots,n/2-1$. Different from radix-2 Cooley-Tukey algorithm, there are extra $n/2$ additions in this case. These additional additions don't cost much. For a fast NTT algorithm, one can continue with the similar radix-2 ($\log \frac{n}{3}$)-level FFT trick in $\mathbb{Z}_q[x]/( x^{n/2} - \zeta_{1}) $ and $ \mathbb{Z}_q[x]/( x^{n/2} - \zeta_{2}) $, as in the power-of-two cyclotomic rings above, until the leaf nodes are of the form $\mathbb{Z}_q[x]/(x^3 - \psi_{n}^{j})$ instead of linear terms. The inverse transform can be obtained by inverting the trick mentioned above, where Gentleman-Sande butterflies are used in the radix-2 steps. The point-wise multiplication is performed about the corresponding polynomials of degree 2 in each $\mathbb{Z}_q[x]/(x^3 - \psi_{n}^{j})$. Detailedly, the CRT map can be described as follows:
\begin{equation*}
	\mathbb{Z}_{q}[x]/(x^{n} - x^{n/2} + 1) \  \cong  \prod\limits_{ j \in \mathbb{Z}_{n}^{\times} }{ \mathbb{Z}_{q}[x]/(x^{3} - \psi_{n}^{j}) },
\end{equation*}
where $\mathbb{Z}_{n}^{\times}$ is the group of invertible elements of $\mathbb{Z}_{n}$.

\subsection{Splitting Polynomial Ring over $\mathbb{Z}_q[x]/(x^n - x^{n/2} + 1)$}

The method based on splitting polynomial ring can be generalized to the ring $\mathbb{Z}_q[x]/(x^n - x^{n/2} + 1)$ where $n=3\cdot 2^e$ and $q$ is a prime number, based on which Liang et al.~\cite{ntt-inscrypt20} proposed a generalized, modular and parallelizable NTT method referred to as Generalized 3-NTT (G3-NTT for simplicity). Similarly, let  $\alpha,\beta$ be non-negative integer. The general $\alpha$-round G3-NTT with $\beta$ levels cropped is essentially based on the following isomorphism.
\begin{align*}
		\begin{split}
			\Psi_{\alpha,3}: \mathbb{Z}_q[x]/(x^n - x^{n/2} + 1) \  &\cong  \  \left( \mathbb{Z}_q[y]/(y^{\frac{n}{3 \cdot 2^\alpha}} - y^{\frac{n}{3 \cdot 2^{\alpha+1}}} + 1) \right)  [x] / (x^{3 \cdot 2^\alpha} - y ) \\
			\boldsymbol{a}=\sum\limits_{i=0}^{n-1}{a_i x^i} \  &\mapsto   \  \Psi_{\alpha,3}(\boldsymbol{a})= \sum\limits_{i=0}^{3 \cdot 2^\alpha-1}{ \left(  \sum\limits_{j=0}^{\frac{n}{3 \cdot 2^\alpha}-1}{a_{3 \cdot 2^\alpha \cdot j + i } y^j} \right)  x^i}
		\end{split}
\end{align*}
where $ y^{\frac{n}{3 \cdot 2^\alpha}} - y^{\frac{n}{3 \cdot 2^{\alpha+1}}} + 1$ is  the $\frac{n}{2^\alpha}$-th cyclotomic polynomial of degree $\frac{n}{3 \cdot 2^\alpha}$. Similar to NTTRU~\cite{nttru}, there is a CRT map as follows: $\mathbb{Z}_q[y]/(y^{\frac{n}{3 \cdot 2^\alpha}} - y^{\frac{n}{3 \cdot 2^{\alpha+1}}} + 1) \cong \mathbb{Z}_q[y]/(y^{\frac{n}{3 \cdot 2^{\alpha+1}}}  -\zeta_{1} ) \times \mathbb{Z}_q[y]/(y^{\frac{n}{3 \cdot 2^{\alpha+1}}} - \zeta_{2})$ where $\zeta_{1} + \zeta_{2}=1, \zeta_{1} \cdot \zeta_{2}=1$. It turns out that radix-2 truncated-NTTs can be performed in $\mathbb{Z}_q[y]/(y^{\frac{n}{3 \cdot 2^{\alpha+1}}}  -\zeta_{1} ) $ and $ \mathbb{Z}_q[y]/(y^{\frac{n}{3 \cdot 2^{\alpha+1}}} - \zeta_{2})$. If there are $\beta$ levels to be cropped, $\beta=0,1,\ldots,\log \frac{n}{3 \cdot 2^\alpha} -1 $, the modulus $q$ can be chosen to satisfy only $q \equiv 1 \ (\bmod \frac{n}{2^{\alpha+\beta}})$ such that the primitive $\frac{n}{2^{\alpha+\beta}}$-th root of unity $\psi_{n/2^{\alpha+\beta}}$ exits. The leaf nodes of CRT tree map are degree-$2^\beta$ polynomials, e.g., $\mathbb{Z}_q[y]/(y^{2^{\beta}} - \psi_{n/2^{\alpha+\beta}} ) $. They choose $\zeta_{1} = \psi_{n/2^{\alpha+\beta}}^{n/(6 \cdot 2^{\alpha+\beta})}$ and $\zeta_{2} = \zeta_{1}^5$. One-iteration Karatsuba algorithm can be used in a same way as H-NTT. Given appropriate and fixed $(n,q)$, the computational complexity of G3-NTT can reach its optimization if $\alpha=0,\beta=0$~\cite{ntt-inscrypt20}.

\section{Some Skills}

Those lattice-based schemes fully utilizes the advantages of NTT mentioned in section~\ref{sec-ntt-advantages} to improve their efficiency. 

As for MLWE-based schemes such as Kyber, Dilithium, etc, at a high level, their basic operations can be described as matrix-vector polynomial multiplication $\mathbf{{A}}^{T} \mathbf{{r}}$ and vector-vector polynomial multiplication $\mathbf{{s}}^{T}\mathbf{{u}}$, where $\mathbf{{A}} \in \mathcal{R}_q^{k \times k}, \mathbf{{r}}, \mathbf{{s}}, \mathbf{{u}} \in \mathcal{R}_q^{k \times 1}$. Those schemes directly generate the public key term $\hat{\mathbf{A}}$ already in the NTT domain by rejection sampling, instead of generating ${\mathbf{A}}$ followed by applying forward transform on each element. It can save $k^2$ forward transforms. The linearity of NTT can lead to $\text{INTT}\left( \hat{\mathbf{A}}^{T} \circ  \text{NTT}(\mathbf{{r}} ) \right) $ where there are only $k$ forward transforms and $k$ inverse transforms. Once  the forward transform result $\hat{\mathbf{s}}$ is computed, $\hat{\mathbf{s}}$ can be stored or transmitted without any extra requirement of storage, for following use in multiple polynomial multiplications. Using $\hat{\mathbf{s}}$ that is stored or transmitted in advance, one can compute $\mathbf{{s}}^{T} \mathbf{{u}}$ by $\text{INTT}\left( \hat{ \mathbf{{s}}}^{T} \circ  \text{NTT}(\mathbf{{u}}) \right) $, within only $k$ forward transforms and $k$ inverse transforms.

As for MLWR-based schemes, the linearity of NTT is helpful in the NTT-based implementation of MLWR-based schemes such as Saber. The number of forward transforms and inverse transforms in Saber's matrix-vector multiplication $\mathbf{{A}} \mathbf{{s}} $ can be reduced from $2k^2$ and $k^2$ to $k^2+k$ and $k$, respectively, while the number of inverse transforms $\mathbf{{b}}^{T} \mathbf{{s}} $ in vector-vector multiplication can be reduced from $k$ to $1$, where $\mathbf{{A}} \in \mathcal{R}_q^{k \times k}, \mathbf{{b}}, \mathbf{{s}}\in \mathcal{R}_q^{k \times 1}$.

\section{Radix-2 Fast Number Theoretic Transform From FFT Perspectives}\label{sec-radix-2-ntt-from-fft-perspectives}

In this section, we will describe NTT algorithms from FFT perspectives. We follow the same notations and make the same requirements on parameters as in section~\ref{sec-ntt-definition}. The basic principle of fast NTT algorithms is using ``divide and conquer'' skill to divide the $n$-point NTT into two $n/2$-point NTTs, based on the periodicity and symmetry of the primitive root of unity. This section will first introduce the properties of the primitive roots of unity, and then the radix-2 fast NTT algorithms. The properties of primitive roots of unity in NTT are similar to those of twiddle factors in FFT. The primitive $n$-th root of unity $\omega_n$ in $\mathbb{Z}_q$ has the following properties:
\begin{equation}\label{equ-root-property}
\begin{split}
\text{periodicity:   }  \omega_n^{k+n}=    \omega_n^{k}  \\ 
\text{symmetry:   }  \omega_n^{k+n/2}=  -\omega_n^{k}
\end{split}
\end{equation}
where $k$ is a non-negative integer. It is trivial that the primitive $2n$-th root of unity  $\psi_{2n}$ shares the similar properties if $\psi_{2n}$ exists.

\subsection{Cooley-Tukey Algorithm for CC-based NTT}\label{sec-cooley-tukey-algorithm}

\subsubsection{Cooley-Tukey algorithm for $\mathsf{NTT}$}

The fast algorithms of $\mathsf{NTT}$ are introduced first. Based on the parity of the indexes of the coefficients $a_i$ in ${\boldsymbol{a}}$, the terms in the summation in formula (\ref{equ-classic-ntt}) can be separated into two parts, for $j=0,1,\ldots,n-1$:
\begin{align*} 
\begin{split}
\hat{{a}}_{j} &= \sum_{i=0}^{n/2-1}{{a}_{2i} \omega_n^{2ij}} +  \sum_{i=0}^{n/2-1}{{a}_{2i+1} \omega_n^{(2i+1)j}}\bmod q \\
&= \sum_{i=0}^{n/2-1}{{a}_{2i} (\omega_n^2)^{ij}} +  \omega_n^{j} \sum_{i=0}^{n/2-1}{{a}_{2i+1} (\omega_n^2)^{ij}}\bmod q.
\end{split}
\end{align*}

Based on the periodicity and symmetry of the primitive root of unity in fomula (\ref{equ-root-property}), for $j=0,1,\ldots,n/2-1$, we get: 
\begin{align}\label{equ-classic-ntt-ct-init}
\begin{split}
\hat{{a}}_{j} =& \sum_{i=0}^{n/2-1}{{a}_{2i} (\omega_n^2)^{ij}} +  \omega_n^{j} \sum_{i=0}^{n/2-1}{{a}_{2i+1} (\omega_n^2)^{ij}}\bmod q\\
\hat{{a}}_{j+n/2} = &\sum_{i=0}^{n/2-1}{{a}_{2i} (\omega_n^2)^{ij}} -  \omega_n^{j} \sum_{i=0}^{n/2-1}{{a}_{2i+1} (\omega_n^2)^{ij}}\bmod q.
\end{split}
\end{align}

Let $\hat{{a}}_{j}'=\sum\limits_{i=0}^{n/2-1}{{a}_{2i} (\omega_n^2)^{ij}} \bmod q $, $\hat{{a}}_{j}''=\sum\limits_{i=0}^{n/2-1}{{a}_{2i+1} (\omega_n^2)^{ij}} \bmod q, j=0,1,\ldots,n/2-1$. Formula  (\ref{equ-classic-ntt-ct-init}) can be rewritten as:
\begin{align}\label{equ-classic-ntt-ct-iteration}
\begin{split}
\hat{{a}}_{j} &=     \hat{{a}}_{j}' +  \omega_n^{j}  \hat{{a}}_{j}'' \bmod q            ,\\               
\hat{{a}}_{j+n/2} &=  \hat{{a}}_{j}' -  \omega_n^{j}  \hat{{a}}_{j}'' \bmod q, j=0,1,\ldots,n/2-1.
\end{split}
\end{align}

One can learn from the definition of $\mathsf{NTT}$ that, $\hat{{a}}_{j}'$ and $\hat{{a}}_{j}''$ can be computed by $n/2$-point $\mathsf{NTT}$, from the even-indexed and the odd-indexed coefficients of $\boldsymbol{a}$, respectively. Formula (\ref{equ-classic-ntt-ct-iteration}) shows that, the original $n$-point $\mathsf{NTT}$ can be divided into two  $n/2$-point $\mathsf{NTT}$s by ``divide-and-conquer'' method. After getting the $n/2$-point $\mathsf{NTT}$ results $\hat{{a}}_{j}'$ and $\hat{{a}}_{j}''$, the original $n$-point $\hat{\boldsymbol{a}}$ can be easily achieved by multiplying $\omega_n^{j}$ and simple additions/subtractions. This kind of ``divide and conquer'' skill can also be applied to compute $\hat{{a}}_{j}'$ and $\hat{{a}}_{j}''$. Since $n$ is a power of two, it can be separated down to 2-point $\mathsf{NTT}$s. Such fast $\mathsf{NTT}$ algorithm was first proposed by Cooley and Tukey~\cite{ct65}, and named as radix-2 Cooley-Tukey $\mathsf{NTT}$ algorithm, or radix-2 CT $\mathsf{NTT}$ algorithm for short. By the term of FFT, it is also called radix-2 decimation-in-time $\mathsf{NTT}$, or CT decimation-in-time $\mathsf{NTT}$. The process of deriving $\hat{{a}}_{j}$ and $\hat{{a}}_{j+n/2}$ from $\hat{{a}}_{j}'$ and $\hat{{a}}_{j}''$ is named Cooley-Tukey butterfly, or CT butterfly for short, which is illustrated in Figure~\ref{fig-ct-gs-butterfly-a}. 

Note that, in the CT $\mathsf{NTT}$ derived from formula (\ref{equ-classic-ntt-ct-iteration}), the coefficients of the input polynomials are indexed under bit-reversed order, while the coefficients of the output polynomials are indexed under natural order. In this paper, we follow the notations as used in~\cite{intt-merge-psi-POG15} which denotes this kind of CT $\mathsf{NTT}$ by $\mathsf{NTT}_{bo \rightarrow no}^{CT}$ where the subscripts $bo \rightarrow no$ indicates the input coefficients are under bit-reversed order and output coefficients are under natural order. The signal flow of $\mathsf{NTT}_{bo \rightarrow no}^{CT}$ for $n=8$ can be seen in Figure~\ref{fig-classic-ntt-ct-gs-signal-flow-a} in Appendix~\ref{sec-signal-flow}. Adjust the input to natural order, then the Cooley-Tukey butterflies in the signal flow is changed elsewhere, as in Figure~\ref{fig-classic-ntt-ct-gs-signal-flow-c}. The output will be under bit-reversed order. This new transform is denoted by $\mathsf{NTT}_{no \rightarrow bo}^{CT}$. Obviously, the two transforms shares the same number of Cooley-Tukey butterflies, and therefore the same complexity.

\subsubsection{Cooley-Tukey algorithm for $\mathsf{INTT}$}

Cooley-Tukey butterfly can also be applied to compute $\mathsf{INTT}$. In contrast to $\mathsf{NTT}$, there is extra multiplications by a scale factor $n^{-1}$ in $\mathsf{INTT}$.

With neglecting $n^{-1}$, the terms in the summation of formula (\ref{equ-classic-intt}) are the same as those in formula (\ref{equ-classic-ntt}), except replacing $\omega_n$ with $\omega_n^{-1}$. Therefore, the procedure of applying Cooley-Tukey butterfly to compute $\mathsf{INTT}$ is basically the same as that in the $\mathsf{NTT}$ case, except replacing $\omega_n$ with $\omega_n^{-1}$ in each step. For the convenience of understanding, its brief computing process is given here. The terms in the summation of formula (\ref{equ-classic-intt}) are divided into two parts based on the parity of the index of $\hat{{a}}_j$. That is, for $i=0,1,\ldots,n-1$,

\begin{equation}\label{equ-classic-intt-ct-init}
{{a}}_{i} =  \sum_{j=0}^{n/2-1}{\hat{a}_{2j} (\omega_n^2)^{-ij}} +  \omega_n^{-i} \sum_{j=0}^{n/2-1}{\hat{a}_{2j+1} (\omega_n^2)^{-ij}}   \bmod q .
\end{equation}

Let ${{a}}_{i}'= \sum\limits_{j=0}^{n/2-1}{\hat{a}_{2j} (\omega_n^2)^{-ij}} \bmod q $, ${{a}}_{i}''=\sum\limits_{j=0}^{n/2-1}{\hat{a}_{2j+1} (\omega_n^2)^{-ij}}  \bmod q, i=0,1,\ldots,n/2-1$. Formula (\ref{equ-classic-intt-ct-init}) can be rewritten as:

\begin{align}\label{equ-classic-intt-ct-iteration}
\begin{split}
{{a}}_{i} =     & {{a}}_{i}' +  \omega_n^{-i}  {{a}}_{i}'' \bmod q                         \\
{{a}}_{i+n/2} = & {{a}}_{i}' -  \omega_n^{-i}  {{a}}_{i}'' \bmod q, i=0,1,\ldots,n/2-1.
\end{split}
\end{align}

Therefore, the computing of $n$-point $\mathsf{INTT}$ can be divided into two $n/2$-point $\mathsf{INTT}$s via formula (\ref{equ-classic-intt-ct-iteration}), which can be done down to 2-point $\mathsf{INTT}$s finally. This kind of $\mathsf{INTT}$ is called radix-2 Cooley-Tukey $\mathsf{INTT}$ algorithm, or radix-2 decimation-in-time $\mathsf{INTT}$. In this paper it is denoted by $\mathsf{INTT}_{bo \rightarrow no}^{CT}$.  Its signal flow for $n=8$ is illustrated in Figure~\ref{fig-classic-intt-ct-gs-signal-flow-a}. Adjust its input to natural order, and then the output is changed under bit-reversed order. This new transform is denoted by $\mathsf{INTT}_{no \rightarrow bo}^{CT}$, the signal flow of which for $n=8$ can be seen in Figure~\ref{fig-classic-intt-ct-gs-signal-flow-c}.

\subsection{Gentlemen-Sande Algorithm for CC-based NTT}\label{sec-gentlemen-sande-algorithm}

\subsubsection{Gentlemen-Sande algorithm for $\mathsf{NTT}$}

Gentlemen-Sande algorithm separates the coefficients of ${\boldsymbol{a}}$ into the upper half and the lower half. Specifically, for $j=0,1,\ldots,n-1$, the formula (\ref{equ-classic-ntt}) is separated into: 
\begin{align}\label{equ-classic-ntt-gs-split}
\begin{split}
\hat{{a}}_{j} &= \sum_{i=0}^{\frac{n}{2}-1}{{a}_{i} \omega_n^{ij}} +  \sum_{i=\frac{n}{2}}^{n-1}{{a}_{i} \omega_n^{ij}}\bmod q \\
&= \sum_{i=0}^{\frac{n}{2}-1}{{a}_{i} \omega_n^{ij}} +   \sum_{i=0}^{\frac{n}{2}-1}{{a}_{i+\frac{n}{2}} \omega_n^{(i+\frac{n}{2})j}}\bmod q.
\end{split}
\end{align}

Based on the periodicity and symmetry of the primitive root of unity (see formula (\ref{equ-root-property})), the terms $\hat{{a}}_{j}$ continue to be dealt with according to the parity of index $j$. For $j=0,1,\ldots,\frac{n}{2}-1$:
\begin{align*}
\begin{split}
\hat{{a}}_{2j} &= \sum_{i=0}^{\frac{n}{2}-1}{{a}_{i} \omega_n^{2ij}} + (-1)^{2j} \cdot \sum_{i=0}^{\frac{n}{2}-1}{{a}_{i+\frac{n}{2}} \omega_n^{2ij}}\bmod q \\
&= \sum_{i=0}^{\frac{n}{2}-1}{  ({a}_{i} + {a}_{i+\frac{n}{2}})  (\omega_n^2)^{ij}}  \bmod q, \\
\hat{{a}}_{2j+1} &= \sum_{i=0}^{\frac{n}{2}-1}{{a}_{i} \omega_n^{(2j+1)i}} + (-1)^{2j+1} \cdot \sum_{i=0}^{\frac{n}{2}-1}{{a}_{i+\frac{n}{2}} \omega_n^{(2j+1)i}}\bmod q \\
&= \sum_{i=0}^{\frac{n}{2}-1}{ \left[ ({a}_{i} - {a}_{i+\frac{n}{2}}) \cdot \omega_n^i \right] (\omega_n^2)^{ij}}  \bmod q .\\
\end{split}
\end{align*}

Let ${{b}}_{i}'= {a}_{i} + {a}_{i+\frac{n}{2}} \bmod q $, ${{b}}_{i}''=  ({a}_{i} - {a}_{i+\frac{n}{2}}) \cdot \omega_n^i  \bmod q, i=0,1,\ldots, \frac{n}{2} -1$. The above formula can be rewritten as:

\begin{align}\label{equ-classic-ntt-gs-iteration}
\begin{split}
\hat{{a}}_{2j} &=    \sum_{i=0}^{\frac{n}{2}-1}{  {{b}}_{i}'  (\omega_n^2)^{ij}} \bmod q            , \\                    
\hat{{a}}_{2j+1} &=  \sum_{i=0}^{\frac{n}{2}-1}{ {{b}}_{i}'' (\omega_n^2)^{ij}}  \bmod q, j=0,1,\ldots, \frac{n}{2} -1.
\end{split}
\end{align}

One can learn from the definition of $\mathsf{NTT}$ that, formula (\ref{equ-classic-ntt-gs-iteration}) is exact the $n/2$-point $\mathsf{NTT}$s with respect to ${{b}}_{i}',{{b}}_{i}'',i=0,1,\ldots, \frac{n}{2} -1$. Thus, after deriving ${{b}}_{i}',{{b}}_{i}'',i=0,1,\ldots, \frac{n}{2} -1$ from  $a_i,i=0,1,\ldots ,n-1$, the original $n$-point $\mathsf{NTT}$ with respect to ${a}_{i}$ is transformed into  $n/2$-point $\mathsf{NTT}$s with respect to ${{b}}_{i}',{{b}}_{i}''$. Similarly, the $n/2$-point $\mathsf{NTT}$ in formula (\ref{equ-classic-ntt-gs-iteration}) can be tranformed into $n/4$-point $\mathsf{NTT}$s, and down to 2-point $\mathsf{NTT}$s. This kind of fast algorithm was first proposed by Gentlemen and Sande~\cite{gs66}, and named as radix-2 Gentlemen-Sande $\mathsf{NTT}$ algorithm, or radix-2 GS $\mathsf{NTT}$ algorithm for short, or radix-2 decimation-in-frequency $\mathsf{NTT}$, or else GS decimation-in-frequency $\mathsf{NTT}$. The process of deriving ${{b}}_{i}'$ and ${{b}}_{i}''$ from ${a}_{i}$ and ${a}_{i+\frac{n}{2}}$ is referred to as Gentlemen-Sande butterfly, or GS butterfly for short (see in Figure~\ref{fig-ct-gs-butterfly-b}). Such GS $\mathsf{NTT}$ algorithm is denoted by $\mathsf{NTT}_{no \rightarrow bo}^{GS}$. Its signal flow for $n=8$ is shown in Figure~\ref{fig-classic-ntt-ct-gs-signal-flow-d}. Adjust the input to bit-reversed order and it outputs under natural order, as shown in Figure~\ref{fig-classic-ntt-ct-gs-signal-flow-b}. The new transform is denoted by $\mathsf{NTT}_{bo \rightarrow no}^{GS}$.

\subsubsection{Gentlemen-Sande algorithm for $\mathsf{INTT}$}

Gentleman-Sande butterfly can be similarly applied to compute $\mathsf{INTT}$, by neglecting $n^{-1}$ and replacing $\omega_n$ with $\omega_n^{-1}$ in $\mathsf{NTT}$. That is, for $i=0,1,\ldots,\frac{n}{2}-1$,
\begin{align}\label{equ-classic-intt-gs-init}
\begin{split}
{{a}}_{2i} =& \sum_{j=0}^{\frac{n}{2}-1}{  (\hat{a}_{j} + \hat{a}_{j+\frac{n}{2}})  (\omega_n^2)^{-ij}} \bmod q \\
{{a}}_{2i+1}=& \sum_{i=0}^{\frac{n}{2}-1}{ \left[ (\hat{a}_{j} - \hat{a}_{j+\frac{n}{2}}) \cdot \omega_n^{-j} \right] (\omega_n^2)^{-ij}} \bmod q .
\end{split}
\end{align}

Let ${\hat{b}}_{j}'= \hat{a}_{j} + \hat{a}_{j+\frac{n}{2}} \bmod q $, ${\hat{b}}_{j}''=  (\hat{a}_{j} - \hat{a}_{j+\frac{n}{2}}) \cdot \omega_n^{-j}  \bmod q, j=0,1,\ldots, \frac{n}{2} -1$. Formula (\ref{equ-classic-intt-gs-init}) can be rewritten as: 
\begin{align}\label{equ-classic-intt-gs-iteration}
\begin{split}
{{a}}_{2i} =   & \sum_{j=0}^{\frac{n}{2}-1}{ \hat{{b}}_{j}'  (\omega_n^2)^{-ij}} \bmod q                                \\
{{a}}_{2i+1} = & \sum_{j=0}^{\frac{n}{2}-1}{\hat{{b}}_{j}'' (\omega_n^2)^{-ij}}  \bmod q, i=0,1,\ldots, \frac{n}{2} -1.
\end{split}
\end{align}

Similarly, $n$-point $\mathsf{INTT}$ can be divided into two $n/2$-point $\mathsf{INTT}$s according to formula (\ref{equ-classic-intt-gs-iteration}), and down to 2-point $\mathsf{INTT}$s. This kind of $\mathsf{INTT}$ is called radix-2 Gentlemen-Sande $\mathsf{INTT}$ algorithm, or radix-2 decimation-in-frequency $\mathsf{INTT}$, which is denoted by $\mathsf{INTT}_{bo \rightarrow no}^{GS}$. Its signal flow for $n=8$ is shown in Figure~\ref{fig-classic-intt-ct-gs-signal-flow-d}. Adjust its input to natural order, and then the output will be under bit-reversed order. This new transform is denoted by $\mathsf{INTT}_{no \rightarrow bo}^{GS}$, the signal flow of which for $n=8$ is shown in Figure~\ref{fig-classic-intt-ct-gs-signal-flow-b}.

\subsection{Radix-2 Fast NWC-based NTT}

In this paper, the process of multiplying the coefficients by $\psi_{2n}^{i}$ before forward transform in formula (\ref{equ-classic-ntt-psi}) is referred to as pre-processing, while the process of multiplying the coefficients by $\psi_{2n}^{-i}$ after inverse transform in formula (\ref{equ-classic-intt-psi})  is referred to as post-processing. 

Fast algorithms for negative wrapped convolution-based (NWC-based) NTT such as $\mathsf{NTT}^{\psi}$ and $\mathsf{INTT}^{\psi^{-1}}$, can be constructed by using radix-2 CT/GS $\mathsf{NTT}$/$\mathsf{INTT}$ algorithm with pre-processing and post-processing, according to formula (\ref{equ-classic-ntt-psi})(\ref{equ-classic-intt-psi}). However, such construction requires extra point-wise multiplication with $\boldsymbol{\psi}$/$\boldsymbol{\psi}^{-1}$ besides radix-2 algorithms, resulting with $n$ extra multiplications. In fact, these additional multiplications are not necessary. Roy et al.~\cite{ntt-merge-psi-RVM14} integrate  the pre-processing about $\boldsymbol{\psi}$ into $\mathsf{NTT}_{bo \rightarrow no}^{CT}$. P$\ddot{\text{o}}$ppelmann et al.~\cite{intt-merge-psi-POG15} integrate the post-processing about $\boldsymbol{\psi}^{-1}$ into $\mathsf{INTT}_{bo \rightarrow no}^{GS}$. Furthermore, Zhang et al.~\cite{newhope-zhangneng20} integrate $\boldsymbol{\psi}^{-1}$ and $n^{-1}$ into  $\mathsf{INTT}_{bo \rightarrow no}^{GS}$. They will be  introduced ad follows.

\subsubsection{Cooley-Tukey algorithm for $\mathsf{NTT}^{\psi}$}

Roy et al.~\cite{ntt-merge-psi-RVM14} take advantage of Cooley-Tukey algorithm. According to the definition of  $\mathsf{NTT}^{\psi}$, the coefficients of $\hat{\boldsymbol{a}}$ can be written as:
\begin{equation}\label{equ-classic-ntt-psi-coefficient}
\hat{{a}}_{j} = \sum_{i=0}^{n-1}{{a}_{i} \psi_{2n}^{i} \omega_n^{ij}} \bmod q , j=0,1,\ldots,n-1.
\end{equation}

Divide the summation into two parts based on the parity of the index of $a_i$, and for $j=0,1,\ldots,n-1$, we get:
\begin{align}
\begin{split}
\hat{{a}}_{j} =& \sum_{i=0}^{n/2-1}{{a}_{2i} \omega_n^{2ij} \psi_{2n}^{2i}} +  \sum_{i=0}^{n/2-1}{{a}_{2i+1} \omega_n^{(2i+1)j} \psi_{2n}^{2i+1} }\bmod q \\
=& \sum_{i=0}^{n/2-1}{{a}_{2i} (\omega_n^2)^{ij}} (\psi_{2n}^2)^{i}  +  \omega_n^{j} \psi_{2n} \sum_{i=0}^{n/2-1}{{a}_{2i+1} (\omega_n^2)^{ij} (\psi_{2n}^2)^{i} }\bmod q.
\end{split}
\end{align}

Based on the periodicity and symmetry of the primitive root of unity, for $j=0,1,\ldots,n/2-1$, we get 
\begin{align}
\begin{split}
\hat{{a}}_{j} =& \sum_{i=0}^{n/2-1}{{a}_{2i} (\omega_n^2)^{ij}} (\psi_{2n}^2)^{i}  +  \omega_n^{j} \psi_{2n} \sum_{i=0}^{n/2-1}{{a}_{2i+1} (\omega_n^2)^{ij} (\psi_{2n}^2)^{i} }\bmod q\\
\hat{{a}}_{j+n/2} = & \sum_{i=0}^{n/2-1}{{a}_{2i} (\omega_n^2)^{ij}} (\psi_{2n}^2)^{i}  -  \omega_n^{j} \psi_{2n} \sum_{i=0}^{n/2-1}{{a}_{2i+1} (\omega_n^2)^{ij} (\psi_{2n}^2)^{i} }\bmod q.
\end{split}
\end{align}

Let $\hat{{a}}_{j}'=\sum\limits_{i=0}^{n/2-1}{{a}_{2i} (\omega_n^2)^{ij}} (\psi_{2n}^2)^{i} \bmod q $, $\hat{{a}}_{j}''= \sum\limits_{i=0}^{n/2-1}{{a}_{2i+1} (\omega_n^2)^{ij} (\psi_{2n}^2)^{i} }$ $\bmod q, j=0,1,\ldots,n/2-1$. With $\omega_n^{j} \psi_{2n} = \psi_{2n}^{2j+1}$, the above formula can be rewritten as, for $j=0,1,\ldots,n/2-1$:
\begin{align}\label{equ-classic-ntt-psi-ct-iteration}
\begin{split}
\hat{{a}}_{j} =      \hat{{a}}_{j}' +  \psi_{2n}^{2j+1}  \hat{{a}}_{j}'' \bmod q       , \ \                  
\hat{{a}}_{j+n/2} =  \hat{{a}}_{j}' -  \psi_{2n}^{2j+1}  \hat{{a}}_{j}'' \bmod q.
\end{split}
\end{align}

One can see that, $\hat{{a}}_{j}'$ and $\hat{{a}}_{j}''$ can be obtained via exact $n/2$-point $\mathsf{NTT}^{\psi}$s. The following analysis is the same as that of radix-2 CT $\mathsf{NTT}$ in section~\ref{sec-cooley-tukey-algorithm}, so for briefness the redundant analysis is omitted here. Such fast algorithm of  $\mathsf{NTT}^{\psi}$ is called radix-2 CT $\mathsf{NTT}^{\psi}$, and  is denoted by $\mathsf{NTT}_{bo \rightarrow no}^{CT,\psi}$. Its signal flow for $n=8$ is shown in Figure~\ref{fig-classic-ntt-intt-psi-signal-flow-a}. Adjust the input and output order, and we get $\mathsf{NTT}_{no \rightarrow bo}^{CT,\psi}$. Its signal flow for $n=8$ is shown in Figure~\ref{fig-classic-ntt-intt-psi-signal-flow-c}.

\subsubsection{Gentleman-Sande algorithm for $\mathsf{INTT}^{\psi^{-1}}$}

P$\ddot{\text{o}}$ppelmann et al.~\cite{intt-merge-psi-POG15} progress the integration with $\boldsymbol{\psi}^{-1}$ by using Gentleman-Sande algorithm. According to the definition of $\mathsf{INTT}^{\psi^{-1}}$, the coefficients of $\boldsymbol{a}$ can be written as:
\begin{equation}\label{equ-classic-intt-psi-coefficient}
{{a}}_{i} = n^{-1}  \psi_{2n}^{-i} \sum_{j=0}^{n-1}{\hat{a}_{j} \omega_n^{-ij}} \bmod q , i=0,1,\ldots,n-1.
\end{equation}

With neglecting $n^{-1}$, the summation can be divided into the upper half and the lower half with respect to the index of $\hat{a}_{j}$. For $i=0,1,\ldots,n-1$,
\begin{align*}
\begin{split}
{{a}}_{i} &= \psi_{2n}^{-i} \left( \sum_{j=0}^{\frac{n}{2}-1}{\hat{a}_{j} \omega_n^{-ij}} +  \sum_{j=\frac{n}{2}}^{n-1}{\hat{a}_{j} \omega_n^{-ij}}   \right) \\
&= \psi_{2n}^{-i}  \left[   \sum_{j=0}^{\frac{n}{2}-1}{\hat{a}_{j} \omega_n^{ij}} +   \sum_{j=0}^{\frac{n}{2}-1}{\hat{a}_{j+\frac{n}{2}} \omega_n^{-i(j+\frac{n}{2})}}  \right] \bmod q.
\end{split}
\end{align*}

Based on the periodicity and symmetry of the primitive root of unity, for $ i=0,1,\ldots,\frac{n}{2}-1$ we have
\begin{align*}
\begin{split}
{{a}}_{2i}	=&  \psi_{2n}^{-2i} \left[ \sum_{j=0}^{\frac{n}{2}-1}{\hat{a}_{j} \omega_n^{-2ij}} + (-1)^{2i} \cdot \sum_{j=0}^{\frac{n}{2}-1}{\hat{a}_{j+\frac{n}{2}} \omega_n^{-2ij}}  \right] \bmod q \\
=& (\psi_{2n}^{2})^{-i}    \sum_{j=0}^{\frac{n}{2}-1}{  (\hat{a}_{j} + \hat{a}_{j+\frac{n}{2}})  (\omega_n^2)^{-ij}}   \bmod q , \\
{{a}}_{2i+1} =& \psi_{2n}^{-(2i+1)} \left[   \sum_{j=0}^{\frac{n}{2}-1}{\hat{a}_{j} \omega_n^{-(2i+1)j}} + (-1)^{2i+1} \cdot \sum_{j=0}^{\frac{n}{2}-1}{\hat{a}_{j+\frac{n}{2}} \omega_n^{-(2i+1)j}}  \right] \bmod q \\
=&(\psi_{2n}^{2})^{-i}   \sum_{j=0}^{\frac{n}{2}-1}{ \left[ (\hat{a}_{j} - \hat{a}_{j+\frac{n}{2}}) \cdot \omega_n^{-j} \psi_{2n}^{-1}  \right] (\omega_n^2)^{-ij}}    \bmod q .\\
\end{split}
\end{align*}

Since $\omega_n^{-j} \psi_{2n}^{-1} = \psi_{2n}^{-(2j+1)}$, letting  ${\hat{b}}_{j}'= \hat{a}_{j} + \hat{a}_{j+\frac{n}{2}} \bmod q $, ${\hat{b}}_{j}''=  (\hat{a}_{j} - \hat{a}_{j+\frac{n}{2}}) \cdot \psi_{2n}^{-(2j+1)}  \bmod q, j=0,1,\ldots, \frac{n}{2} -1$, the above formula can be rewritten as, for $i=0,1,\ldots, \frac{n}{2} -1$:
\begin{align}\label{equ-classic-intt-psi-gs-iteration}
\begin{split}
{{a}}_{2i} =    (\psi_{2n}^{2})^{-i} \sum_{j=0}^{\frac{n}{2}-1}{ \hat{{b}}_{j}'  (\omega_n^2)^{-ij}} \bmod q   , \\                          
{{a}}_{2i+1} =  (\psi_{2n}^{2})^{-i} \sum_{j=0}^{\frac{n}{2}-1}{\hat{{b}}_{j}'' (\omega_n^2)^{-ij}}  \bmod q.
\end{split}
\end{align}

Similar to the analysis in section~\ref{sec-gentlemen-sande-algorithm}, computing $n$-point $\mathsf{INTT}^{\psi^{-1}}$ can be transformed into two $n/2$-point $\mathsf{INTT}^{\psi^{-1}}$s with respect to $\hat{b}_{j}',\hat{b}_{j}''$. Such fast algorithm for $\mathsf{INTT}^{\psi^{-1}}$ is named as radix-2 GS $\mathsf{INTT}^{\psi^{-1}}$, and denoted by $\mathsf{INTT}_{no \rightarrow bo}^{GS,\psi^{-1}}$. Its signal flow for $n=8$ is shown in Figure~\ref{fig-classic-ntt-intt-psi-signal-flow-d}. Adjust the input/output order and get $\mathsf{INTT}_{bo \rightarrow no}^{GS,\psi^{-1}}$ whose signal flow for $n=8$ is shown in Figure~\ref{fig-classic-ntt-intt-psi-signal-flow-b}.

Zhang et al.~\cite{newhope-zhangneng20} noticed that $\boldsymbol{\psi}^{-1}$ and $n^{-1}$ can both be integrated into  $\mathsf{INTT}_{bo \rightarrow no}^{GS}$. Thus, $n^{-1}$ is no longer neglected, and one can learn from formula (\ref{equ-classic-intt-psi-coefficient}), for $i=0,1,\ldots, \frac{n}{2} -1$:
\begin{align}
\begin{split}
{{a}}_{2i} =  (\frac{n}{2})^{-1}  (\psi_{2n}^{2})^{-i} \sum_{j=0}^{\frac{n}{2}-1}{ \hat{{b}}_{j}'  (\omega_n^2)^{-ij}} \bmod q , \\                                
{{a}}_{2i+1} = (\frac{n}{2})^{-1}  (\psi_{2n}^{2})^{-i} \sum_{j=0}^{\frac{n}{2}-1}{\hat{{b}}_{j}'' (\omega_n^2)^{-ij}}  \bmod q.
\end{split}
\end{align}
where ${\hat{b}}_{j}'= ( \hat{a}_{j} + \hat{a}_{j+\frac{n}{2}}) / 2 \bmod q $, ${\hat{b}}_{j}''=  ( \hat{a}_{j} - \hat{a}_{j+\frac{n}{2}} )  /2 \cdot \psi_{2n}^{-(2j+1)}  \bmod q, j=0,1,\ldots, \frac{n}{2} -1$. Different from formula (\ref{equ-classic-intt-psi-gs-iteration}), when computing $\hat{b}_{j}'$ and $\hat{b}_{j}''$ , the scale factor 2 will be dealt with directly, by using addition and displacement (i.e., ``$>>$'') to compute $x/2 \bmod q$. When $x$ is even, $x/2 \equiv (x>>1) \bmod q$. When $x$ is odd, $x/2 \equiv (x>>1) + (q+1)/2 \bmod q$.

\end{document}

%% file: table/tab-raidx-2-fast-ntt.tex
\begin{table}[H]
	\centering
	\caption{Radix-2 fast NTT algorithms}
	\label{tab-raidx-2-fast-ntt}
	\tabcolsep 25pt 
	\begin{tabular}{ccc}
		\hline
		        Transforms          &                                   Cooley-Tukey algorithm                                   &                                       Gentleman-Sande algorithm                                        \\ \hline
		      $\mathsf{NTT}$        &      $\mathsf{NTT}_{bo \rightarrow no}^{CT}$, $\mathsf{NTT}_{no \rightarrow bo}^{CT}$      &            $\mathsf{NTT}_{bo \rightarrow no}^{GS}$, $\mathsf{NTT}_{no \rightarrow bo}^{GS}$            \\
		      $\mathsf{INTT}$       &     $\mathsf{INTT}_{bo \rightarrow no}^{CT}$, $\mathsf{INTT}_{no \rightarrow bo}^{CT}$     &           $\mathsf{INTT}_{bo \rightarrow no}^{GS}$, $\mathsf{INTT}_{no \rightarrow bo}^{GS}$           \\
		   $\mathsf{NTT}^{\psi}$    & $\mathsf{NTT}_{bo \rightarrow no}^{CT,\psi}$, $\mathsf{NTT}_{no \rightarrow bo}^{CT,\psi}$ &                                          \diagbox[dir=SW]{}{}                                          \\
		$\mathsf{INTT}^{\psi^{-1}}$ &                                    \diagbox[dir=SW]{}{}                                    & $\mathsf{INTT}_{bo \rightarrow no}^{GS,\psi^{-1}}$, $\mathsf{INTT}_{no \rightarrow bo}^{GS,\psi^{-1}}$ \\ \hline
	\end{tabular}
\end{table}

%% file: table/tab-ntt-complexity.tex
\begin{table}[H]
	\centering
	\caption{Multiplication complexities of NTT algorithms. $\ddag,\natural \in \{ \text{CT,GS}\} $. $\boldsymbol{\psi}_{bo}$ and $\boldsymbol{\psi}^{-1}_{bo}$ mean that the coefficients are under bit-reversed order. $n$ is the length of NTT.}
	\label{tab-ntt-complexity}
	\tabcolsep 15pt 
	\begin{tabular}{cc}
		\hline
		                                                                                        {NTT algorithms}                                                                                         & {Multiplication complexities} \\ \hline
		                                                      $\mathsf{NTT}$, $\mathsf{INTT}$, $\mathsf{NTT}^{\psi}$, $\mathsf{INTT}^{\psi^{-1}}$                                                        &           $O(n^2)$            \\
		    $\mathsf{NTT}^{\natural}_{no \rightarrow bo}$, $\mathsf{NTT}^{\natural}_{bo \rightarrow no}$, $\mathsf{NTT}^{CT,\psi}_{no \rightarrow bo}$, $\mathsf{NTT}^{CT,\psi}_{bo \rightarrow no}$     &    $\frac{1}{2} n \log n$     \\
		$\mathsf{INTT}^{\ddag}_{no \rightarrow bo}$, $\mathsf{INTT}^{\ddag}_{bo \rightarrow no}$, $\mathsf{INTT}^{GS,\psi^{-1}}_{no \rightarrow bo}$, $\mathsf{INTT}^{GS,\psi^{-1}}_{bo \rightarrow no}$ &  $\frac{1}{2} n  \log n + n$  \\
		                       $\mathsf{NTT}^{\natural}_{no \rightarrow bo} \circ \boldsymbol{\psi}$, $\mathsf{NTT}^{\natural}_{bo \rightarrow no} \circ \boldsymbol{\psi}_{bo} $                        &  $\frac{1}{2} n  \log n + n$  \\
		                   $\boldsymbol{\psi}^{-1}_{bo} \circ  \mathsf{INTT}^{\ddag}_{no \rightarrow bo}$, $\boldsymbol{\psi}^{-1} \circ  \mathsf{INTT}^{\ddag}_{bo \rightarrow no}$                     & $\frac{1}{2} n  \log n + 2n$  \\ \hline
	\end{tabular}
\end{table}

%% file: figure/weaken-restriction-methods.tex
\usetikzlibrary{arrows,positioning,shapes.multipart}

\begin{figure}[!t]
	\center
	\begin{tikzpicture}[scale=0.7]


\node[fill = white!20,rounded corners,draw = white] (left) at (0,0) {weaken NTT parameter conditions};

\node[fill = white!20,rounded corners,draw = white] (2a) at (5,3) {};
\node[fill = white!20,rounded corners,draw = white, right = 0.05 cm of 2a] {incomplete FFT trick \cite{kyber-nist-round3,ntt-inscrypt20,newhope-compact,nttru,ntt-in-saber-CHK+21}};

\node[fill = white!20,rounded corners,draw = white] (2b) at (5,0) {};
\node[fill = white!20,rounded corners,draw = white, right = 0.05 cm of 2b] {splitting polynomial ring \cite{ptntt-inscrypt18,ptntt-eprint19,kntt-icics21,ntt-inscrypt20}};

\node[fill = white!20,rounded corners,draw = white] (2c) at (5,-3) {};
\node[fill = white!20,rounded corners,draw = white, right = 2.15 cm of 2c] (3left) {};
\node[fill = white!20,rounded corners,draw = white, right = 0.05 cm of 2c] (2large) {large modulus};

\draw [black, solid, semithick, >=stealth, ->] (left.east) -- (2a.west);
\draw [black, solid, semithick, >=stealth, ->] (left.east) -- (2b.west);
\draw [black, solid, semithick, >=stealth, ->] (left.east) -- (2c.west);


\node[fill = white!20,rounded corners,draw = white] (3a) at (9.5,-1.5) {};
\node[fill = white!20,rounded corners,draw = white, right = 0.05 cm of 3a] {NTT-friendly large prime \cite{ntt-in-saber-CHK+21,ntt-in-saber-FSS20,ntt-in-saber-FVR+21}};

\node[fill = white!20,rounded corners,draw = white] (3b) at (9.5,-3) {};
\node[fill = white!20,rounded corners,draw = white, right = 0.05 cm of 3b] {residue number system \cite{ntt-in-saber-CHK+21,ntt-in-saber-FSS20,multi-moduli-NTT-ACC+21}};

\node[fill = white!20,rounded corners,draw = white] (3c) at (9.5,-4.5) {};
\node[fill = white!20,rounded corners,draw = white, right = 0.05 cm of 3c] {composite-modulus ring \cite{multi-moduli-NTT-ACC+21}};

\draw [black, solid, semithick, >=stealth, ->] (3left.east) -- (3a.west);
\draw [black, solid, semithick, >=stealth, ->] (3left.east) -- (3b.west);
\draw [black, solid, semithick, >=stealth, ->] (3left.east) -- (3c.west);

\end{tikzpicture}
	\vspace{2ex}
	\caption{Overview of methods to weaken NTT parameter conditions}
	\label{fig-weaken-restriction-methods}
\end{figure}

%% file: table/tab-parameter-in-cryptosystem.tex
\begin{table*}[h]
	\footnotesize
	\centering
	\setlength{\tabcolsep}{1mm}
	\caption{Parameter sets of algebraically-structural lattice-based schemes in NIST PQC. Recommended parameter sets of NTRU Prime are given here. Kyber KEM, Dilithium signature and Falcon signature are standardized by NIST~\cite{nist-to-be-standardized}. Saber KEM and NTRU KEM were NIST PQC Round 3 finalists. NTRU Prime KEM was a alternate candidate in NIST PQC Round 3.}
	\label{tab-parameter-in-cryptosystem}
	\begin{tabular*}{\textwidth}{ccccccc}
		\hline
		\multicolumn{2}{ c }{ {Schemes}}                                                                      &           $n$           &   $q$   &                    Rings                     &                               Types                                &                                                                                                                                       Methods \& Algorithms                                                                                                                                       \\ \hline
		  \multirow{2}{*}{Kyber}    &                         Round 1~\cite{kyber-nist-round1}                         &           256           &  7681   &  \multirow{2}{*}{$\mathbb{Z}_q[x]/(x^n+1)$}  &      \makecell[c]{NTT-friendly\\$q \equiv 1 \ (\bmod \ 2n)$}       &                                                                                                                $n$-point full NWC-based NTT~\cite{kyber-nist-round1,kyber-BDK+18}                                                                                                                 \\ \cline{2-4}\cline{6-7}
		                            & \makecell[c]{Round 2~\cite{kyber-nist-round2}\\Round 3~\cite{kyber-nist-round3}} &           256           &  3329   &                                              &       \makecell[c]{NTT-friendly\\$q \equiv 1 \ (\bmod \ n)$}       &                                                                          \makecell[c]{Incomplete FFT trick~\cite{kyber-nist-round2,kyber-nist-round3}\\Splitting polynomial ring~\cite{ntt-inscrypt20,ptntt-inscrypt18}}                                                                          \\ \hline
		         Dilithium          &                       Round 3~\cite{dilithium-nist-round3}                       &           256           & 8380417 &          $\mathbb{Z}_q[x]/(x^n+1)$           &      \makecell[c]{NTT-friendly\\$q \equiv 1 \ (\bmod \ 2n)$}       &                                                                                                                     $n$-point full NWC-based NTT~\cite{dilithium-nist-round3}                                                                                                                     \\ \hline
		          Falcon            &                        Round 3~\cite{falcon-nist-round3}                         & \makecell[c]{512\\1024} &  12289  &          $\mathbb{Z}_q[x]/(x^n+1)$           &      \makecell[c]{NTT-friendly\\$q \equiv 1 \ (\bmod \ 2n)$}       &                                                                                                                      $n$-point full NWC-based NTT~\cite{falcon-nist-round3}                                                                                                                       \\ \hline
		           Saber            &                         Round 3~\cite{saber-nist-round3}                         &           256           &  8192   &          $\mathbb{Z}_q[x]/(x^n+1)$           &           \makecell[c]{NTT-unfriendly\\power-of-two $q$}           &                                                                                      Method based on large modulus~\cite{ntt-in-saber-CHK+21,ntt-in-saber-FSS20,ntt-in-saber-FVR+21,multi-moduli-NTT-ACC+21}                                                                                      \\ \hline
		   \multirow{3}{*}{NTRU}    &                 \multirow{3}{*}{Round 3~\cite{ntru-nist-round3}}                 & \makecell[c]{509\\677}  &  2048   &  \multirow{3}{*}{$\mathbb{Z}_q[x]/(x^n-1)$}  &    \multirow{3}{*}{ \makecell[c]{NTT-unfriendly\\prime $n$}  }     &                                                   \multirow{3}{*}{ \tabincell{c}{Power-of-two $n'$ + Method based on large modulus~\cite{ntt-in-saber-FVR+21}\\Good's trick (+ Method based on large modulus)~\cite{ntt-in-saber-CHK+21} }   }                                                    \\ \cline{3-4}
		                            &                                                                                  &           701           &  8192   &                                              &                                                                    &                                                                                                                                                                                                                                                                                                   \\ \cline{3-4}
		                            &                                                                                  &           821           &  4096   &                                              &                                                                    &                                                                                                                                                                                                                                                                                                   \\ \hline
		\multirow{3}{*}{NTRU Prime} &              \multirow{3}{*}{Round 3~\cite{ntru-prime-nist-round3}}              &           653           &  4621   & \multirow{3}{*}{$\mathbb{Z}_q[x]/(x^n-x-1)$} & \multirow{3}{*}{ \makecell[c]{NTT-unfriendly\\prime $n$ and $q$} } & \multirow{3}{*}{ \tabincell{c}{Power-of-two $n'$ + Method based on large modulus~\cite{ntt-in-ntru-ACC+21}\\Good's trick (+ Method based on large modulus)~\cite{ntt-in-ntru-ACC+21,ntt-in-ntru-prime-PMT21}\\Sch$\ddot{\text{o}}$nhage’s trick + Nussbaumer’s trick~\cite{ntt-in-ntru-BBC21}}  } \\ \cline{3-4}
		                            &                                                                                  &           761           &  4591   &                                              &                                                                    &                                                                                                                                                                                                                                                                                                   \\ \cline{3-4}
		                            &                                                                                  &           857           &  5167   &                                              &                                                                    &                                                                                                                                                                                                                                                                                                   \\ \hline
	\end{tabular*}
\end{table*}

%% file: NTT-survey-ACM.bbl
\newcommand{\etalchar}[1]{$^{#1}$}
\begin{thebibliography}{BCLvV17}

\bibitem[AA16]{mlwr-AA16}
Jacob Alperin{-}Sheriff and Daniel Apon.
\newblock Dimension-preserving reductions from {LWE} to {LWR}.
\newblock {\em {IACR} Cryptol. ePrint Arch.}, page 589, 2016.

\bibitem[AB75]{ntt-AB75}
Ramesh Agarwal and Sidney Burrus.
\newblock Number theoretic transforms to implement fast digital convolution.
\newblock {\em Proceedings of the IEEE}, 63(4):550--560, 1975.

\bibitem[ABC19]{newhope-compact}
Erdem Alkim, Yusuf~Alper Bilgin, and Murat Cenk.
\newblock Compact and simple {RLWE} based key encapsulation mechanism.
\newblock In {\em {LATINCRYPT} 2019}, volume 11774, pages 237--256. Springer,
  2019.

\bibitem[ABD{\etalchar{+}}17]{kyber-nist-round1}
Roberto Avanzi, Joppe Bos, L{\'e}o Ducas, Eike Kiltz, Tancr{\`e}de Lepoint,
  Vadim Lyubashevsky, John~M Schanck, Peter Schwabe, Gregor Seiler, and Damien
  Stehl{\'e}.
\newblock Supporting documentation: Crystals-kyber: Algorithm specifications
  and supporting documentation (version 1.0).
\newblock {\em NIST PQC}, 2017.

\bibitem[ABD{\etalchar{+}}19]{kyber-nist-round2}
Roberto Avanzi, Joppe Bos, L{\'e}o Ducas, Eike Kiltz, Tancr{\`e}de Lepoint,
  Vadim Lyubashevsky, John~M Schanck, Peter Schwabe, Gregor Seiler, and Damien
  Stehl{\'e}.
\newblock Supporting documentation: Crystals-kyber: Algorithm specifications
  and supporting documentation (version 2.0).
\newblock {\em NIST PQC}, 2019.

\bibitem[ABD{\etalchar{+}}20]{kyber-nist-round3}
Roberto Avanzi, Joppe Bos, L{\'e}o Ducas, Eike Kiltz, Tancr{\`e}de Lepoint,
  Vadim Lyubashevsky, John~M. Schanck, Peter Schwabe, Gregor Seiler, and Damien
  Stehl{\'e}.
\newblock Supporting documentation: Crystals-kyber: Algorithm specifications
  and supporting documentation (version 3.0).
\newblock {\em NIST PQC}, 2020.

\bibitem[ACC{\etalchar{+}}21]{ntt-in-ntru-ACC+21}
Erdem Alkim, Dean~Yun{-}Li Cheng, Chi{-}Ming~Marvin Chung, H{\"{u}}lya Evkan,
  Leo~Wei{-}Lun Huang, Vincent Hwang, Ching{-}Lin~Trista Li, Ruben Niederhagen,
  Cheng{-}Jhih Shih, Julian W{\"{a}}lde, and Bo{-}Yin Yang.
\newblock Polynomial multiplication in {NTRU} prime comparison of optimization
  strategies on cortex-m4.
\newblock {\em TCHES}, 2021(1):217--238, 2021.

\bibitem[ACC{\etalchar{+}}22]{multi-moduli-NTT-ACC+21}
Amin Abdulrahman, Jiun{-}Peng Chen, Yu{-}Jia Chen, Vincent Hwang, Matthias~J.
  Kannwischer, and Bo{-}Yin Yang.
\newblock Multi-moduli ntts for saber on cortex-m3 and cortex-m4.
\newblock {\em TCHES}, 2022(1):127--151, 2022.

\bibitem[BBC{\etalchar{+}}20]{ntru-prime-nist-round3}
Daniel~J. Bernstein, Billy~Bob Brumley, Ming-Shing Chen, Chitchanok
  Chuengsatiansup, Tanja Lange, Adrian Marotzke, Bo-Yuan Peng, Nicola Tuveri,
  Christine van Vredendaal, and Bo-Yin Yang.
\newblock Ntru prime: round 3.
\newblock {\em NIST PQC}, 2020.

\bibitem[BBCT22]{ntt-in-ntru-BBC21}
Daniel~J. Bernstein, Billy~Bob Brumley, Ming{-}Shing Chen, and Nicola Tuveri.
\newblock {OpenSSLNTRU}: Faster post-quantum {TLS} key exchange.
\newblock In {\em USENIX Security 2022}, 2022.

\bibitem[BCLvV17]{ntru-prime-BCLV17}
Daniel~J. Bernstein, Chitchanok Chuengsatiansup, Tanja Lange, and Christine van
  Vredendaal.
\newblock {NTRU} prime: Reducing attack surface at low cost.
\newblock In {\em {SAC} 2017}, volume 10719, pages 235--260. Springer, 2017.

\bibitem[BDK{\etalchar{+}}18]{kyber-BDK+18}
Joppe~W. Bos, L{\'{e}}o Ducas, Eike Kiltz, Tancr{\`{e}}de Lepoint, Vadim
  Lyubashevsky, John~M. Schanck, Peter Schwabe, Gregor Seiler, and Damien
  Stehl{\'{e}}.
\newblock {CRYSTALS} - kyber: {A} cca-secure module-lattice-based {KEM}.
\newblock In {\em EuroS{\&}P 2018}, pages 353--367, 2018.

\bibitem[BDK{\etalchar{+}}20]{dilithium-nist-round3}
Shi Bai, Leo Ducas, Eike Kiltz, Tancrède Lepoint, Vadim Lyubashevsky, Peter
  Schwabe, Gregor Seiler, and Damien Stehlé.
\newblock Supporting documentation: Crystals-dilithium: Algorithm
  specifications and supporting documentation.
\newblock {\em NIST PQC}, 2020.

\bibitem[Ber01]{ber01}
Daniel~J. Bernstein.
\newblock Multidigit multiplication for mathematicians.
\newblock \url{http://cr.yp.to/papers.html\#m3}, 2001.

\bibitem[BGV14]{fhe-bgv12}
Zvika Brakerski, Craig Gentry, and Vinod Vaikuntanathan.
\newblock (leveled) fully homomorphic encryption without bootstrapping.
\newblock {\em {ACM} Trans. Comput. Theory}, 6(3):13:1--13:36, 2014.

\bibitem[BMD{\etalchar{+}}20]{saber-nist-round3}
Andrea Basso, Jose Maria~Bermudo Mera, Jan-Pieter D'Anvers, Angshuman Karmakar,
  Sujoy~Sinha Roy, Michiel~Van Beirendonck, and Frederik Vercauteren.
\newblock Supporting documentation: Saber: Mod-lwr based kem (round 3
  submission).
\newblock {\em NIST PQC}, 2020.

\bibitem[BPR12]{lwr-BPR12}
Abhishek Banerjee, Chris Peikert, and Alon Rosen.
\newblock Pseudorandom functions and lattices.
\newblock In {\em {EUROCRYPT} 2012}, volume 7237, pages 719--737, 2012.

\bibitem[Bra12]{fhe-bra12}
Zvika Brakerski.
\newblock Fully homomorphic encryption without modulus switching from classical
  gapsvp.
\newblock In {\em {CRYPTO} 2012}, volume 7417, pages 868--886, 2012.

\bibitem[CA69]{cook69}
S.~A. Cook and S.~O. Aanderaa.
\newblock On the minimum computation time of functions.
\newblock In {\em Transactions of the American Mathematical Society}, volume
  142, pages 291--314, 1969.

\bibitem[CCF67]{sk-fft}
W.~T. Cochran, J.~W. Cooley, and D.~L. Favin.
\newblock What is the fast fourier transform?
\newblock In {\em {IEEE} Proc.}, volume~55, pages 1664--1674, 1967.

\bibitem[CDH{\etalchar{+}}20]{ntru-nist-round3}
Cong Chen, Oussama Danba, Jeffrey Hoffstein, Andreas Hulsing, Joost Rijneveld,
  John~M. Schanck, and Tsunekazu Saito.
\newblock Ntru submission.
\newblock {\em NIST PQC}, 2020.

\bibitem[CG99]{fft-book-chu00}
Eleanor Chu and Alan George.
\newblock {\em Inside the FFT black box: serial and parallel fast Fourier
  transform algorithms}.
\newblock CRC press, 1999.

\bibitem[CHK{\etalchar{+}}21]{ntt-in-saber-CHK+21}
Chi{-}Ming~Marvin Chung, Vincent Hwang, Matthias~J. Kannwischer, Gregor Seiler,
  Cheng-Jhih Shih, and Bo-Yin Yang.
\newblock {NTT} multiplication for ntt-unfriendly rings new speed records for
  saber and {NTRU} on cortex-m4 and {AVX2}.
\newblock {\em TCHES}, 2021(2):159--188, 2021.

\bibitem[CT65]{ct65}
James~W Cooley and John~W Tukey.
\newblock An algorithm for the machine calculation of complex fourier series.
\newblock {\em Mathematics of computation}, 19(90):297--301, 1965.

\bibitem[DKL{\etalchar{+}}18]{dilithium-CHES18}
L{\'{e}}o Ducas, Eike Kiltz, Tancr{\`{e}}de Lepoint, Vadim Lyubashevsky, Peter
  Schwabe, Gregor Seiler, and Damien Stehl{\'{e}}.
\newblock Crystals-dilithium: {A} lattice-based digital signature scheme.
\newblock {\em TCHES}, 2018(1):238--268, 2018.

\bibitem[DKRV18]{saber-DKR+18}
Jan{-}Pieter D'Anvers, Angshuman Karmakar, Sujoy~Sinha Roy, and Frederik
  Vercauteren.
\newblock Saber: Module-lwr based key exchange, cpa-secure encryption and
  cca-secure {KEM}.
\newblock In {\em {AFRICACRYPT} 2018}, volume 10831, pages 282--305, 2018.

\bibitem[FBR{\etalchar{+}}22]{ntt-in-saber-FVR+21}
Tim Fritzmann, Michiel~Van Beirendonck, Debapriya~Basu Roy, Patrick Karl,
  Thomas Schamberger, Ingrid Verbauwhede, and Georg Sigl.
\newblock Masked accelerators and instruction set extensions for post-quantum
  cryptography.
\newblock {\em TCHES}, 2022(1):414--460, 2022.

\bibitem[FHK{\etalchar{+}}20]{falcon-nist-round3}
Pierre-Alain Fouque, Jeffrey Hoffstein, Paul Kirchner, Vadim Lyubashevsky,
  Thomas Pornin, Thomas Prest, Thomas Ricosset, Gregor Seiler, William Whyte,
  and Zhenfei Zhang.
\newblock Falcon: Fast-fourier lattice-based compact signatures over ntru.
\newblock {\em NIST PQC}, 2020.

\bibitem[FO99]{fo-transform-FO99}
Eiichiro Fujisaki and Tatsuaki Okamoto.
\newblock Secure integration of asymmetric and symmetric encryption schemes.
\newblock In {\em {CRYPTO} 1999}, volume 1666, pages 537--554. Springer, 1999.

\bibitem[FO13]{fo-transform-FO13}
Eiichiro Fujisaki and Tatsuaki Okamoto.
\newblock Secure integration of asymmetric and symmetric encryption schemes.
\newblock {\em J. Cryptol.}, 26(1):80--101, 2013.

\bibitem[FSS20]{ntt-in-saber-FSS20}
Tim Fritzmann, Georg Sigl, and Johanna Sep{\'{u}}lveda.
\newblock {RISQ-V:} tightly coupled {RISC-V} accelerators for post-quantum
  cryptography.
\newblock {\em TCHES}, 2020(4):239--280, 2020.

\bibitem[F{\"{u}}r09]{faster-integer-mul-Fur09}
Martin F{\"{u}}rer.
\newblock Faster integer multiplication.
\newblock {\em {SIAM} J. Comput.}, 39(3):979--1005, 2009.

\bibitem[FV12]{fhe-fv12}
Junfeng Fan and Frederik Vercauteren.
\newblock Somewhat practical fully homomorphic encryption.
\newblock {\em {IACR} Cryptol. ePrint Arch.}, 2012:144, 2012.

\bibitem[Gau05]{gauss1805}
Carl~Friedrich Gauss.
\newblock Theoria interpolationis methodo nova tractata.
\newblock 1805.

\bibitem[Goo51]{good-trick51}
Irving~J Good.
\newblock Random motion on a finite abelian group.
\newblock In {\em Mathematical proceedings of the cambridge philosophical
  society}, volume~47, pages 756--762. Cambridge University Press, 1951.

\bibitem[GS66]{gs66}
W.~Morven Gentleman and G.~Sande.
\newblock Fast fourier transforms: for fun and profit.
\newblock In {\em {AFIPS} '66}, volume~29, pages 563--578, 1966.

\bibitem[HHK17]{fo-transform-HHK17}
Dennis Hofheinz, Kathrin H{\"{o}}velmanns, and Eike Kiltz.
\newblock A modular analysis of the fujisaki-okamoto transformation.
\newblock In {\em {TCC} 2017}, volume 10677, pages 341--371. Springer, 2017.

\bibitem[HPA21]{sca-sok-ct-rsa-HPA21}
James Howe, Thomas Prest, and Daniel Apon.
\newblock Sok: How (not) to design and implement post-quantum cryptography.
\newblock In {\em {CT-RSA} 2021}, volume 12704, pages 444--477. Springer, 2021.

\bibitem[HPS98]{ntru-HPS98}
Jeffrey Hoffstein, Jill Pipher, and Joseph~H. Silverman.
\newblock {NTRU:} {A} ring-based public key cryptosystem.
\newblock In {\em ANTS-III}, volume 1423, pages 267--288. Springer, 1998.

\bibitem[Knu14]{schoolbook}
Donald~E Knuth.
\newblock {\em Art of computer programming, volume 2: Seminumerical
  algorithms}.
\newblock Addison-Wesley Professional, 2014.

\bibitem[KO62]{karatsuba62}
Anatolii~Alekseevich Karatsuba and Yu~Ofman.
\newblock Multiplication of many-digital numbers by automatic computers.
\newblock {\em Doklady Akademii Nauk}, 145(2):293--294, 1962.

\bibitem[Koc96]{sca-timing-attack-Kocher96}
Paul~C. Kocher.
\newblock Timing attacks on implementations of diffie-hellman, rsa, dss, and
  other systems.
\newblock In {\em {CRYPTO} 1996}, volume 1109, pages 104--113. Springer, 1996.

\bibitem[LPR10]{rlwe-LPR10}
Vadim Lyubashevsky, Chris Peikert, and Oded Regev.
\newblock On ideal lattices and learning with errors over rings.
\newblock In {\em {EUROCRYPT} 2010}, volume 6110, pages 1--23, 2010.

\bibitem[LS15]{mlwe-LS15}
Adeline Langlois and Damien Stehl{\'{e}}.
\newblock Worst-case to average-case reductions for module lattices.
\newblock {\em Des. Codes Cryptogr.}, 75(3):565--599, 2015.

\bibitem[LS19]{nttru}
Vadim Lyubashevsky and Gregor Seiler.
\newblock {NTTRU:} truly fast {NTRU} using {NTT}.
\newblock {\em TCHES}, 2019(3):180--201, 2019.

\bibitem[LSS{\etalchar{+}}20]{ntt-inscrypt20}
Zhichuang Liang, Shiyu Shen, Yuantao Shi, Dongni Sun, Chongxuan Zhang, Guoyun
  Zhang, Yunlei Zhao, and Zhixiang Zhao.
\newblock Number theoretic transform: Generalization, optimization, concrete
  analysis and applications.
\newblock In {\em Inscrypt 2020}, volume 12612, pages 415--432, 2020.

\bibitem[Moe76]{incomplete-ntt-moenck76}
Robert~T. Moenck.
\newblock Practical fast polynomial multiplication.
\newblock In {\em {SYMSAC} 1976}, pages 136--148. {ACM}, 1976.

\bibitem[NDR{\etalchar{+}}19]{pqc-implementations-a-survey-NDRRB19}
Hamid Nejatollahi, Nikil~D. Dutt, Sandip Ray, Francesco Regazzoni, Indranil
  Banerjee, and Rosario Cammarota.
\newblock Post-quantum lattice-based cryptography implementations: {A} survey.
\newblock {\em {ACM} Comput. Surv.}, 51(6):129:1--129:41, 2019.

\bibitem[NIS16]{nist-round-1-submissions}
NIST.
\newblock Post-quantum cryptography, round 1 submissions.
\newblock
  \url{https://csrc.nist.gov/Projects/Post-Quantum-Cryptography/round-1-submissions},
  2016.

\bibitem[NIS19]{nist-round-2-submissions}
NIST.
\newblock Post-quantum cryptography, round 2 submissions.
\newblock
  \url{https://csrc.nist.gov/Projects/Post-Quantum-Cryptography/round-2-submissions},
  2019.

\bibitem[NIS20]{nist-round-3-submissions}
NIST.
\newblock Post-quantum cryptography, round 3 submissions.
\newblock
  \url{https://csrc.nist.gov/Projects/Post-Quantum-Cryptography/round-3-submissions},
  2020.

\bibitem[NIS22]{nist-to-be-standardized}
NIST.
\newblock Pqc standardization process: Announcing four candidates to be
  standardized, plus fourth round candidates.
\newblock
  \url{https://csrc.nist.gov/News/2022/pqc-candidates-to-be-standardized-and-round-4},
  2022.

\bibitem[Nus80]{nus80}
H.~Nussbaumer.
\newblock Fast polynomial transform algorithms for digital convolution.
\newblock {\em {IEEE} Transactions on Acoustics, Speech, and Signal Processing
  1980}, 28(2):205--215, 1980.

\bibitem[PMT{\etalchar{+}}21]{ntt-in-ntru-prime-PMT21}
Bo-Yuan Peng, Adrian Marotzke, Ming{-}Han Tsai, Bo{-}Yin Yang, and Ho{-}Lin
  Chen.
\newblock Streamlined ntru prime on fpga.
\newblock {\em {IACR} Cryptol. ePrint Arch.}, page 1444, 2021.

\bibitem[POG15]{intt-merge-psi-POG15}
Thomas P{\"{o}}ppelmann, Tobias Oder, and Tim G{\"{u}}neysu.
\newblock High-performance ideal lattice-based cryptography on 8-bit atxmega
  microcontrollers.
\newblock In {\em {LATINCRYPT} 2015}, volume 9230, pages 346--365, 2015.

\bibitem[Pol71]{ntt-Pol71}
John~M Pollard.
\newblock The fast fourier transform in a finite field.
\newblock {\em Mathematics of computation}, 25(114):365--374, 1971.

\bibitem[Reg09]{lwe-regev09}
Oded Regev.
\newblock On lattices, learning with errors, random linear codes, and
  cryptography.
\newblock {\em J. {ACM}}, 56(6):34:1--34:40, 2009.

\bibitem[RPBC20]{sca-ntt-RPBC20}
Prasanna Ravi, Romain Poussier, Shivam Bhasin, and Anupam Chattopadhyay.
\newblock On configurable {SCA} countermeasures against single trace attacks
  for the {NTT}.
\newblock In {\em {SPACE} 2020}, volume 12586, pages 123--146. Springer, 2020.

\bibitem[RVM{\etalchar{+}}14]{ntt-merge-psi-RVM14}
Sujoy~Sinha Roy, Frederik Vercauteren, Nele Mentens, Donald~Donglong Chen, and
  Ingrid Verbauwhede.
\newblock Compact ring-lwe cryptoprocessor.
\newblock In {\em {CHES} 2014}, volume 8731, pages 371--391, 2014.

\bibitem[Sch77]{schonhage-trick77}
Arnold Sch{\"{o}}nhage.
\newblock Schnelle multiplikation von polynomen {\"{u}}ber k{\"{o}}rpern der
  charakteristik 2.
\newblock {\em Acta Informatica}, 7:395--398, 1977.

\bibitem[Sei18]{seiler18}
Gregor Seiler.
\newblock Faster {AVX2} optimized {NTT} multiplication for ring-lwe lattice
  cryptography.
\newblock {\em {IACR} Cryptol. ePrint Arch.}, 2018:39, 2018.

\bibitem[SZS80]{ntt-book-sun80}
Qi~Sun, Desun Zheng, and Zhongqi Shen.
\newblock {\em Fast Number Theoretic Transform}.
\newblock China Science Press, 1980.

\bibitem[Too63]{toom63}
L.~Toom.
\newblock The complexity of a scheme of functional elements realizing the
  multiplication of integers.
\newblock {\em Soviet Mathematics Doklady}, 3(4):714--716, 1963.

\bibitem[Was97]{cyclotomic-fields}
Lawrence~C. Washington.
\newblock {\em Introduction to cyclotomic fields}.
\newblock Graduate Texts in Mathematics 83, Springer-Verlag, 1997.

\bibitem[Win76]{wfta}
S.~Winograd.
\newblock On computing the discrete fourier transform.
\newblock {\em National Academy of Sciences}, 73(4):1005--1006, 1976.

\bibitem[Win96]{convolution96}
Franz Winkler.
\newblock {\em Polynomial Algorithms in Computer Algebra}.
\newblock Springer, 1996.

\bibitem[WP06]{karatsuba06}
Andr{\'{e}} Weimerskirch and Christof Paar.
\newblock Generalizations of the karatsuba algorithm for efficient
  implementations.
\newblock {\em {IACR} Cryptol. ePrint Arch.}, 2006:224, 2006.

\bibitem[ZLP19]{ptntt-eprint19}
Yiming Zhu, Zhen Liu, and Yanbin Pan.
\newblock When {NTT} meets karatsuba: Preprocess-then-ntt technique revisited.
\newblock {\em {IACR} Cryptol. ePrint Arch.}, page 1079, 2019.

\bibitem[ZLP21]{kntt-icics21}
Yiming Zhu, Zhen Liu, and Yanbin Pan.
\newblock When {NTT} meets karatsuba: Preprocess-then-ntt technique revisited.
\newblock In {\em {ICICS} 2021}, volume 12919, pages 249--264, 2021.

\bibitem[ZXZ{\etalchar{+}}18]{ptntt-inscrypt18}
Shuai Zhou, Haiyang Xue, Daode Zhang, Kunpeng Wang, Xianhui Lu, Bao Li, and
  Jingnan He.
\newblock Preprocess-then-ntt technique and its applications to kyber and
  newhope.
\newblock In {\em Inscrypt 2018}, volume 11449, pages 117--137, 2018.

\bibitem[ZYC{\etalchar{+}}20]{newhope-zhangneng20}
Neng Zhang, Bohan Yang, Chen Chen, Shouyi Yin, Shaojun Wei, and Leibo Liu.
\newblock Highly efficient architecture of newhope-nist on {FPGA} using
  low-complexity {NTT/INTT}.
\newblock {\em TCHES}, 2020(2):49--72, 2020.

\end{thebibliography}
